\documentclass[12pt,twoside,fleqn]{article}
\usepackage{epsfig}
\usepackage{cite}
\def\lsim{\raise0.3ex\hbox{$<$\kern-0.75em\raise-1.1ex\hbox{$\sim$}}}
\def\gsim{\raise0.3ex\hbox{$>$\kern-0.75em\raise-1.1ex\hbox{$\sim$}}}
\makeatletter
\@addtoreset{equation}{section}
\@addtoreset{figure}{section}
\makeatother

\newcommand{\bq}{\begin{equation}}
\newcommand{\eq}{\end{equation}}
\newcommand{\bqa}{\begin{eqnarray}}
\newcommand{\eqa}{\end{eqnarray}}
\newcommand{\bqas}{\begin{eqnarray*}}
\newcommand{\eqas}{\end{eqnarray*}}
\newcommand{\bdm}{\begin{displaymath}}
\newcommand{\edm}{\end{displaymath}}

\newcommand{\bfmut}{\hbox{\boldmath $\mu^2$\unboldmath}}
\begin{document}
\thispagestyle{empty}
\mbox{} \hfill BI-TP 2005/02\\[0mm]
\mbox{} \hfill SWAT/05/422\\
\begin{center}
{\Huge \bf 
Thermodynamics of Two Flavor \\
QCD to Sixth Order in \\[1mm]
Quark Chemical Potential}

\vspace*{1.0cm}
{\large 
\bf C.R.~Allton\rlap,$^{\rm a}$ M. D\"oring\rlap,$^{\rm b}$ 
S.~Ejiri\rlap,$^{\rm b}$
S.J.~Hands\rlap,$^{\rm a}$ O.~Kaczmarek\rlap,$^{\rm b}$
F.~Karsch\rlap,$^{\rm b}$ 
E.~Laermann\rlap,$^{\rm b}$ and
K. Redlich$^{\rm b,c}$
} 
\vspace*{0.6cm}

{\normalsize
$^{\rm a}$ Department of Physics, University of Wales Swansea,
          Singleton Park,\\ Swansea SA2 8PP, U.K. \\
$^{\rm b}$ Fakult\"at f\"ur Physik, 
Universit\"at Bielefeld, D-33615 Bielefeld, Germany. \\
$^{\rm c}$ Institute of Theoretical Physics, University of Wroclaw,\\
PL-50204 Wroclaw, Poland.\\
}
\end{center}
\vspace*{0.6cm}
\centerline{\large ABSTRACT}
\baselineskip 20pt
\noindent
We present results of a simulation of two flavor QCD on a $16^3\times4$
lattice using p4-improved staggered fermions with bare quark mass 
$m/T=0.4$.  Derivatives of the thermodynamic grand canonical partition
function  $Z(V,T,\mu_u,\mu_d)$ with respect to 
chemical potentials $\mu_{u,d}$ for different quark flavors are calculated
up to sixth order, enabling estimates of 
the pressure and the quark number density as well as the chiral condensate 
and various 
susceptibilities as functions of $\mu_q = (\mu_u +\mu_d)/2$ 
via Taylor series expansion. Furthermore, we analyze baryon as well as
isospin fluctuations and discuss the
relation between the radius of convergence of the Taylor series
and the chiral critical point in the QCD phase diagram. We argue that
bulk thermodynamic observables do not, at present, provide direct
evidence for the existence of a chiral critical point in the QCD 
phase diagram.
Results are compared to high temperature perturbation theory as well
as a hadron resonance gas model. 

\vfill
\noindent
\mbox{}PACS numbers: 12.38.Gc, 12.38.Mh\\
\baselineskip 15pt

\section{Introduction}
\label{sec:intro}

The thermodynamics of strongly interacting matter has been studied 
extensively in lattice calculations at vanishing quark chemical
potential, $\mu_q$ (or baryon $\mu_B\equiv 3\mu_q$) \cite{review}. 
Current lattice calculations strongly suggest that the transition from 
the hadronic low temperature phase to the high temperature phase is a 
continuous, non-singular but rapid transition happening in a narrow temperature 
interval around the transition temperature $T_0\simeq 170$~MeV. 
Recent advances in the development of techniques for lattice calculations 
at non-zero quark chemical potential $\mu_q$ 
\cite{Fodor1,us1,Gavai1,Lombardo1,Philipsen1} have also enabled the first
exploratory studies of the QCD phase diagram and bulk thermodynamics in a regime 
of small $\mu_q$, {\it i.e.} for $\mu_q/T\ \lsim \ 1$ and $T\ \gsim \ 0.8\ T_0$.

Guided by phenomenological models which suggest that 
at low temperature and non-zero quark chemical potential 
the low and high density 
regions will be separated by a first order phase transition
it has been speculated \cite{Stephanov}
that a $2^{nd}$ order phase transition point,
the chiral critical point, exists in the interior of the QCD phase diagram, 
at which the line of first order transitions ends. For smaller
values of $\mu_q/T$ the low and high temperature regime will then be only
separated by a crossover transition. The first exploratory studies at
non-zero quark chemical potential indeed gave evidence for  such a chiral
critical point \cite{Fodor1}, although subsequent investigations made clear that at 
present any quantitative statement about the location \cite{Where,Fodor2}
and maybe even about the existence of such a chiral critical point is premature.
The first calculations of the baryonic contribution to the pressure in
strongly interacting matter 
\cite{us1,FKS,us2,Fodoreos,Gavaieos,Lombardoeos} 
also suggested that the transition and the
thermodynamic behaviour in the high-T phase resemble the picture found
previously at zero net baryon density. In this case ($\mu_q=0$) the
transition occurs over a narrow temperature interval. Beyond a region
$T \in [T_0,1.5 T_0]$ where there are large deviations from ideal gas
behaviour, thermodynamic observables rapidly approach the high-T ideal
gas
limit; for instance, the pressure agrees with the Stefan-Boltzmann
prediction to within $\sim 20\%$.
In the low temperature hadronic phase it has been found that a hadron resonance
gas model provides an astonishingly good description of basic features of the 
$T$ and $\mu_q$-dependence of thermodynamic observables \cite{KRT}.   

Maybe one of the largest changes compared to the thermodynamics at 
$\mu_q\equiv 0$ has been observed in the temperature dependence of the
quark number and isovector susceptibilities \cite{us2}. For $\mu_q=0$
these observables show a similar temperature dependence. They have been found 
to change rapidly at the transition temperature but continue to increase 
monotonically at larger temperatures 
\cite{Gottlieb,suscept,Gupta02,suscept2,Gavai04}. 
For $\mu_q > 0$, however,
the quark number susceptibility develops a pronounced peak at the transition 
temperature while the isovector susceptibility continues to show a temperature 
dependence similar to that found at $\mu_q = 0$. Such a behavior, indeed, is
expected to occur in QCD in the vicinity of a $2^{nd}$ order phase transition
point \cite{Kunihiro}. Susceptibilities thus may provide the most direct
evidence for the existence of a $2^{nd}$ order phase transition in the QCD
phase diagram. Finding the characteristic volume and/or quark mass dependent
universal scaling behavior\footnote{The $2^{nd}$ order critical point in the
QCD phase diagram is expected to belong to the universality class of the 
3-dimensional Ising model.} of susceptibilities would undoubtedly establish the
existence of a chiral critical point. 

In this paper we want to extend our previous study of thermodynamics at
non-zero quark chemical potential \cite{us1,us2}, which is based on a 
Taylor series expansion around $\mu_q=0$,  to the $6^{th}$ 
order\footnote{Some results on the radius of convergence
have been reported recently in an $8^{th}$ order Taylor expansion for two flavor
QCD \cite{Gavai04}.}. Going to higher orders in the expansion is of 
particular importance for the analysis of higher derivatives of the 
density of the grand potential expressed in units of the 
temperature\footnote{We will call this in the following the grand potential
although conventionally the extensive quantity $VT^4\Omega$ is called the
grand potential.}, $\Omega (V,T,\mu)\equiv (V T^3)^{-1}\ln Z$, {\it i.e.}
for an analysis of generalized susceptibilities. Accordingly our main
emphasis will be to further analyze the properties of quark number and
isovector susceptibilities, relate them to diagonal and non-diagonal
flavor susceptibilities, calculate the chiral susceptibility and discuss to 
what extent the pronounced peaks
found in some of these susceptibilities give evidence for the existence of a 
chiral critical point in the QCD phase diagram. 

This paper is organized as follows.  We start in the next section by summarizing
basic results on QCD thermodynamics at non-zero quark chemical potential 
obtained in high temperature perturbation theory \cite{Blaizot,Vuorinen} 
and expectations based
on properties of the hadron resonance gas model at low temperature 
\cite{Hagedorn,pbm}.
In section 3 we present results on the calculation of various
thermodynamic observables obtained from a Taylor expansion up to $6^{th}$
order in $\mu_q/T$. In section 4 we analyze the convergence properties 
of the Taylor series. Section 5 is devoted to a discussion of the reweighting
approach to QCD  thermodynamics at $\mu_q \ne 0$ and its comparison to the 
Taylor expansion approach. In section 6 we give our conclusions. An Appendix
contains details on the Taylor expansion of various observables studied here.

\section{Thermodynamics at low and high temperature}
\label{sec:hrgm}

In this section we want to briefly discuss the
density or chemical potential dependence of thermodynamic
observables in the asymptotic high temperature regime as well as at
low temperatures in the hadronic phase of QCD. On the former we have information
from high temperature perturbation theory which
recently has been extended to ${\cal O}(g^6\ln g)$ \cite{Vuorinen} 
also for non-vanishing quark chemical potentials $\mu_f$, $f=u,~d,..$, and 
as for $\mu_f\equiv 0$ \cite{Kajantie} is thus now known to all 
perturbatively calculable orders. Although a comparison
of the perturbative expansion with lattice calculations at $\mu_f =0$  
suggests that quantitative agreement cannot be expected at temperatures
close to the transition temperature, $T_0$, we can gain
useful insight into the structure of the Taylor expansion used in
lattice calculations to study thermodynamics at $\mu_f \ne 0$.

In the low temperature hadronic phase a systematic QCD based analysis
is difficult and one generally has to rely on calculations
within the framework of effective low energy theories or to 
phenomenological approaches. 
Here we will focus on a discussion of the properties of a hadron
resonance gas model, which recently has been compared to lattice results
at non-zero quark chemical potential quite successfully \cite{KRT}
and which also is known to describe experimental results on the 
chemical freeze-out of particle ratios observed in heavy ion collisions
rather well \cite{pbm}.
 
\subsection{High temperature perturbation theory}
\label{subsec:pert}

In the infinite temperature limit the grand potential of QCD\footnote{We
suppress here the volume dependence of $\Omega$. Perturbative
calculations are performed in the thermodynamic limit. The volume
dependence, however, has to be analyzed more carefully in lattice
calculations.},
$\Omega (T,\mu) \equiv \ln {\cal Z}(V,T,\mu) /VT^3$, which is equivalent
to the pressure in units of $T^4$, approaches that of a free 
quark-gluon gas (Stefan-Boltzmann (SB) gas),
\begin{eqnarray}
\frac{p_{SB}}{T^4} = \Omega^{(0)} (T,\mu) =  \frac{8 \pi^2}{45} + 
 \sum_{f=u,d,..} \left[\frac{7 \pi^2}{60} +
\frac{1}{2}  \left(\frac{\mu_f}{T}\right)^2 
+ \frac{1}{4 \pi^2} \left(\frac{\mu_f}{T}\right)^4 
\right] \quad ,
\label{eq:free}
\end{eqnarray}
where the first term gives the contribution of the gluon sector and
the sum over the fermion sector extends over $n_f$ different flavors.
In Eq.~(\ref{eq:free}) we only gave the result for massless quarks and
gluons. Also in the following we will restrict our discussion of
perturbative results to the case of QCD with massless quarks.
In this section we also use $\mu$ to denote the entire set of $n_f$ different
chemical potentials $(\mu_u,~\mu_d,...)$. We also introduce the shorthand 
notation $\bfmut \equiv \sum_{f} \left(\mu_f /  T\right)^2$. 

The additive structure of contributions arising from gluons
and the different fermion flavor sectors persists at ${\cal O}(g^2)$.
Only starting at ${\cal O}(g^3)$ is a coupling among the different partonic 
sectors introduced. At this order it is induced through the non-vanishing 
electric (Debye) mass term $m_E$
\cite{Vuorinen},
\begin{eqnarray}
\Omega (T,\mu) = \Omega^{(0)} (T,\mu) + g^2\; \Omega^{(2)} (T,\mu)
+ g^3\; \Omega^{(3)} (T,\mu) + {\cal O}(g^4)\quad ,
\end{eqnarray}
with
\begin{eqnarray}
\Omega^{(2)} (T,\mu)
\hspace*{-0.2cm}&=&\hspace*{-0.2cm} 
-\left( \frac{1}{6} + 
\frac{5 n_f}{72} + \frac{1}{4 \pi^2} \bfmut
+\frac{1}{8 \pi^4}\sum_{f=u,d,..} \left(\frac{\mu_f}{T}\right)^4 
\right) \quad , \nonumber \\
\Omega^{(3)} (T,\mu)  
\hspace*{-0.2cm}&=&\hspace*{-0.2cm} \frac{1}{6 \pi} \left( 
\frac{m_E}{gT}\right)^3 = 
\frac{1}{6 \pi} \left( 1 +\frac{n_f}{6} + \frac{1}{2 \pi^2} 
\bfmut \right)^{3/2} \; .
\label{pg3}
\end{eqnarray}
The electric mass term introduces
a dependence of a given quark flavor sector on changes in another sector,
{\it i.e.} the quark number density in a flavor sector $\ell$,
\begin{equation}
\frac{n_\ell}{T^3} = \frac{\partial \Omega (T,\mu)}{\partial \mu_\ell/T} \quad ,
\end{equation}
depends on the other quark chemical potentials only at ${\cal O}(g^3)$. 
This is also reflected 
in the structure of diagonal and non-diagonal susceptibilities
\cite{Gavai1,Gottlieb,suscept,Gupta02,suscept2,Gavai04}. 
\begin{equation}
\frac{\chi_{ff}(T,\mu)}{T^2} = 
\frac{\partial^{2} \Omega (T,\mu)}{\partial (\mu_f/T)^2} \quad ,\quad
\frac{\chi_{fk}(T,\mu)}{T^2} = 
\frac{\partial^{2} \Omega (T,\mu)}{\partial (\mu_f/T)
\partial (\mu_k/T)} \quad .
\label{sus}
\end{equation}
The diagonal susceptibilities are non-zero in the ideal gas
limit and, moreover, the leading order perturbative term stays non-zero 
also in the limit of vanishing quark chemical potential,
\begin{eqnarray}
\frac{\chi_{ff}(T,\mu)}{T^2} \hspace{-0.1cm}&=&\hspace{-0.1cm}
1 + \frac{3}{\pi^2}\left(\frac{\mu_f}{T}\right)^2
+ {\cal O}(g^2) \quad . 
\label{sus_lead_d}
\end{eqnarray}
The non-diagonal susceptibilities, however, receive non-zero contributions
only at ${\cal O}(g^3)$. The leading perturbative
contribution is positive and inversely proportional to the electric 
screening mass. However, it vanishes in the limit of vanishing 
chemical potentials. In this case the first non-zero contribution
arises at ${\cal O}(g^6 \ln 1/g)$ \cite{Blaizot},  
\begin{eqnarray}
\frac{\chi_{fk}(T,\mu)}{T^2} \hspace{-0.1cm}&=&\hspace{-0.1cm}
\frac{g^3}{2 \pi^5} \left( 1 +\frac{n_f}{6} + \frac{1}{2 \pi^2}
\bfmut \right)^{-1/2} 
\frac{\mu_\ell}{T} \frac{\mu_k}{T} 
+ {\cal O}(g^4)\quad , \\
\frac{\chi_{fk}(T,0)}{T^2} \hspace{-0.1cm}&\simeq&\hspace{-0.1cm} 
- \frac{5}{144 \pi^6}\; g^6 \ln 1/g \quad .
\label{sus_lead_nd}
\end{eqnarray}

As we are going to discuss lattice calculations at non-zero chemical
potential which are based on a Taylor expansion of the grand potential
$\Omega$ in terms of $\mu_f/T$ it also is instructive to consider a 
Taylor expansion of the perturbative series for $\Omega$. 
While perturbative terms up to ${\cal O}(g^2) $ only contribute
to the series up to ${\cal O}(\mu^4)$ higher order terms in the
expansion start receiving non-vanishing contributions at ${\cal O}(g^3)$.
These again arise from the electric mass term, {\it i.e.} from an expansion of 
$m_E/gT$ in powers of $\bfmut$;
\begin{eqnarray}
\Omega^{(3)} &=& \frac{1}{6\pi}\left( 1 +\frac{n_f}{6}\right)^{3/2}
+ \frac{1}{8\pi^3}\left( 1 +\frac{n_f}{6}\right)^{1/2} 
\bfmut 
+ \frac{1}{64\pi^5}\left( 1 +\frac{n_f}{6}\right)^{-1/2}
\left( \bfmut \right)^2 \nonumber \\[2mm]
&~&- \frac{1}{768\pi^7}\left( 1 +\frac{n_f}{6}\right)^{-3/2}
\left( \bfmut \right)^3
+{\cal O}(\mu^8) \quad .
\end{eqnarray}
Note that the expansion coefficients up to and including  ${\cal O}(\mu^4)$
are positive. Starting at ${\cal O}(\mu^6)$ they alternate in sign.
From the sign of the $\mu^6$-term it follows via Eq.~(\ref{sus}) that the 
leading perturbative contribution to the expansion coefficient 
of diagonal as well as non-diagonal susceptibilities at ${\cal O}(\mu^4)$ 
is negative, with the latter being an order of magnitude smaller\footnote{It
has also recently
been pointed out in the context of a large-$N_c$ expansion that
flavor non-diagonal contributions to the free energy are suppressed by
$O(1/N_c^2)$ \cite{Nc}.}.

Rather than discussing the thermodynamics of QCD at non vanishing
quark chemical potential in terms of chemical potentials related to
quark numbers in different flavor channels, it is convenient
to introduce chemical potentials which are related to
conserved quantum numbers considered at low energy,
{\it i.e.} quark or baryon number and isospin. In the case
of two flavor QCD, which we are going to analyze in our lattice
calculations, we thus introduce also the quark chemical potential, 
$\mu_q = (\mu_u+\mu_d)/2$ and the isovector chemical potential\footnote{This
definition of the isovector chemical potential differs from that used in
\cite{us2} by a factor of 2.} 
$\mu_I = (\mu_u-\mu_d)/2$. In analogy to Eq.~(\ref{sus}) we then can
introduce quark number and isovector susceptibilities, 
\begin{eqnarray}
\frac{\chi_q}{T^2} 
=  \frac{\partial^{2} \Omega}{\partial (\mu_q/T)^2}
\equiv 2 \left( \chi_{uu} + \chi_{ud} \right)
\quad , \quad
\frac{\chi_I}{T^2} 
= \frac{\partial^{2} \Omega}{\partial (\mu_I/T)^2} 
\equiv 2 \left( \chi_{uu} - \chi_{ud} \right) \quad ,
\label{eq:chiqi}
\end{eqnarray}
where in the second equality we have assumed degenerate $(u,d)$-quark
masses.

\subsection{Low temperature hadron resonance gas}

The success in describing particle abundance ratios observed in
heavy ion experiments at varying beam energies in terms of equilibrium
properties of a hadron resonance gas model \cite{pbm} begs a
comparison of this model for the low temperature hadronic phase with
lattice QCD calculations. Indeed this led to astonishingly good 
agreement \cite{KRT}.

In the hadron resonance gas (HRG) model it is  assumed that for $T<T_0$ the 
QCD partition function can be approximated by that of a non-interacting 
gas of hadron resonances, either bosonic mesons or fermionic 
baryons. This, however, does not mean that interactions in dense
hadronic matter have been ignored;
in the spirit of Hagedorn's bootstrap model \cite{Hagedorn} 
the inclusion of
heavy resonances as stable particles also takes care of the interaction
among the hadrons in the dense gas at low temperature. 

The partition
function of the  hadron resonance gas may be split into mesonic and
baryonic contributions,   
\begin{eqnarray}
\ln{\cal Z}_{HRG}(T,V,\mu_q,\mu_I) 
=\sum_{i\in\;mesons}\hspace{-3mm} \ln{\cal Z}^{M}_{m_i}(T,V,\mu_q,\mu_I)
+\hspace{-3mm} 
\sum_{i\in\;baryons}\hspace{-3mm} \ln{\cal Z}^{B}_{m_i}(T,V,\mu_q,\mu_I)\; ,
\label{eq:ZHRG}
\end{eqnarray}
where 
\begin{equation}
\ln{\cal Z}^{M/B}_{m_i}(T,V\mu_q,\mu_I)
=\mp{V\over{2\pi^2}}\int_0^\infty dk k^2
\ln(1\mp z_ie^{-\varepsilon_i/T}) \quad ,
\label{eq:ZMB}
\end{equation}
with energies $\varepsilon_i^2=k^2+m_i^2$ and fugacities 
\begin{equation}
z_i=\exp\left((3B_i\mu_q+2I_{3i}\mu_I)/T\right) \quad .
\label{eq:fuga}
\end{equation}
Here $B_i$ is the baryon number and $I_{3i}$ denotes the third component of 
the isospin of the species in question. The upper sign in Eq.~(\ref{eq:ZMB})
refers to bosons, and the lower sign to fermions. Note that with this 
convention anti-particles must be counted separately in Eq.~(\ref{eq:ZHRG})
with fugacity $z_i^{-1}$, and self-conjugate species such as 
$\pi^0, \eta^\prime$ have $z=1$. If the logarithms are expanded in powers 
of fugacity, the integral over momenta, $k$, can be performed. This yields
\begin{equation}
\ln{\cal Z}^{M/B}_{m_i}=\frac{VTm_i^2}{2\pi^2} \sum_{\ell=1}^\infty
\left\{\matrix{1\cr(-1)^{\ell+1}}\right\}\ell^{-2} K_2\left(\frac{\ell m_i}{
T}\right) z_i^{\ell} \quad ,
\label{eq:Zmi}
\end{equation}
where the upper factor in braces applies to bosons and the lower to fermions,
and $K_2$ is a modified Bessel function. For large argument, {\it i.e.} for
$m_i \gg T$ the Bessel function can be approximated by 
$K_2(x)\sim\sqrt{\pi/2x}\; {\rm e}^{-x} (1+15/8x +{\cal O}(x^{-2}))$. 
Terms with $\ell\geq2$ in the series given in Eq.~(\ref{eq:Zmi}) thus are 
exponentially suppressed. For temperatures and quark chemical
potentials less than a typical scale of about $200~$MeV it generally 
suffices to keep the first term in the sum appearing in Eq.~(\ref{eq:Zmi}).
The only species for which this step would need further justification is
the pion; clearly for realistic pion masses more care must be taken 
when evaluating the sum over $\ell$. This, however, does not influence the
$\mu_q$-dependence of the hadron resonance gas; $B_i=0$ for mesons 
and their contribution thus is independent of $\mu_q$. At   
$\mu_I=0$ one readily derives,
\begin{equation}
{p(T,\mu_q)\over T^4}= \frac{1}{VT^3} \ln{\cal Z}_{HRG}\biggr|_{\mu_I=0}
\simeq G(T)+F(T)\cosh \left( \frac{3\mu_q}{T} \right) \quad ,
\label{eq:pHRG}
\end{equation}
with $G(T)=(VT^3)^{-1}\sum_{i\in mesons} \ln {\cal Z}^M_{m_i}$ and
$F(T)$ given by the Boltzmann approximation to the fermion partition
function of baryons,
\begin{eqnarray}
F(T) &=& {1\over\pi^2}\sum_{i\in\;baryons}\left({m_i\over T}\right)^2
K_2\left({m_i\over T}\right)\quad .
\label{eq:Phip}
\end{eqnarray}
Note that each term in the sum for $F$ now counts both baryon and
anti-baryon. 
As the meson sector of the partition functions is independent of $\mu_q$
the mesonic component does not contribute to $\chi_q$,
\begin{equation}
{\chi_q\over T^2}=9F(T)\cosh{{3\mu_q}\over T} \quad .
\label{eq:chiHRG}
\end{equation}
Similarly, at $\mu_I = 0$ we find for the isovector susceptibility
\begin{equation}
{\chi_I\over T^2}
=G^I(T) +F^I(T)\cosh{{3\mu_q}\over T}
\label{eq:chiHRGI}
\end{equation}
with
\begin{eqnarray}
G^I(T) &=& {1\over2\pi^2}\sum_{i\in\;mesons}
\sum_{\ell=1}^\infty
(2I_{3i})^2
\left({m_i\over T}\right)^2 K_2\left({\ell m_i\over T}\right) \quad ,\nonumber \\
F^I(T) &=& {1\over \pi^2}\sum_{i\in\;baryons}(2I_{3i})^2
\left({m_i\over T}\right)^2 K_2\left({m_i\over T}\right) \quad .
\label{eq:Phichi}
\end{eqnarray}
where again only in the baryon sector $\ell > 1$ terms have been
neglected.

\noindent
The resonance gas model in the (partial)
Boltzmann approximation given by
Eqs.~(\ref{eq:pHRG})-(\ref{eq:Phichi}) leads to simple predictions 
for the dependence of thermodynamic observables on the quark chemical
potential $\mu_q$. In particular, it predicts that ratios of the
density dependent part of thermodynamic observables are insensitive
to details of the hadronic mass spectrum. 
Nor do they depend explicitly on temperature, but instead only on
the ratio $\mu_q/T$. For instance, one finds 
\begin{equation}
\frac{p(T,\mu_q)-p(T,0)}{\chi_q T^2} = \frac{\cosh(3 \mu_q/T) -1}{9
\cosh(3 \mu_q/T)} \quad , \quad
\frac{n_q}{\mu_q \chi_q}
=\frac{T}{3\mu_q} \tanh \left( \frac{3\mu_q}{T} \right) \quad .
\label{eq:npchiq}
\end{equation}
Similarly ratios of Taylor expansion coefficients of thermodynamic
quantities, $X$, are temperature and spectrum independent. For an
observable $X$, of generic form $X=G^X +F^X \cosh(3 \mu_q/T)$, the
expansion in $\mu_q/T$ is given by 
\begin{equation}
X=\sum_{n=0}^\infty c_n^X (T) (\mu_q/T)^{n}, \hspace{5mm} 
\end{equation}
with 
\begin{equation}
c_0^X=G^X+F^X, \hspace{5mm} c_{2n}^X=\frac{9^n}{(2n)!} F^X 
\end{equation}
and $c_{2n+1}^X=0$.
Hence, ratios are given by
\begin{equation}
\frac{c_{2n+2}^X}{c_{2n}^X}=\frac{9}{(2n+2)(2n+1)} \hspace{5mm}
{\rm for} \ n \geq 1 \quad .
\label{eq:HRGratios}
\end{equation}
These ratios as well as ratios of physical observables calculated within
the resonance gas approximation will be compared to corresponding lattice results
in the following.

\section{Taylor expansion for two flavor QCD}
\label{sec:eos}

The basic concepts of our Taylor expansion approach to QCD
thermodynamics at non-zero quark chemical potential have been introduced
in \cite{us2} where observables have been analyzed 
up to ${\cal O}(\mu_q^4)$. Here we extend the analysis up to the sixth
order in $\mu_q/T$ with significantly improved statistics.

Our calculations have been performed for two flavor
QCD on an $16^3\times4$ lattice with bare quark mass $ma=0.1$
using Symanzik-improved gauge and p4-improved
staggered fermion actions. These parameters are identical to those used
for our analysis of thermodynamic observables 
at non-zero chemical potential up to ${\cal O}(\mu_q^4)$
\cite{us1,us2}. 
The simulation uses the hybrid R molecular dynamics algorithm,
and measurements were performed on equilibrated configurations separated
by 5 units of molecular dynamics time $\tau$. The gauge couplings, 
$\beta=6/g^2$, used for our calculations 
cover the interval [3.52,4.0], which corresponds to a temperature range 
$T/T_{0}\in[0.76,1.98]$, where $T_{0}$ is the pseudocritical temperature
at $\mu=0$ for which we use\footnote{In \cite{us1} we determined as
critical coupling $\beta_c=3.649(2)$. Our current analysis favors a
slightly larger value, $\beta_c=3.655(5)$.} $\beta_c=3.65$. This value is
used to define the temperature scale $T/T_0$. The number of 
configurations generated at each $\beta$-value is given in 
Table~\ref{tab:configs}. 
The third and seventh columns give the sample sizes used in \cite{us2}, 
where coefficients up to $n=4$ were calculated. The numbers of additional
configurations generated for the present study of expansion coefficients
up to $n=6$ are listed in the fourth and eighth columns.
It can be seen that we have increased our 
statistics in the hadronic phase by a factor 4-5 and in the plasma phase
by a factor 3. 

For the calculation of various operator traces we use the
method of noisy estimators. We generally found that expectation values
involving odd derivatives of $\ln {\rm det}M$ with respect to $\mu$ are
noisier and require averages over more random vectors than needed to
estimate expectation values involving only  even
derivatives. Odd derivatives have to appear in even numbers in an
expectation value in order for this to be non-zero. Such 
expectation values behave very much like susceptibilities and receive
their largest contributions in the vicinity of $T_0$. Still their total
contribution to the expansion of e.g. the pressure is found to be small 
up to ${\cal O}(\mu^4)$. It only becomes sizeable at ${\cal O}(\mu^6)$. 
For these reasons we used 100 stochastic noise vectors to estimate operator 
traces on each configuration for $\beta\in[3.60,3.68]$. For other $\beta$ 
values 50 noise vectors were found to suffice.

\begin{table}[t]
\centering
\setlength{\tabcolsep}{1.0pc}
\begin{tabular}{|llll|llll|}
\hline
$\beta$ & $T/T_c$ & \#(2,4) & $\!\!\!\!$\#(2,4,6) &  
$\beta$ & $T/T_c$ & \#(2,4) & $\!\!\!\!$\#(2,4,6)   \\
\hline
3.52  & 0.76    &  1000   &  3500 &  
3.70  & 1.11    &   800   &  2000 \\
3.55  & 0.81    &  1000   &  3500 & 
3.72  & 1.16    &   500   &  2000 \\
3.58  & 0.87    &  1000   &  3500 & 
3.75  & 1.23    &   500   &  1000 \\
3.60  & 0.90    &  1000   &  3800 & 
3.80  & 1.36    &   500   &  1000 \\
3.63  & 0.96    &  1000   &  3500 & 
3.85  & 1.50    &   500   &  1000 \\
3.65  & 1.00    &  1000   &  4000 & 
3.90  & 1.65    &   500   &  1000 \\
3.66  & 1.02    &  1000   &  4000 & 
3.95  & 1.81    &   500   &  1000 \\
3.68  & 1.07  & $\;\,$800 &  3600 & 
4.00  & 1.98    &   500   &  1000 \\
\hline
\end{tabular}
\caption{Sample size at each $\beta$ value.}
\smallskip
\label{tab:configs}
\end{table}

\subsection{Pressure, quark number density and susceptibilities}
\label{subsec:pressure}

To start the discussion of lattice results on thermodynamics
of two flavor QCD for small values of the quark chemical potential we
will present results obtained from a Taylor expansion of the
grand potential $\Omega(T,\mu_u,\mu_d)\equiv
\Omega(T,\mu_q+\mu_I,\mu_q-\mu_I)$ and some of its derivatives.
We will consider expansions in terms of $\mu_q/T$ at fixed, vanishing
$\mu_I$. The pressure is then given by

\begin{table}[h]
\setlength{\tabcolsep}{0.8pc}
\begin{tabular}{|l|lll|lll|}
\hline
$T/T_c$ & $c_2$ & $c_4\times10$ & $c_6\times10^2$ 
        & $c^I_2$ & $c^I_4\times10$ & $c^I_6\times10^2$\\
\hline
0.76 & 0.0243(19) & 0.238(61)  & $\!\!$-1.12(121) 
     & 0.0649(6)  & 0.098(5)  &  0.23(9)  \\
0.81 & 0.0450(20) & 0.377(64)  &  1.98(141) 
     & 0.0874(8)  & 0.140(6)  &  0.44(10) \\
0.87 & 0.0735(23) & 0.506(68)  &  1.69(155) 
     & 0.1206(11) & 0.216(8)  &  0.60(13) \\
0.90 & 0.1015(24) & 0.765(72)  &  2.06(159) 
     & 0.1551(14) & 0.302(12) &  0.83(18) \\
0.96 & 0.2160(31) & 1.491(135) &  4.96(260) 
     & 0.2619(21) & 0.564(23) &  1.47(37) \\
1.00 & 0.3501(32) & 2.133(121) & $\!\!$-5.00(359) 
     & 0.3822(26) & 0.839(28) &  0.26(49) \\
1.02 & 0.4228(33) & 2.258(118) & $\!\!$-4.49(312) 
     & 0.4501(27) & 0.909(28) &  0.02(44) \\
1.07 & 0.5824(23) & 1.417(62)  & $\!\!$-5.73(158) 
     & 0.5972(21) & 0.741(17) & $\!\!$-0.75(26) \\
1.11 & 0.6581(20) & 0.951(39)  & $\!\!$-1.65(62)  
     & 0.6662(18) & 0.618(11) & $\!\!$-0.18(10) \\
1.16 & 0.7091(15) & 0.763(24)  & $\!\!$-0.31(26)  
     & 0.7156(14) & 0.564(6)  & $\!\!$-0.03(4) \\
1.23 & 0.7517(16) & 0.667(23)  & $\!\!$-0.44(23)  
     & 0.7573(13) & 0.527(5)  & $\!\!$-0.06(3) \\
1.36 & 0.7880(11) & 0.572(12)  & $\!\!$-0.09(11)  
     & 0.7906(9)  & 0.495(3)  & $\!\!$-0.03(1) \\
1.50 & 0.8059(10) & 0.539(10)  & $\!\!$-0.17(7)   
     & 0.8076(7)  & 0.477(2)  & $\!\!$-0.05(1) \\
1.65 & 0.8157(8)  & 0.499(7)   & $\!\!$-0.13(8)   
     & 0.8169(7)  & 0.461(2)  & $\!\!$-0.05(1) \\
1.81 & 0.8203(8)  & 0.497(7)   & $\!\!$-0.11(6)   
     & 0.8218(6)  & 0.452(1)  & $\!\!$-0.05(1) \\
1.98 & 0.8230(7)  & 0.473(6)   &  0.03(4)   
     & 0.8250(6)  & 0.441(1)  & $\!\!$-0.03(1) \\
\hline
\end{tabular}
\caption{Taylor expansion coefficients $c_n(T)$ and $c^I_n(T)$.}
\smallskip
\label{tab:c_n}
\end{table}

\begin{equation}
{p\over T^4}\equiv\Omega(T,\mu_q,\mu_q)=
\sum_{n=0}^\infty c_n(T) \left({\mu_q\over T}\right)^n.\label{eq:p}
\end{equation}
CP symmetry implies that the series is even in $\mu_q$, so 
the coefficients $c_n$ are non-zero only for $n$ even, and are defined as
\begin{equation}
c_n (T)={1\over n!}{{\partial^n \Omega}
\over{{\partial(\mu_q/T)^n}}}\biggr\vert_{\mu_q=0}=
{1\over n!}{N_\tau^{3}\over N_\sigma^3}{{\partial^n\ln{\cal Z}}\over
{\partial(\mu N_\tau)^n}}\biggr\vert_{\mu=0} \quad ,
\label{eq:cn}
\end{equation}
where in the second equality we have explicitly specified that the calculations
have been performed on a $N_\sigma^3\times N_\tau$ lattice with dimensionless
quark chemical potential $\mu\equiv\mu_q a$. ${\cal Z}$ denotes the
lattice regularized partition function for two flavor QCD. 
Similarly we calculate the quark number density
\begin{equation}
\frac{n_q(T,\mu_q)}{T^3}=\frac{\partial \Omega(T,\mu_q,\mu_q)}{\partial \mu_q/T}
=2c_2\left(\frac{\mu_q}{T}\right) +4c_4\left(\frac{\mu_q}{T}\right)^3
+6c_6\left(\frac{\mu_q}{T}\right)^5 + \cdots \quad ,
\label{eq:nq}
\end{equation}
as well as the quark and isovector susceptibilities using
Eq.~(\ref{eq:chiqi}),
\begin{eqnarray}
{\chi_q(T,\mu_q)\over T^2}
&=&2c_2+12c_4\left({\mu_q\over T}\right)^2+30c_6\left({\mu_q\over
T}\right)^4+\cdots\label{eq:chiq}\\
{\chi_I(T,\mu_q)\over T^2}
&=&2c^I_2+12c^I_4\left({\mu_q\over T}\right)^2+30c^I_6\left({\mu_q\over
T}\right)^4+\cdots\label{eq:chiI}
\end{eqnarray}
where
\begin{equation}
c^I_n={1\over n!}{\partial^n\; \Omega
(T,\mu_q+\mu_I,\mu_q-\mu_I)
\over{{\partial(\mu_I/T)^2}\partial(\mu_q/T)^{n-2}}}
\biggr\vert_{\mu_q=0,\mu_I=0} \quad.
\end{equation}
Explicit expressions for $c_2$, $c_4$, $c_6$ 
and $c^I_2$, $c^I_4$ and $c^I_6$ are given in the Appendix.
Note that the expansion for the quark number susceptibility
$\chi_q$ given in Eq.~(\ref{eq:chiq}) is a derivative of the grand
potential at $\mu_I\equiv 0$ and thus 
has the same radius of convergence as that of the pressure and quark
number density given by Eqs.~(\ref{eq:p}) and (\ref{eq:nq}). 
The expansion coefficient $c_2^I$ also defines the first term in an
expansion of the pressure at non-zero isospin. This series
may have a different radius of convergence \cite{Kogut}; indeed, since 
the lightest particle carrying non-zero
isospin $I_3$ in the hadronic phase is the pion, we might expect the
expansion to break down in the chiral limit for arbitrarily small $\mu_I$.

\begin{figure}[tb]
\begin{center}
\begin{minipage}[c][5.2cm][c]{5.0cm}
\begin{center}
\epsfig{file=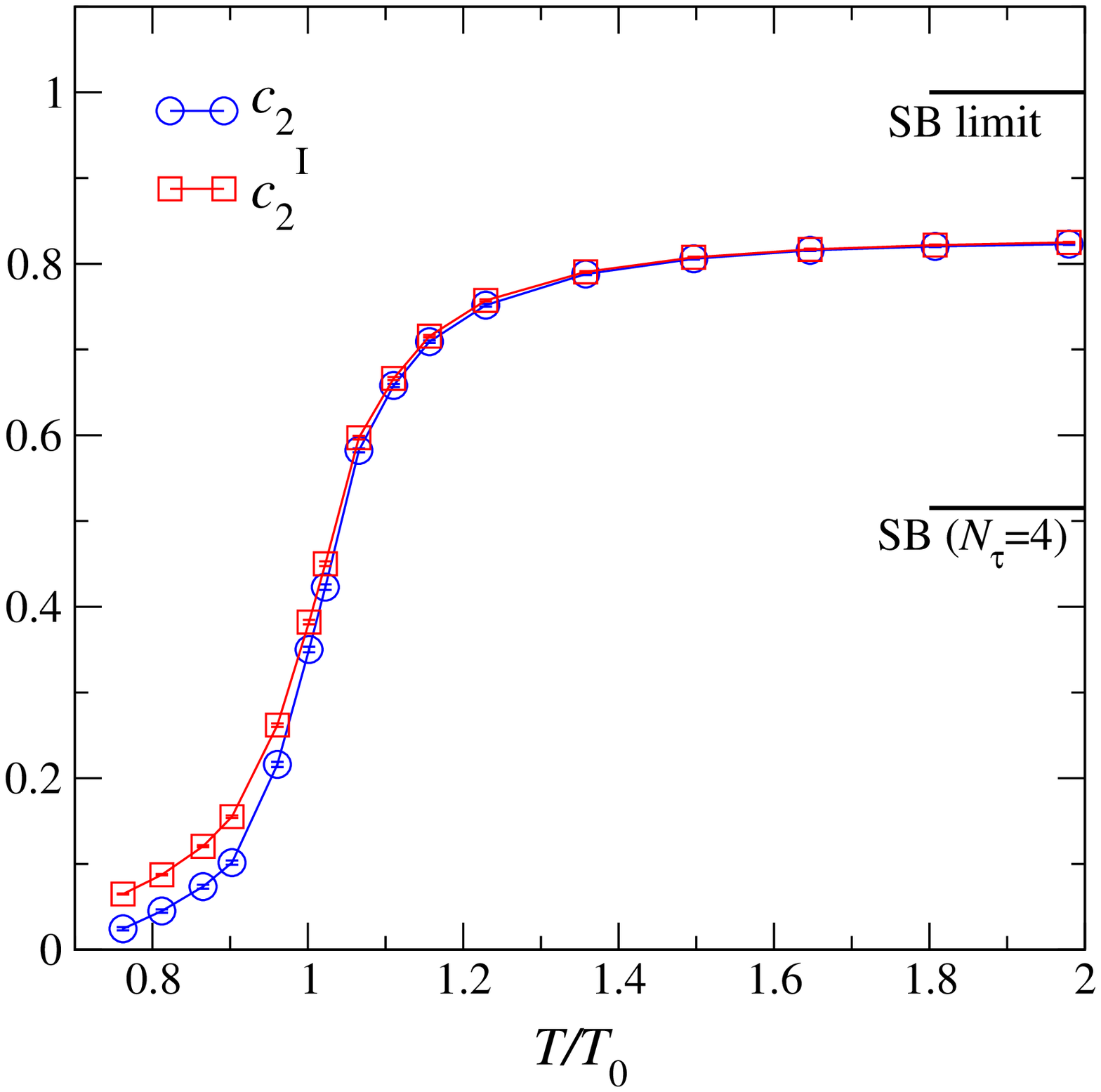, width=5.0cm}\\[-1mm]
\end{center}
\end{minipage}
\begin{minipage}[c][5.2cm][c]{5.0cm}
\begin{center}
\epsfig{file=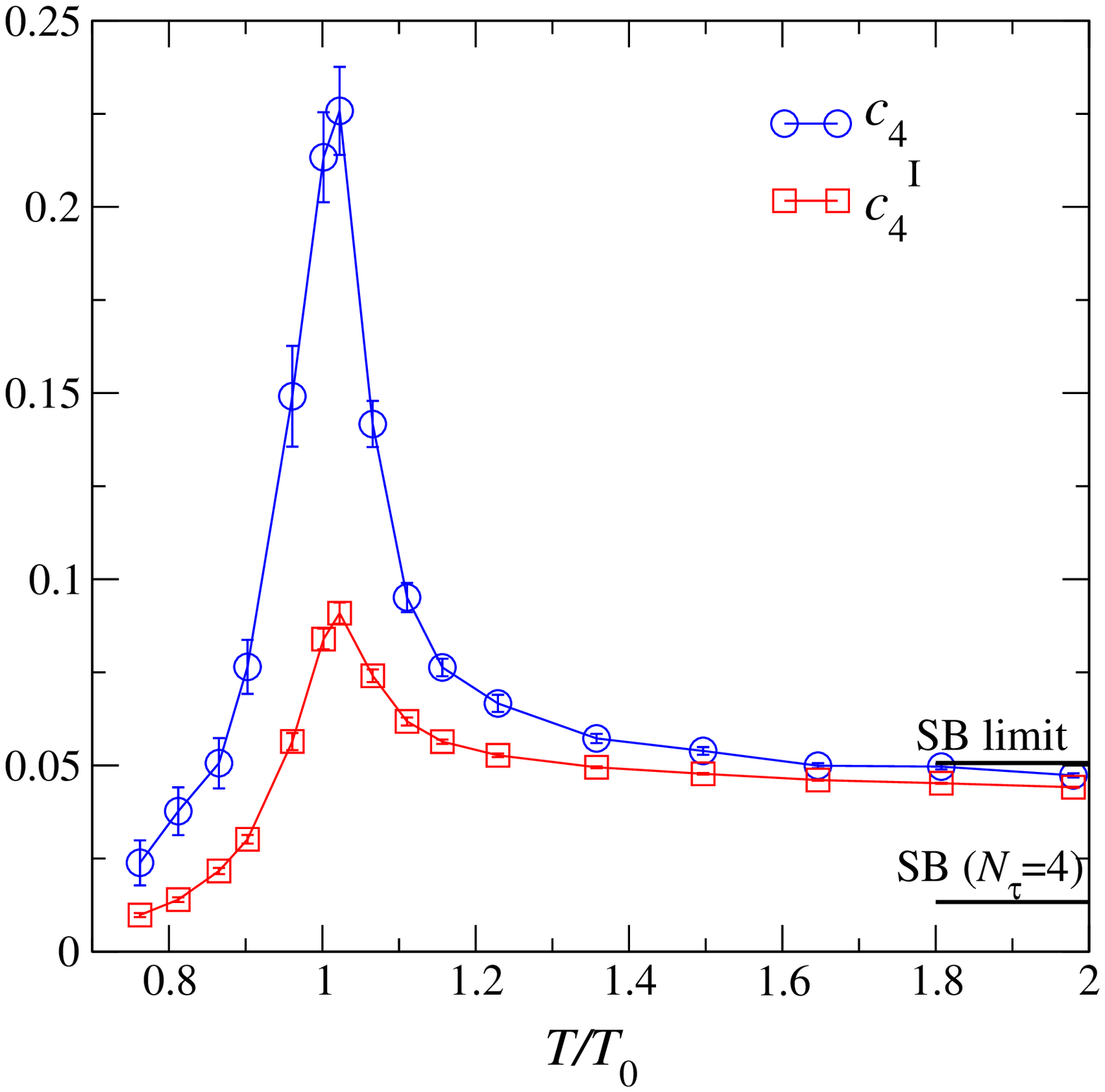, width=5.0cm}\\[-1mm]
\end{center}
\end{minipage}
\begin{minipage}[c][5.2cm][c]{5.0cm}
\begin{center}
\epsfig{file=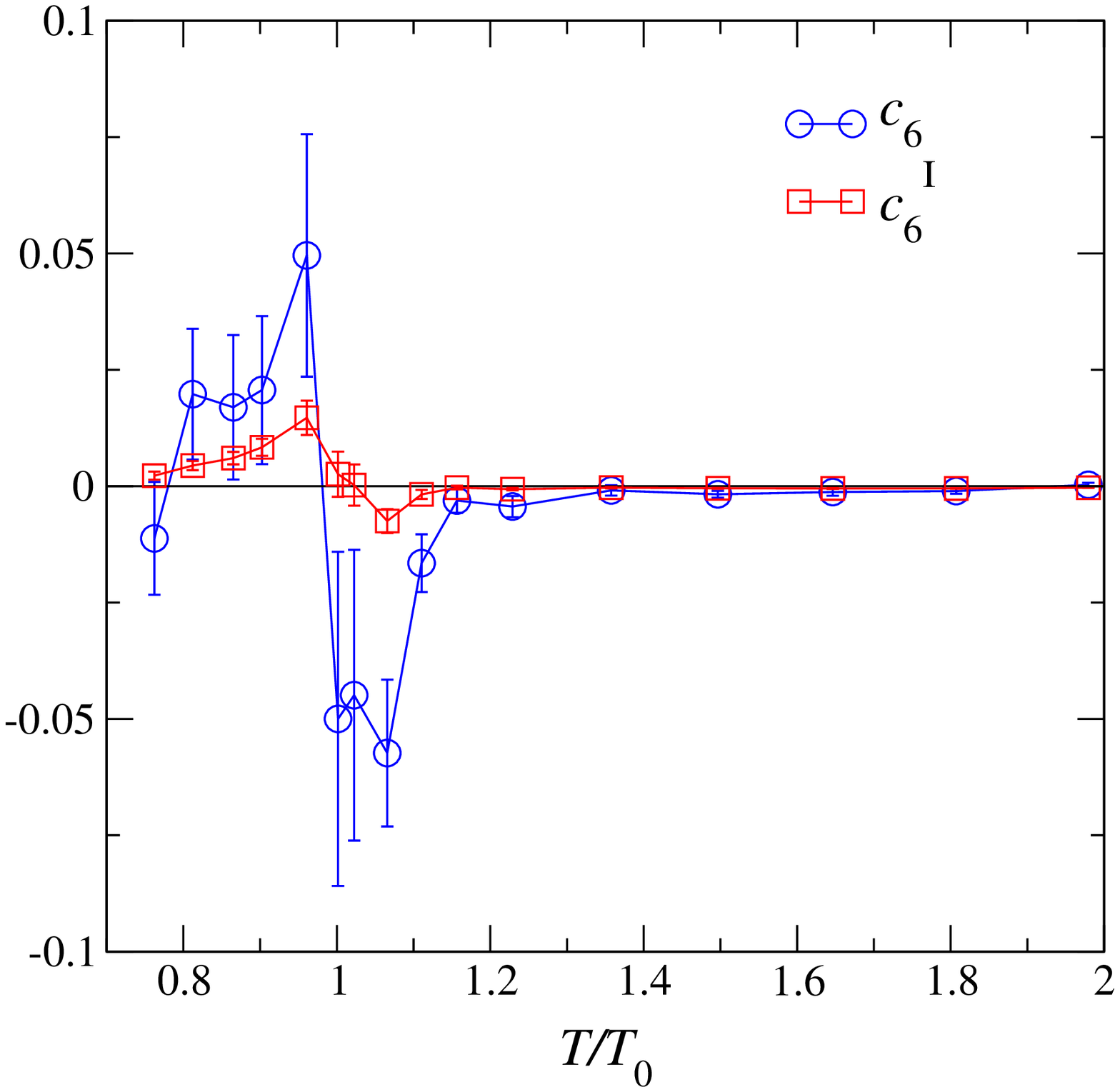, width=5.0cm}\\[-1mm]
\end{center}
\end{minipage}
\caption{The Taylor expansion coefficients $c_n$ and $c^I_n$ for $n=2, 4$ and 6 
as functions of $T/T_{0}$.}
\label{fig:c2c4}
\end{center}
\end{figure}

The coefficient $c_0 (T)$ gives 
the pressure in units of $T^4$ at vanishing baryon density and can be 
calculated using the integral method \cite{KLP}. It is the only expansion
coefficient which also requires lattice calculations at zero temperature.
Higher order terms can be calculated directly from gauge field
configurations generated on finite temperature lattices. They, however, 
require additional derivatives of $\ln\mbox{det}M$,
where $M$ is the quark matrix. They are evaluated at fixed 
temperature, {\it i.e.} fixed gauge coupling $\beta$, by calculating 
combinations of traces of products of 
$\partial^m M/\partial\mu^m$ and $M^{-1}$ (see Appendix).

Results for the Taylor expansion coefficients are listed in 
Table~\ref{tab:c_n}. In Fig.~(\ref{fig:c2c4}) we plot $c_n$  and
$c^I_n$ for $n=2,~4$ and 6 as functions of 
$T$. A comparison with Figs.~3 and 8 of Ref.~\cite{us2} reveals the improvement 
in statistics of the current study. The same features are apparent: 
namely $c_2$ and $c^I_2$ both rise steeply across $T_0$
with $c_2^I> c_2$ as is obvious from the explicit expressions given for
these coefficients in the appendix; they 
reach a plateau at approximately 80\% of the value $n_{\rm f}/2$
predicted in the Stefan-Boltzmann (SB) limit, {\it i.e.} for free 
massless quarks;
$c_4$ rises steeply to peak at $T\simeq T_{0}$ before approaching its
SB limit value $n_{\rm f}/4\pi^2$ from above, whereas the peak in
$c^I_4$ is much less marked\footnote{The difference is largely due to the 
dominance of the disconnected term
$\langle(\partial^2\ln\mbox{det}M/\partial\mu^2)^2\rangle -
\langle \partial^2\ln\mbox{det}M/\partial\mu^2 \rangle^2$ 
which contributes to
$c_4$ with a coefficient three times that of its contribution to $c^I_4$.}.

\begin{figure}[tb]
\begin{center}
\begin{minipage}[c][7.8cm][c]{7.4cm}
\begin{center}
\epsfig{file=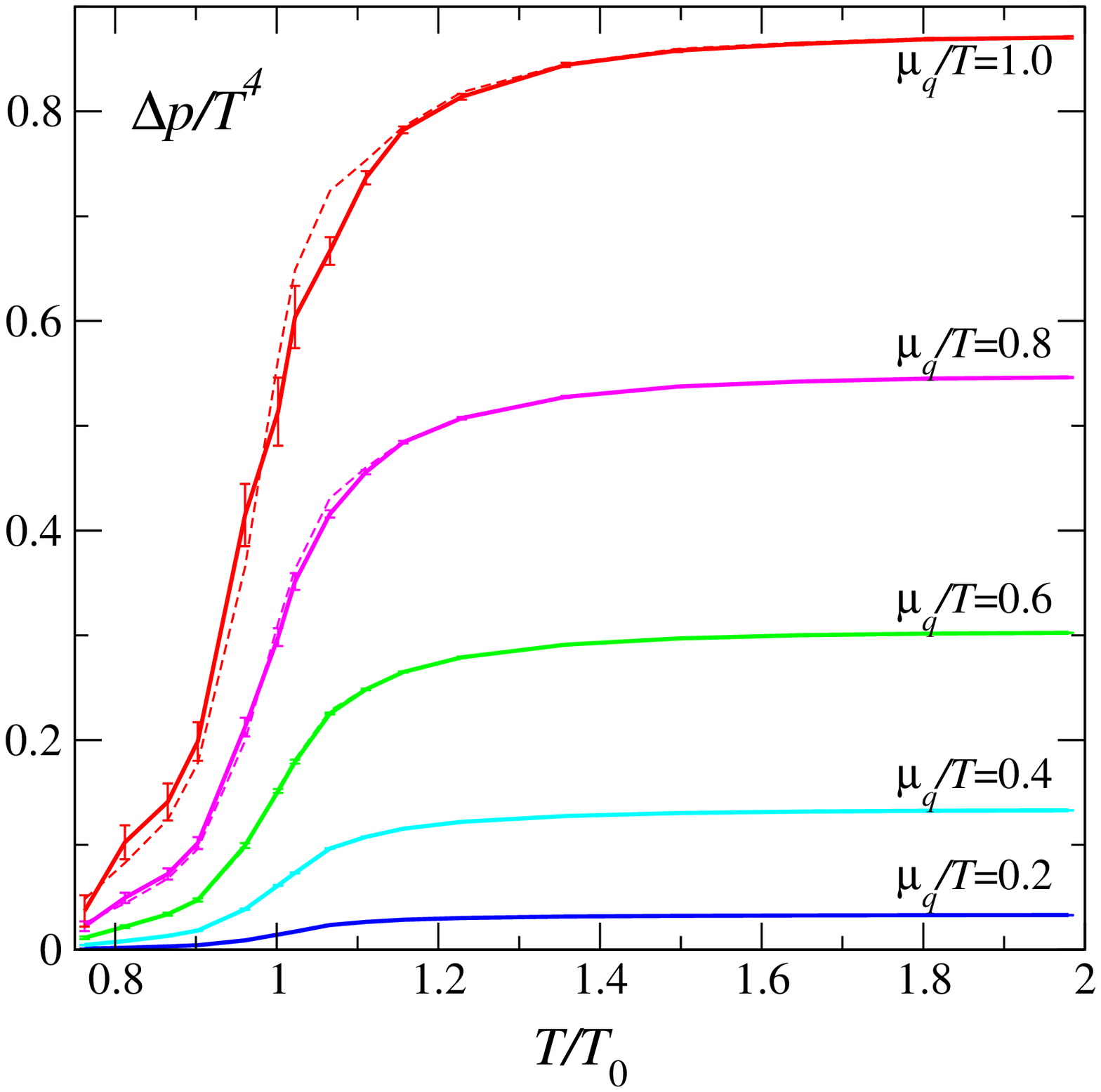, width=7.4cm}\\[-1mm]
(a)
\end{center}
\end{minipage}
\begin{minipage}[c][7.8cm][c]{7.4cm}
\begin{center}
\epsfig{file=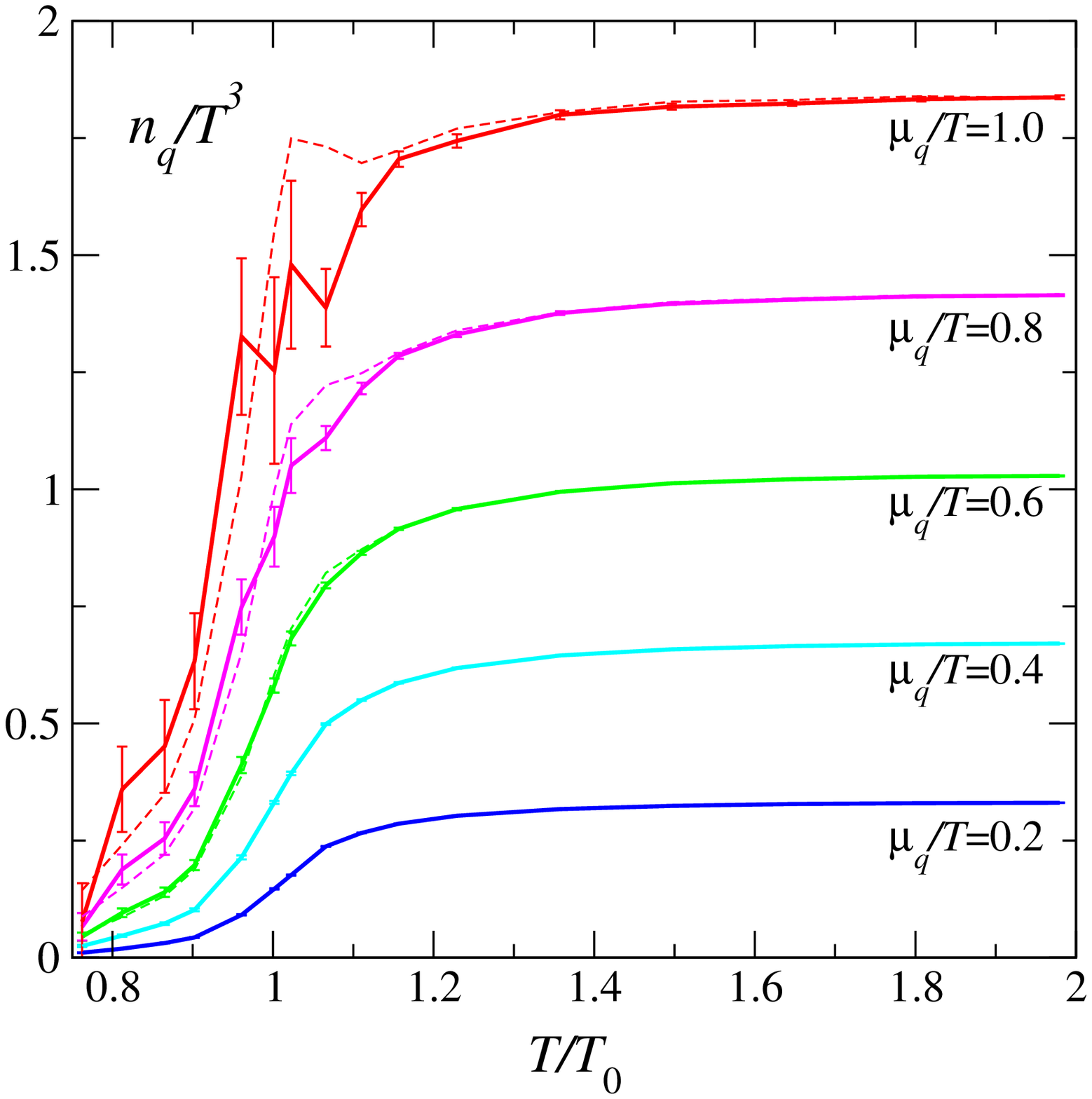, width=7.4cm}\\[-1mm]
(b)
\end{center}
\end{minipage}
\caption{The $\mu_q$ dependent contribution to the pressure (left)
and the quark number density (right) as functions of $T/T_{0}$
for various values of the quark chemical potential calculated
from a Taylor series in $6^{th}$ order. Also shown as dashed lines
are results from a $4^{th}$ order expansion in $\mu_q/T$.}
\label{fig:Deltap}
\end{center}
\end{figure}

As can be seen in Table~\ref{tab:c_n} in the high temperature
phase the $6^{th}$ order expansion coefficients generally are an 
order of magnitude smaller than the $4^{th}$ order coefficients. In the
low temperature phase they are still a factor 3-5 smaller. As a consequence
the $6^{th}$ order contributions to the pressure and quark number density
are small for $\mu_q/T \le 1$. This is seen in 
Fig.~(\ref{fig:Deltap}) which shows $\Delta p/T^4 \equiv
(p(T,\mu_q)-p(T,0))/T^4$ and $n_q/T^3$ 
in the range $0\le \mu_q/T \leq 1$. 
Here we also
show as dashed lines results obtained from a Taylor expansion
which includes only terms up to $4^{th}$ order in $\mu_q/T$. This
suggests that the expansion for the pressure and quark number density
is converging rapidly for $\mu_q/T < 1$. Even for $\mu_q/T=1$ the 
differences between the $4^{th}$ and $6^{th}$-order results are small 
and partly influenced by statistics. We also note that in the high
temperature regime, $T\gsim 1.5 T_0$, our results are compatible with
the continuum extrapolated 
(quenched) results obtained with an unimproved staggered
fermion action \cite{Gavaieos}. This supports the expectation that
deviations from the continuum limit are strongly suppressed with our
improved action.

\begin{figure}[tb]
\begin{center}
\begin{minipage}[c][7.8cm][c]{7.4cm}
\begin{center}
\epsfig{file=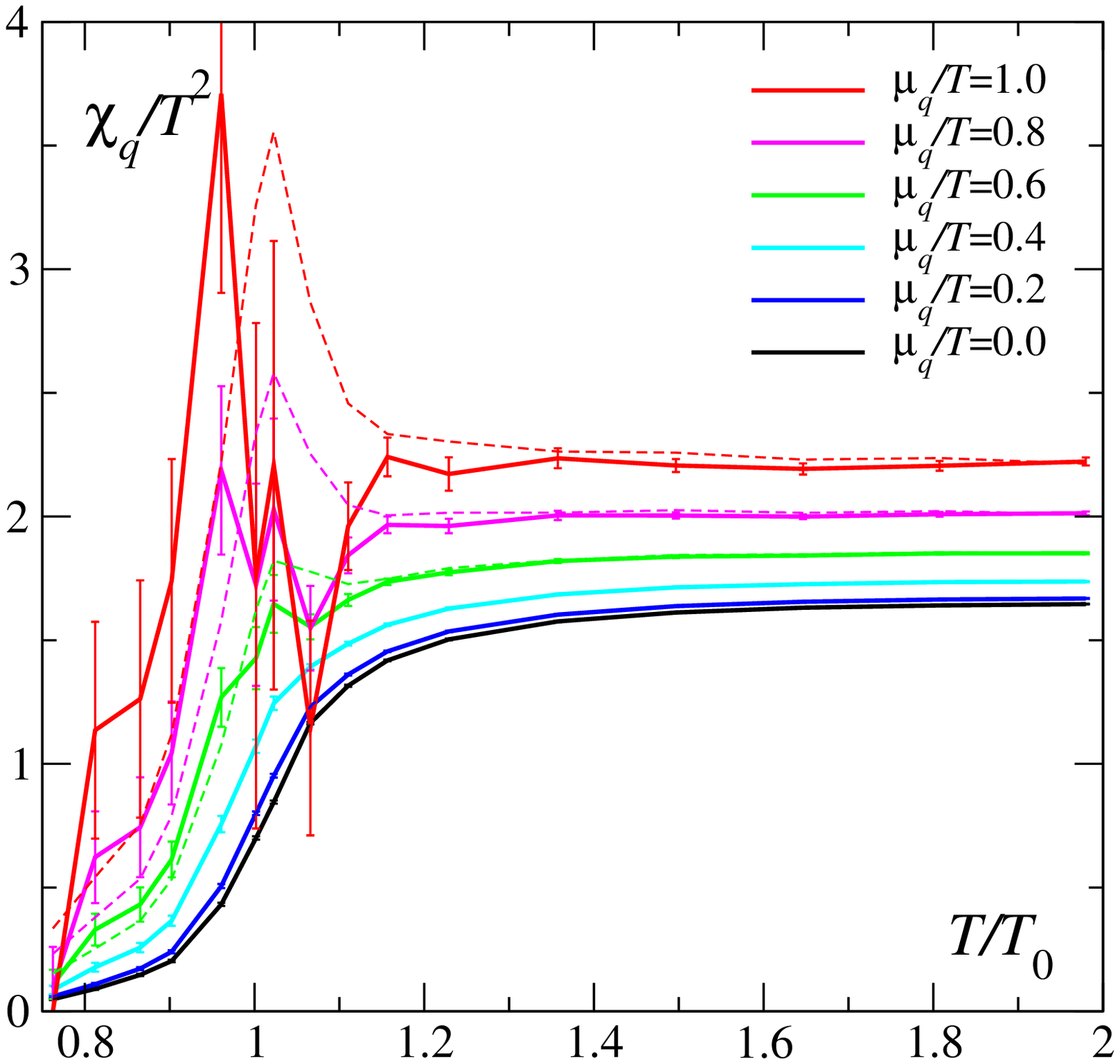, width=7.4cm}\\[-1mm]
(a)
\end{center}
\end{minipage}
\begin{minipage}[c][7.8cm][c]{7.4cm}
\begin{center}
\epsfig{file=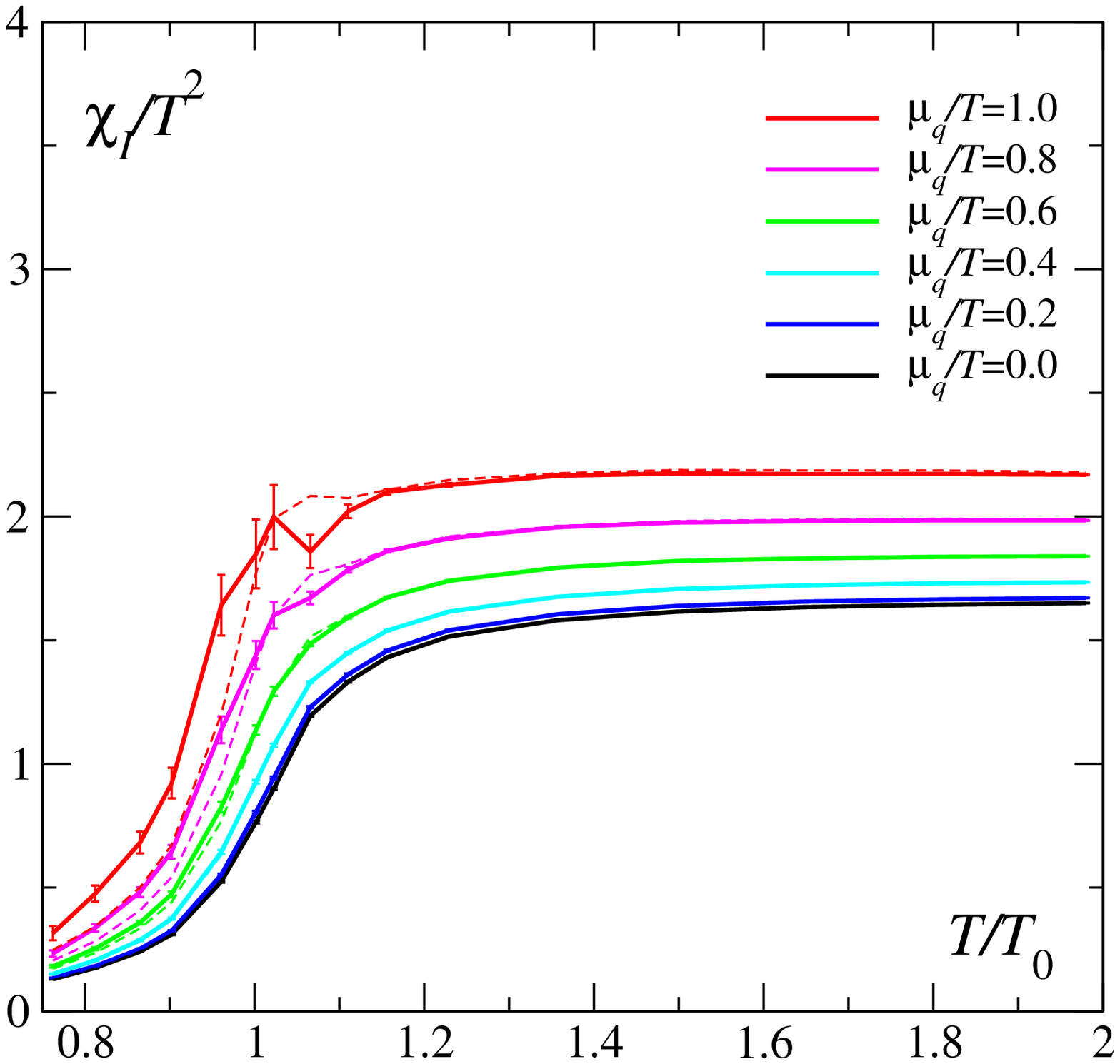, width=7.4cm}\\[-1mm]
(b)
\end{center}
\end{minipage}
\caption{The quark number susceptibility $\chi_q/T^2$ (left) and 
isovector susceptibility $\chi_I/T^2$ (right) as functions of $T/T_0$ for
various $\mu_q/T$ ranging from $\mu_q/T=0$ (lowest curve) rising in
steps of 0.2 to  $\mu_q/T=1$, calculated
from a Taylor series in $6^{th}$ order. Also shown as dashed lines
are results from a $4^{th}$ order expansion in $\mu_q/T$.
}
\label{fig:chiIq}
\end{center}
\end{figure}

Next we turn to a discussion of quark number and isovector susceptibilities 
which are shown in Fig.~(\ref{fig:chiIq}). They have been obtained
using Eqs.~(\ref{eq:chiq}) and (\ref{eq:chiI}). Again we show the corresponding
$4^{th}$-order results as  dashed lines in these figures.
These lines agree with our old results shown as Fig. 9 of \cite{us2}. 
The effect of the new term proportional to $c_6(T)$ is to shift the apparent
maximum in $\chi_q(T)$, arising from the sharply peaked $\mu_q^4$-contribution 
proportional to $c_4(T)$, to lower temperature. 
This suggests that the transition temperature at non-zero $\mu_q$ determined
from the peak position of susceptibilities indeed moves to temperatures
smaller than the transition temperature $T_0$ determined at $\mu_q=0$. The
figure, however, also shows that at least for $T<T_0$ the $6^{th}$-order 
contribution can be sizeable and still suffers from statistical errors. 
Better statistics and the contribution from higher orders in the Taylor 
expansion thus will be needed to get good quantitative results for 
susceptibilities in the hadronic phase.

There is also a pronounced dip in $\chi_q(T)$ for $T/T_{0}\simeq1.05$ which, 
together with the increased error bars makes the presence of a peak in $\chi_q$
less convincing then it is without the inclusion of the $\mu_q^6$-contribution. 
However, the error bars also reflect the problem we have at present in determining 
this additional contribution with sufficient accuracy to include it in the
calculation of higher order derivatives of the partition function. 
On the other side, Fig.~(\ref{fig:chiIq}) confirms that a significant peak
is not present in the isovector channel. 
In fact, if a critical endpoint exists in the ($T,\mu$)-plane of the
QCD phase diagram, this is expected to belong to
the Ising universality class, implying
that exactly one 3$d$ scalar degree of freedom becomes massless at this point.
Since both $\bar\psi\psi$ and $\bar\psi\gamma_0\psi$ are isoscalar and Galilean
scalars, both are candidates to interpolate this massless field, and hence 
we can expect divergent fluctuations in both quark number and chiral
susceptibilities at this point. The latter will be discussed in section 3.3.

The difference in the temperature dependence of $\chi_q$ and $\chi_I$ 
also reflects the strong correlation between
fluctuations in different flavor components. This will become clear from our
discussion of flavor diagonal and non-diagonal susceptibilities  
in the next section.

\subsection{Flavor diagonal and non-diagonal susceptibilities}

Using the relation between quark number and isovector susceptibilities
on the one hand
and diagonal and non-diagonal susceptibilities on the other hand
(Eq.~(\ref{eq:chiqi}))
we also can define expansions for the latter,
\begin{eqnarray}
{\chi_{uu}(T, \mu_q)\over T^2} =
&=&2c^{uu}_2+12c^{uu}_4\left({\mu_q\over T}\right)^2+
30c^{uu}_6\left({\mu_q\over T}\right)^4+\cdots\label{eq:chiuu} \quad ,
\\
{\chi_{ud}(T, \mu_q)\over T^2} =
&=&2c^{ud}_2+12c^{ud}_4\left({\mu_q\over T}\right)^2+
30c^{ud}_6\left({\mu_q\over T}\right)^4+\cdots\label{eq:chiud} \quad ,
\end{eqnarray}
with $c_n^{uu}=(c_n+c_n^I)/4$ and $c_n^{ud}=(c_n-c_n^I)/4$.

\begin{figure}[tb]
\begin{center}
\begin{minipage}[c][9.5cm][c]{5.0cm}
\begin{center}
\epsfig{file=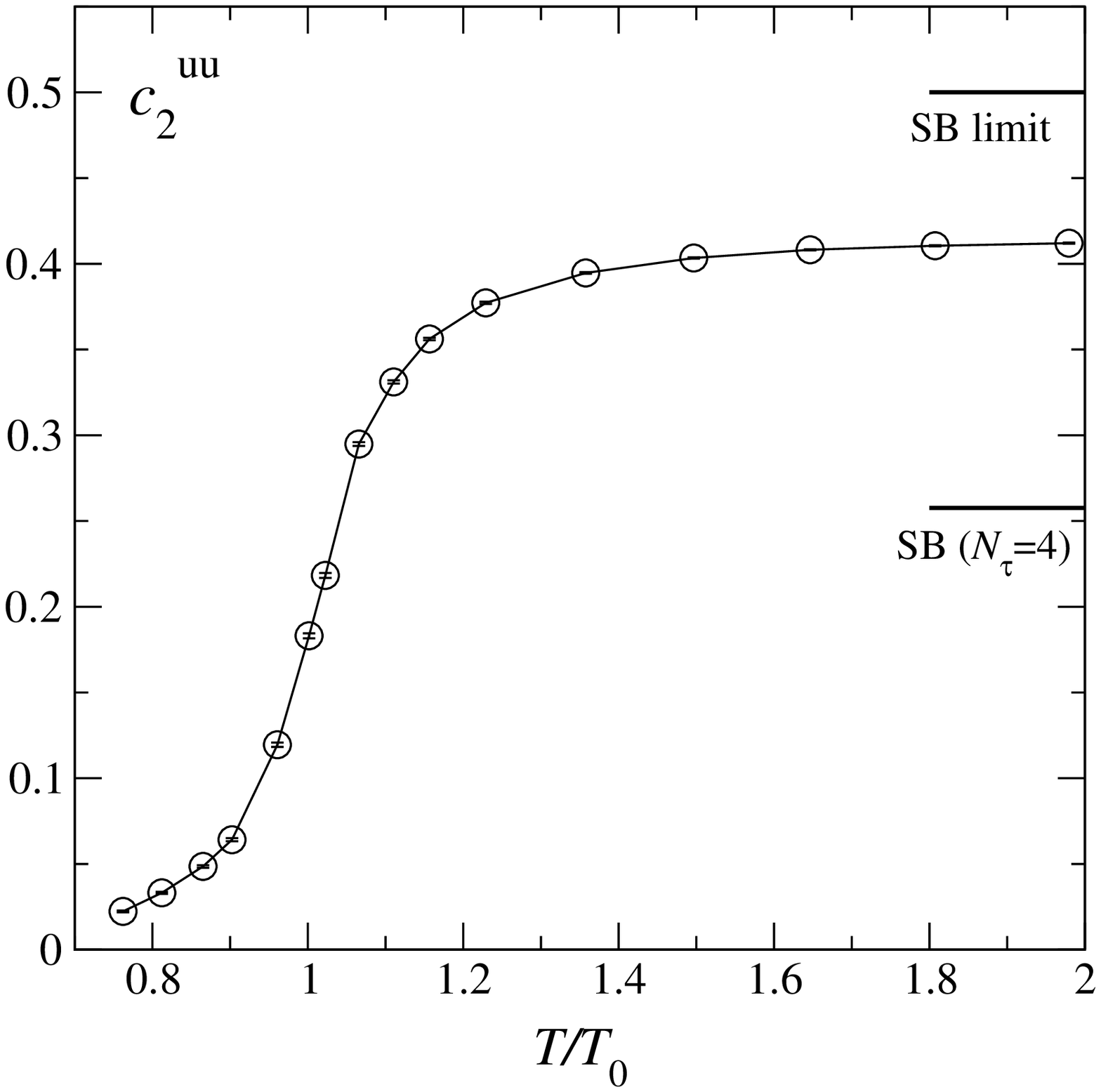, width=5.0cm}\\[-1mm]
\epsfig{file=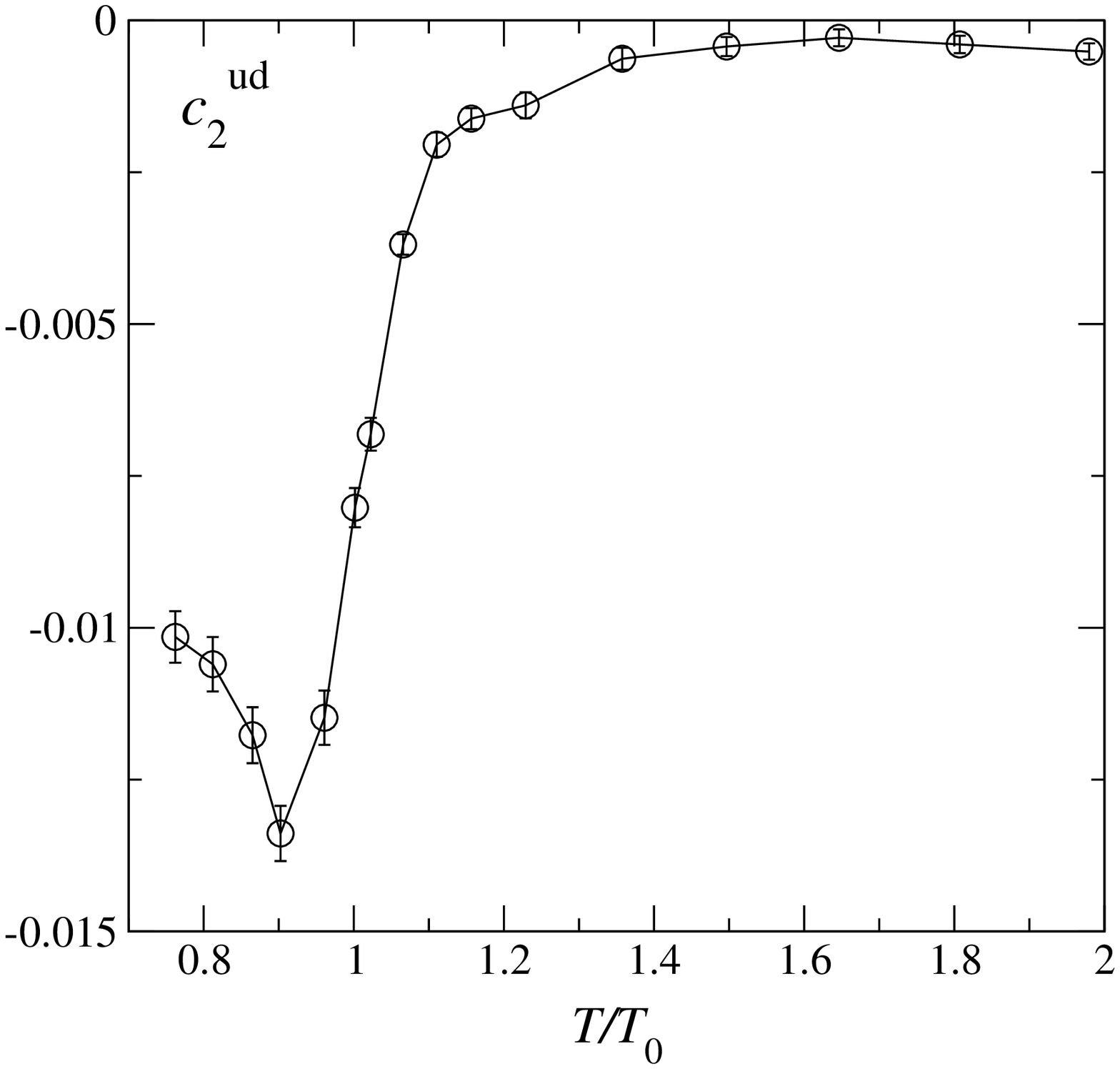, width=5.0cm}\\[-1mm]
\end{center}
\end{minipage}
\begin{minipage}[c][9.5cm][c]{5.0cm}
\begin{center}
\epsfig{file=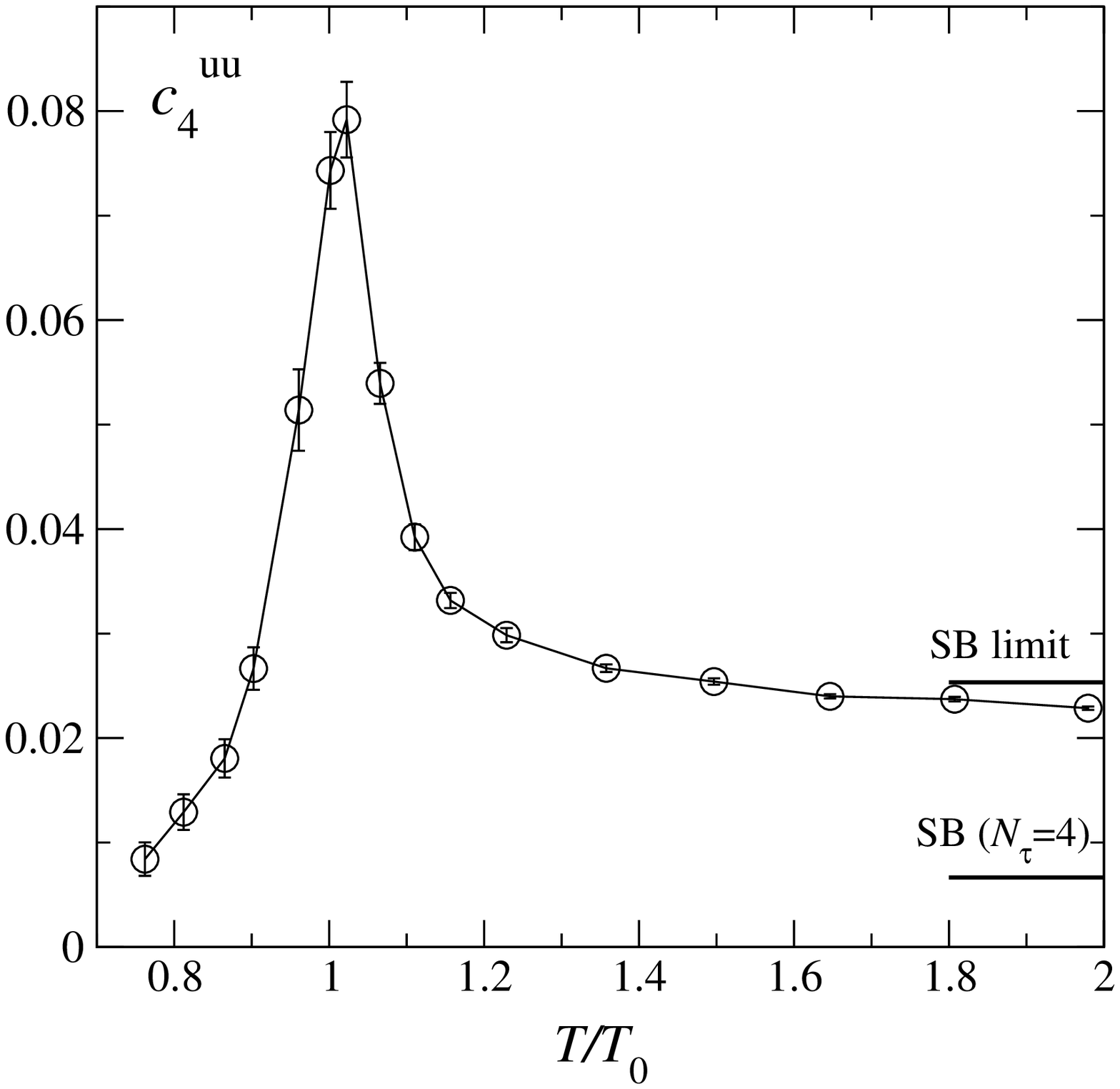, width=5.0cm}\\[-1mm]
\epsfig{file=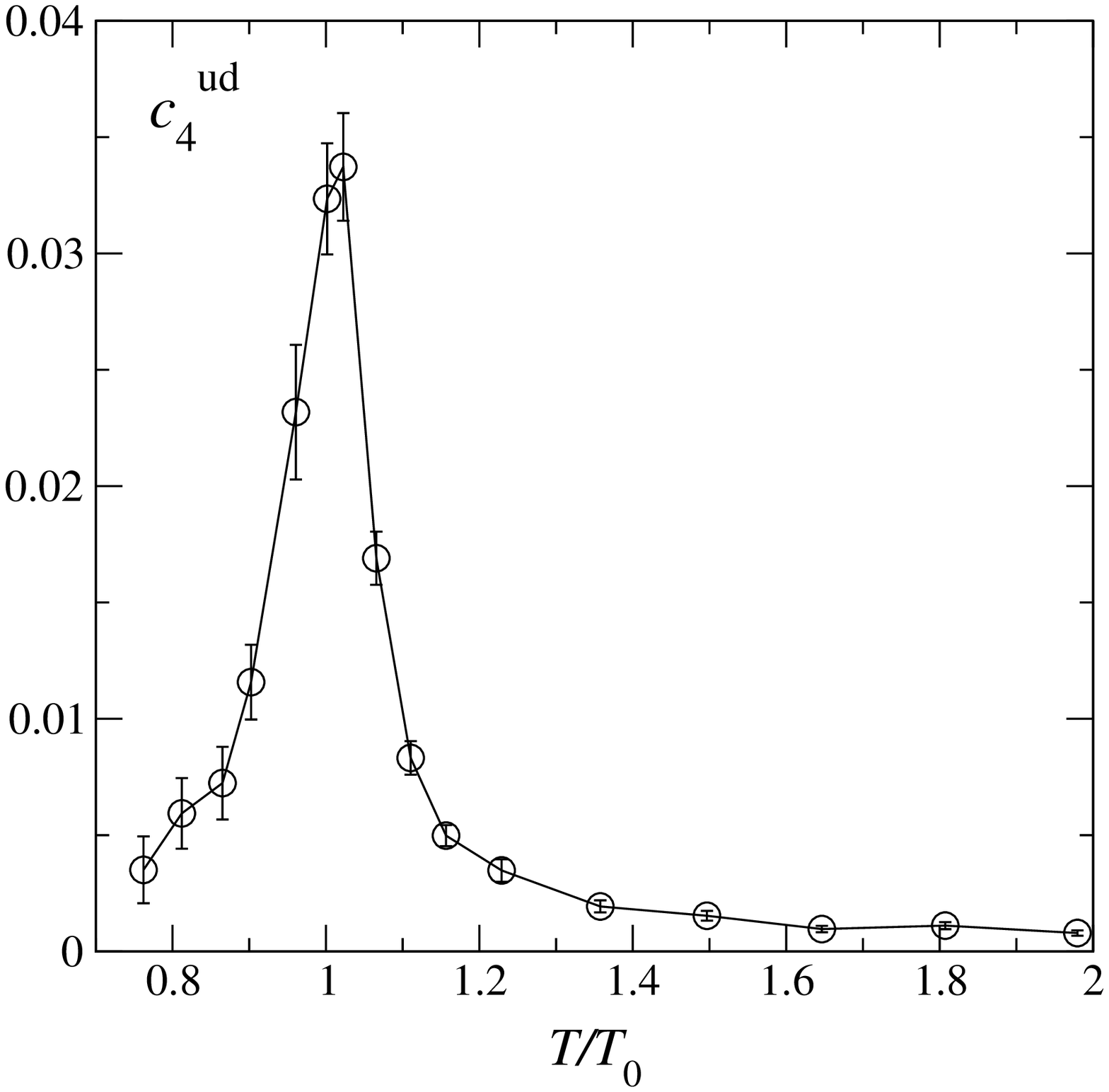, width=5.0cm}\\[-1mm]
\end{center}
\end{minipage}
\begin{minipage}[c][9.5cm][c]{5.0cm}
\begin{center}
\epsfig{file=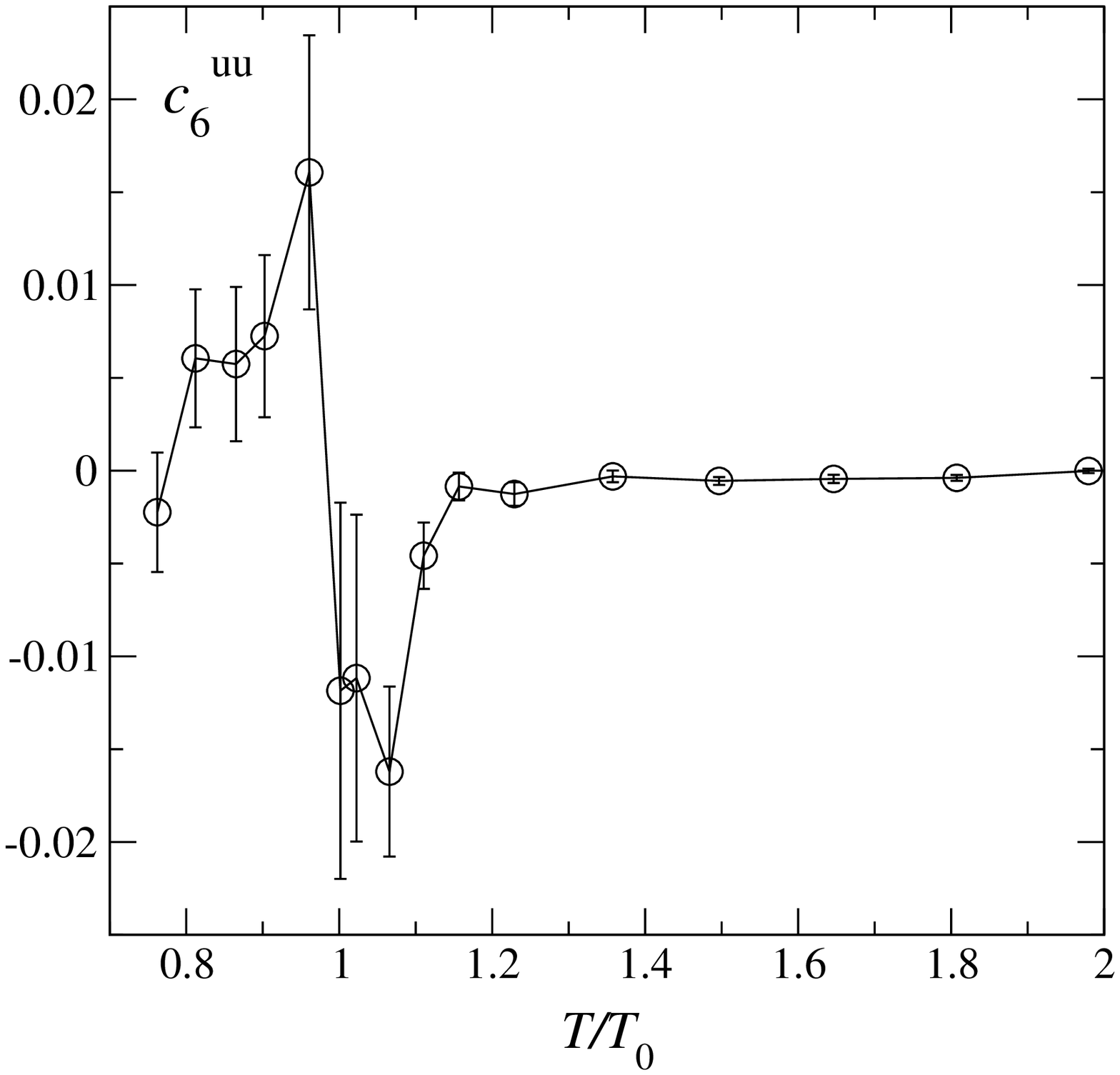, width=5.0cm}\\[-1mm]
\epsfig{file=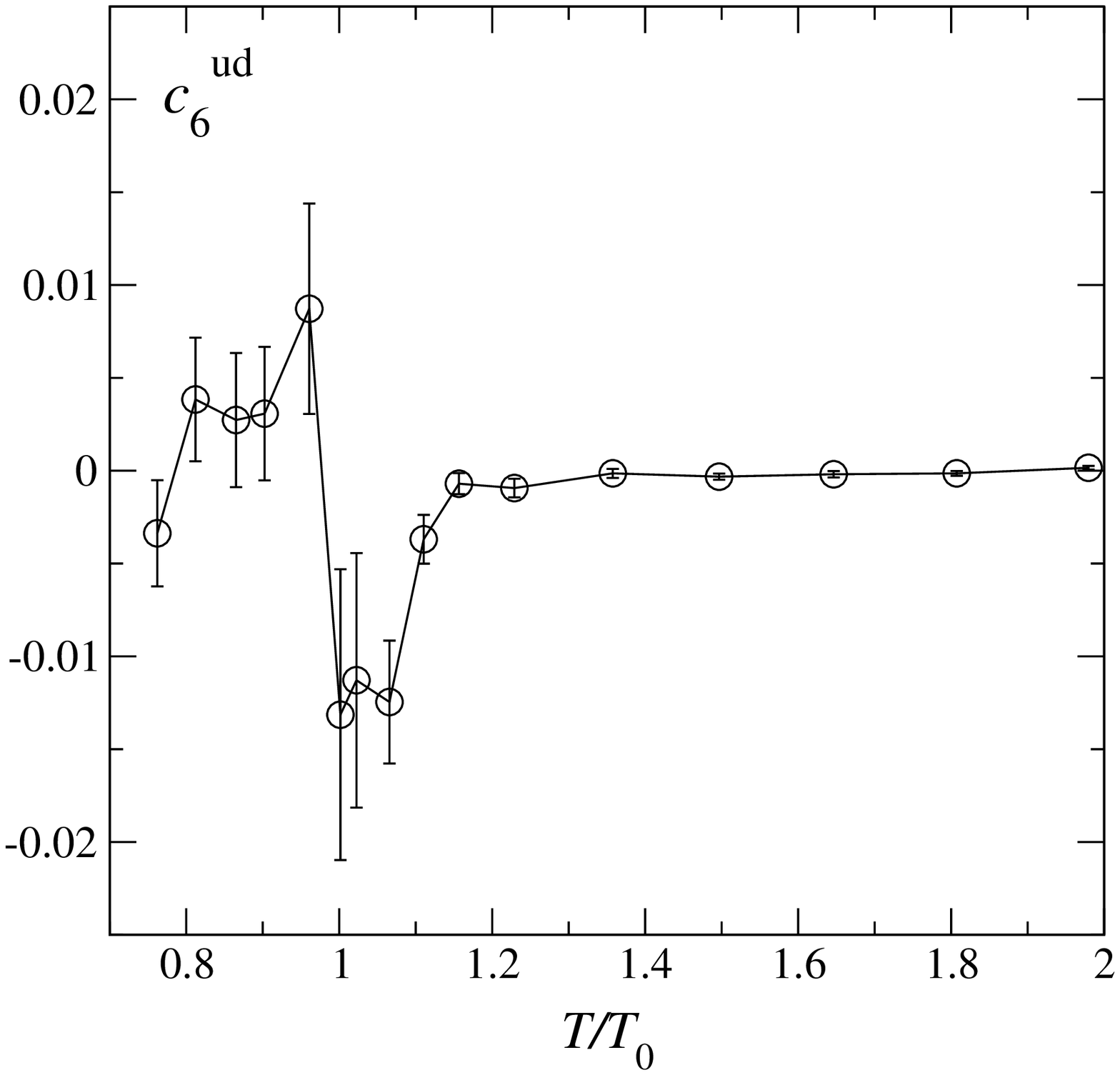, width=5.0cm}\\[-1mm]
\end{center}
\end{minipage}
\caption{The Taylor expansion coefficients $c_n^{uu}$ (upper row) of diagonal and 
$c^{ud}_n$ (lower row) of non-diagonal susceptibilities
for $n=2,~4$ and 6 as functions of $T/T_{0}$.}
\label{fig:uu}
\end{center}
\end{figure}

As discussed in the previous section the expansion coefficients $c_n$ and $c^I_n$ 
become quite similar at high temperature. This was to be expected from the 
discussion of the structure
of the high temperature perturbative expansion given in
section~\ref{subsec:pert} as $c_n$ and $c_n^I$ differ only by 
contributions coming from non-diagonal susceptibilities, which enter
with opposite sign in these two coefficients. It thus is instructive
to analyze directly the expansion coefficients of $\chi_{uu}$ and
$\chi_{ud}$. These are listed in Table~\ref{tab:cuu_n} and plotted in 
Fig.~(\ref{fig:uu}). We note that the errors on these quantities have been
obtained from an independent jackknife analysis and
thus are not simply obtained by adding errors for $c_n$ and $c_n^I$.

\begin{table}[h]
\setlength{\tabcolsep}{0.8pc}
\begin{tabular}{|l|lll|lll|}
\hline
$T/T_c$ & 
$c_2^{uu} \times 10^2$ & $c_4^{uu} \times 10^2$ & $c_6^{uu} \times 10^2$ &
$c_2^{ud} \times 10^2$ & $c_4^{ud} \times 10^2$ & $c_6^{ud} \times 10^2$ \\
\hline
0.76 & $\,\,$2.23(6)  & 0.84(16) & $\!\!$-0.22(32)  
     & $\!\!$-1.015(42) & 0.35(14) & $\!\!$-0.34(29)\\
0.81 & $\,\,$3.31(6)  & 1.29(17) &  0.60(37)  
     & $\!\!$-1.060(45) & 0.59(15) &  0.38(33)\\
0.87 & $\,\,$4.85(8)  & 1.81(18) &  0.57(42)  
     & $\!\!$-1.177(46) & 0.72(16) &  0.27(36)\\
0.90 & $\,\,$6.41(9)  & 2.67(20) &  0.72(44)  
     & $\!\!$-1.339(45) & 1.16(16) &  0.31(36)\\
0.96 & 11.95(12) & 5.14(39) &  1.61(74)  
     & $\!\!$-1.148(45) & 2.32(29) &  0.87(57)\\
1.00 & 18.31(14) & 7.43(37) & $\!\!$-1.19(101) 
     & $\!\!$-0.802(32) & 3.23(24) & $\!\!$-1.31(78)\\
1.02 & 21.82(15) & 7.92(36) & $\!\!$-1.12(88)  
     & $\!\!$-0.681(27) & 3.37(23) & $\!\!$-1.13(69)\\
1.07 & 29.49(11) & 5.39(20) & $\!\!$-1.62(46)  
     & $\!\!$-0.369(17) & 1.69(11) & $\!\!$-1.25(33)\\
1.11 & 33.11(9)  & 3.92(12) & $\!\!$-0.46(18)  
     & $\!\!$-0.205(20) & 0.83(7)  & $\!\!$-0.37(13) \\
1.16 & 35.62(7)  & 3.32(7)  & $\!\!$-0.08(8)   
     & $\!\!$-0.162(17) & 0.50(5)  & $\!\!$-0.07(6) \\
1.23 & 37.73(7)  & 2.98(7)  & $\!\!$-0.13(6)   
     & $\!\!$-0.140(22) & 0.35(5)  & $\!\!$-0.09(5) \\
1.36 & 39.47(5)  & 2.67(4)  & $\!\!$-0.03(3)   
     & $\!\!$-0.063(18) & 0.19(3)  & $\!\!$-0.01(2) \\
1.50 & 40.34(4)  & 2.54(3)  & $\!\!$-0.06(2)   
     & $\!\!$-0.043(16) & 0.15(2)  & $\!\!$-0.03(2) \\
1.65 & 40.81(4)  & 2.40(2)  & $\!\!$-0.04(2)   
     & $\!\!$-0.029(14) & 0.10(1)  & $\!\!$-0.02(2) \\
1.81 & 41.05(3)  & 2.37(2)  & $\!\!$-0.04(2)   
     & $\!\!$-0.040(14) & 0.11(2)  & $\!\!$-0.01(1) \\
1.98 & 41.20(3)  & 2.29(2)  & $\!\!$-0.00(1)   
     & $\!\!$-0.051(13) & 0.08(1)  &  0.02(1) \\
\hline
\end{tabular}
\caption{Taylor expansion coefficients $c_n^{uu}(T)$ and $c_n^{ud}(T)$.}
\smallskip
\label{tab:cuu_n}
\end{table}

Fig.~(\ref{fig:uu}) clearly shows that for $T>T_0$ 
the various expansion coefficients 
rapidly approach the corresponding ideal gas values, which is zero for
all non-diagonal expansion coefficients, $c^{ud}_n$. In fact, as discussed
in section~\ref{subsec:pert} the latter receive contributions
only at ${\cal O}(g^6\ln 1/g^2)$ for $n=2$ and ${\cal O}(g^3)$ for $n>2$.
Moreover, it
is interesting to note that despite the small magnitude of these
contributions the leading order perturbative results correctly
predict the sign of all expansion coefficients 
for $T>T_0$, {\it i.e.} $c_{2,6}^{ud}<0$, $c_4^{ud} >0$, and 
$c_{2,4}^{uu} > 0$, $c_6^{uu} < 0$.  Furthermore, the
order of magnitude for $c_2^{ud}$, {\it i.e.} $|c_2^{ud} | \simeq 5\cdot 10^{-4}$
at $T\simeq 2 T_0$, 
agrees with the perturbative estimate\footnote{A previous analysis
\cite{Gupta02} reported values for $c^{ud}_2$ consistent
with zero for temperatures $T/T_0\ge 1.25$ within an error of $10^{-6}$. 
This has been found subsequently to be incorrect. The corrected
values \cite{Gavai04} are qualitatively
consistent with our findings. (The second panel
of Fig.~10 in \cite{Gavai04} should be compared to twice the value of
$c_2^{ud}$ shown in our Fig.~(\ref{fig:uu}).) 
Similar agreement is found with the results of
\cite{suscept2}. 
} 
\cite{Blaizot}.  

A striking feature of the expansion coefficients $c_n^{uu}$ and
$c_n^{ud}$ is that for $n>2$ they become similar in magnitude close to 
$T_0$.  In fact, $c_4^{uu}$ and $c_4^{ud}$ both have pronounced peaks at 
$T_0$ with $(c_4^{ud}/c_4^{uu})_{\rm peak} \simeq 0.45$ and the expansion 
coefficients for $n=6$ are identical within errors. This suggests that
any divergent piece in $\chi^{uu}$, which could occur when $\mu_q/T$
approaches the radius of convergence of the Taylor expansion, will
show up with identical strength also in $\chi^{ud}$. This, in turn,
implies that the singular behavior will add up constructively in the
quark number susceptibility whereas it can cancel in the isovector
susceptibility giving rise to finite values for $\chi_I$ at such a 
critical point. Even for smaller, non-critical values of $\mu_q/T$, however, 
the rapid
rise of $c_4^{ud}(T)$ for $T\simeq T_0$ is important. It shows that
non-diagonal susceptibilities will become large at non-zero chemical 
potential in the transition region from the low to the high temperature
phase, {\it i.e.} fluctuations in different flavor channels, which are
uncorrelated at high temperature, become strongly correlated in the 
transition region. This correlation is also reflected in the errors of the 
various expansion coefficients, which are of similar size for $c_n^{uu}$ and 
$c_n^{ud}$ but much reduced in the difference, $\chi_I$.

\begin{figure}[tb]
\begin{center}
\begin{minipage}[c][7.8cm][c]{7.4cm}
\begin{center}
\epsfig{file=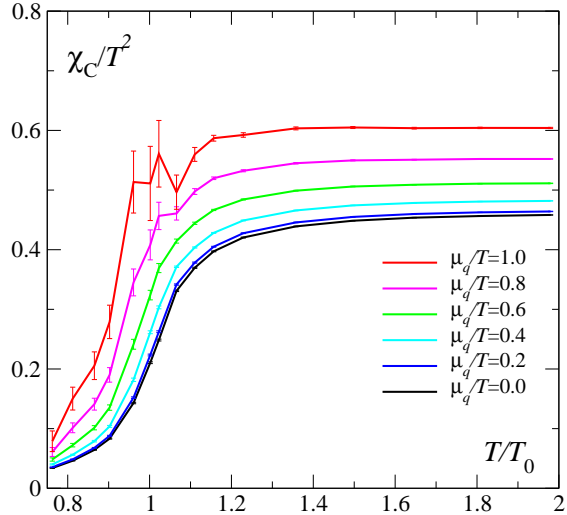, width=7.4cm}\\[-1mm]
\end{center}
\end{minipage}
\caption{The charge susceptibility $\chi_C/T^2$ as a function of $T/T_0$ for
various $\mu_q/T$ ranging from $\mu_q/T=0$ (lowest curve) rising in
steps of 0.2 to  $\mu_q/T=1$, calculated
from a Taylor series in $6^{th}$ order.} 
\label{fig:charge}
\end{center}
\end{figure}

The above considerations also suggest that the electric charge susceptibility,
\begin{eqnarray}
\chi_C=\left({2\over3}{{\partial\;}\over{\partial\mu_u}}
-{1\over3}{{\partial\;}\over{\partial\mu_d}}\right)
\left({2\over3}{{\partial p}\over{\partial\mu_u}}
-{1\over3}{{\partial p}\over{\partial\mu_d}}\right)
&=&\frac{1}{9}\left( 4\chi_{uu}+\chi_{dd} -4\chi_{ud}\right) \nonumber \\
~&=& \frac{1}{4}\left( \chi_I + {1\over 9}\chi_q \right) \quad ,
\end{eqnarray}
will be singular whenever the diagonal and non-diagonal susceptibilities
are singular as the cancellation between the corresponding singular parts 
will be incomplete. We show the charge susceptibility in 
Fig.~(\ref{fig:charge}).
As it is dominated by the contribution from the isovector susceptibility
any possible singular contribution arising from $\chi_q$ will be weak. It thus
may not be too surprising that a peak does not yet show up in $\chi_C$.  

\subsection{Mass derivatives and chiral condensate}
\label{subsec:chiral}

The transition between low and high temperature phases of strongly
interacting matter is expected to be closely related to chiral symmetry
restoration. It is therefore also of interest to analyze the
dependence of the chiral condensate on the quark chemical potential.
We will do so in the framework of a Taylor expansion of the grand
potential,
\begin{equation}
\frac{\langle \bar{\psi} \psi \rangle}{T^3} 
=\left( \frac{N_\tau}{N_{\sigma}}\right)^3 
\frac{\partial \ln{\cal Z}}{\partial m/T}
= \sum_{n=0}^{\infty} c_n^{\bar{\psi}\psi}(T) \left(
\frac{\mu_q}{T}\right)^n \quad ,
\label{eq:pbp}
\end{equation}
with
\begin{equation}
c_n^{\bar{\psi} \psi} = \frac{1}{n!} \left. 
\frac{\partial^n \langle \bar{\psi} \psi \rangle/T^3}{\partial (\mu_q/T)^n} 
\right|_{\mu_q=\mu_I=0} = \frac{1}{n! } \left. 
\frac{\partial^{n+1} \Omega}{\partial (\mu_q/T)^n \partial m/T} 
\right|_{\mu_q=\mu_I=0}
\quad . 
\label{eq:pbpcoef}
\end{equation}
Here we expressed the bare lattice quark masses, $ma$, in units of the
temperature by using $m/T\equiv ma N_\tau$.
For $n>0$ the expansion coefficients of the chiral condensate, 
are directly related to derivatives of the 
expansion coefficients of the grand potential $\Omega$ with respect
to the quark mass, {\it i.e.} $c_n^{\bar{\psi} \psi} = \partial
c_n/\partial (m/T)$. For $n=0$ this holds true up to a contribution 
arising from the normalization of the pressure at $(T=0,\mu_q=0)$.
As such the coefficients $c_n^{\bar{\psi} \psi}$ also provide information on 
the quark mass
dependence of other thermodynamic observables like pressure, number density 
or susceptibilities. For instance, the change of the quark number
susceptibility with quark mass is given by
\begin{equation}
\frac{\partial \chi_q/T^2}{\partial m/T} 
= 2c_2^{\bar{\psi} \psi}+12c_4^{\bar{\psi} \psi}\left({\mu_q\over T}\right)^2+
{\cal O}(\mu_q^4) \quad .
\label{eq:chiqm}
\end{equation}
\begin{figure}[tb]
\begin{center}
\begin{minipage}[c][5.2cm][c]{5.0cm}
\begin{center}
\epsfig{file=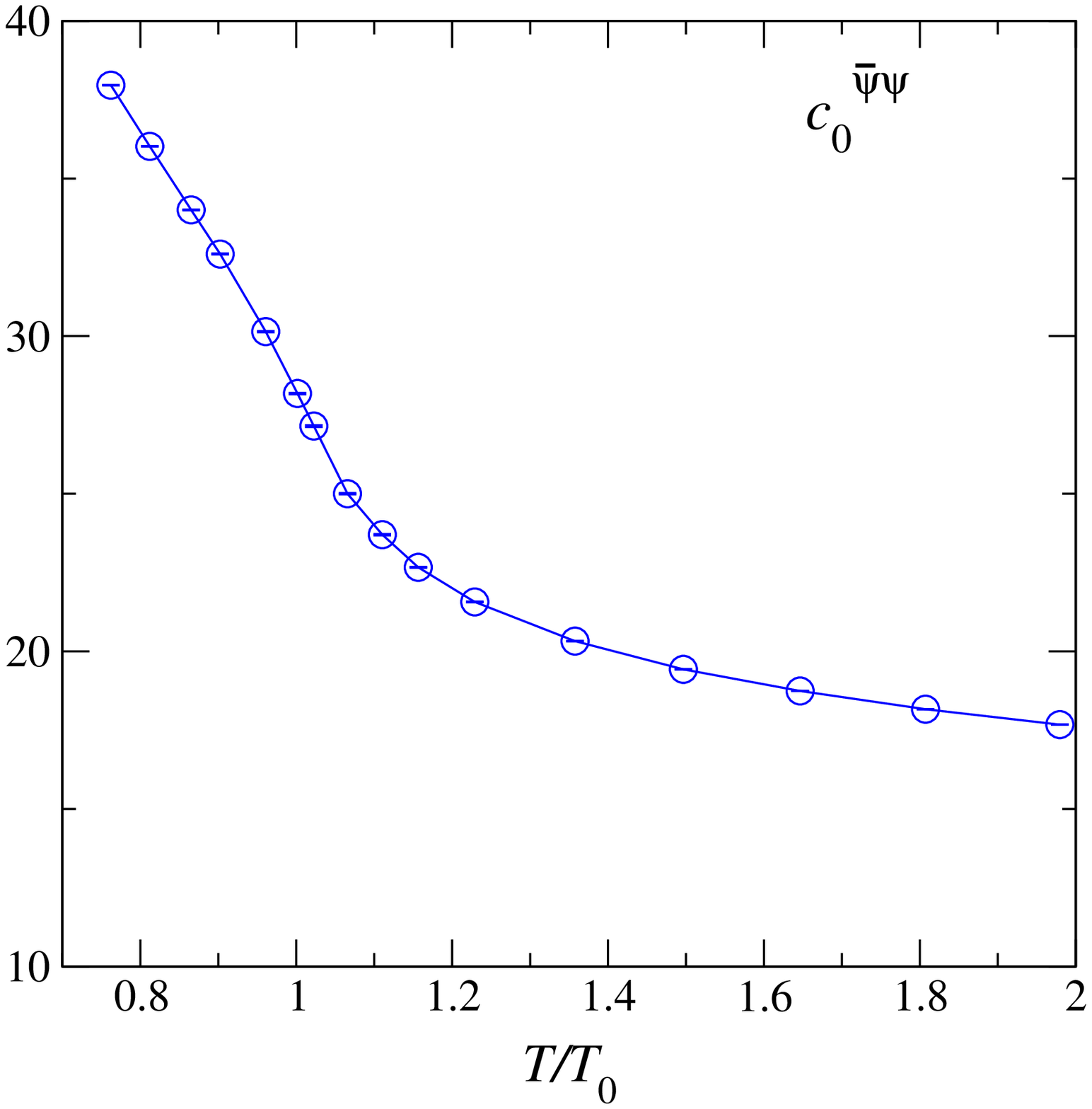, width=5.0cm}\\[-1mm]
\end{center}
\end{minipage}
\begin{minipage}[c][5.2cm][c]{5.0cm}
\begin{center}
\epsfig{file=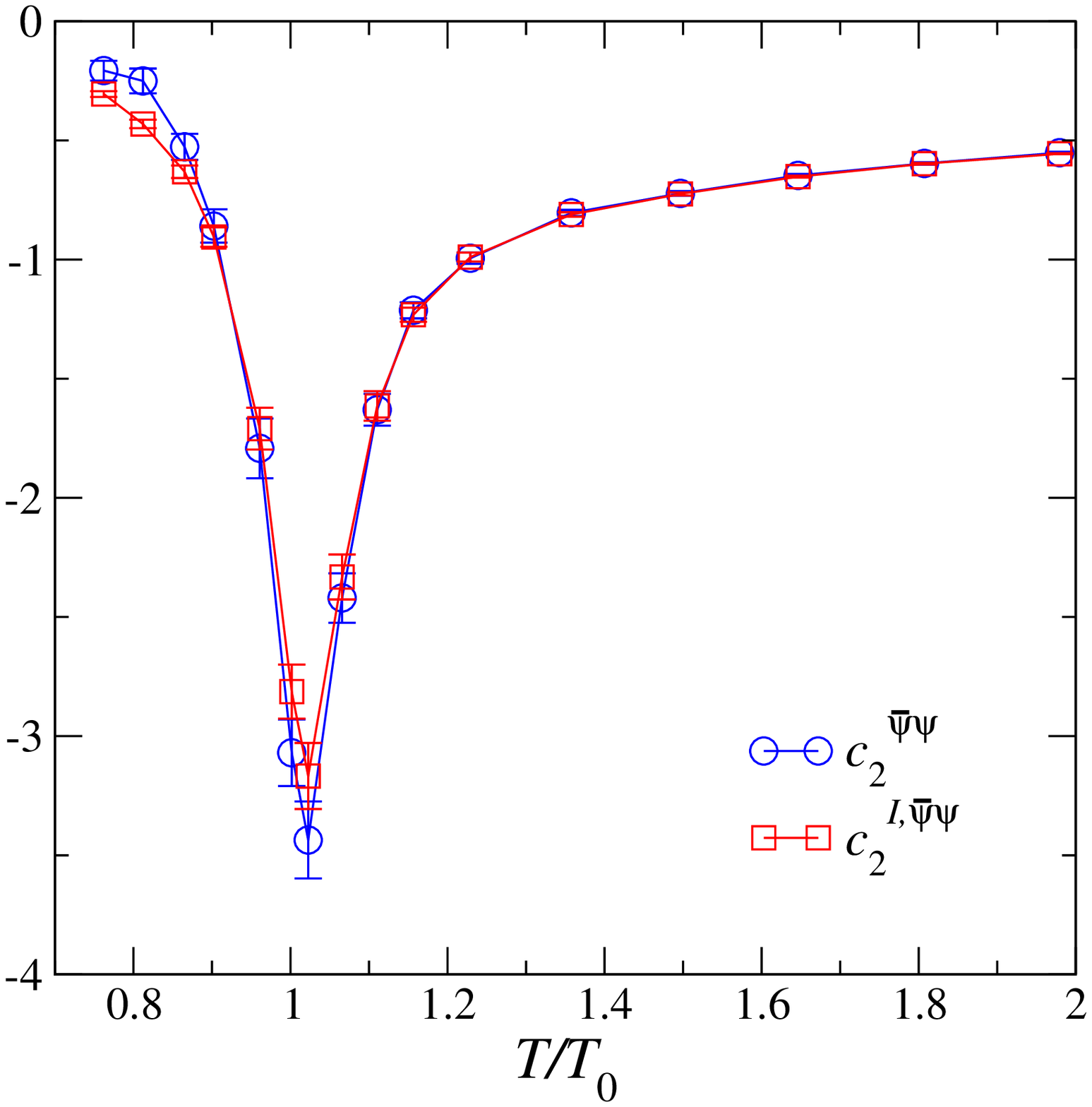, width=5.0cm}\\[-1mm]
\end{center}
\end{minipage}
\begin{minipage}[c][5.2cm][c]{5.0cm}
\begin{center}
\epsfig{file=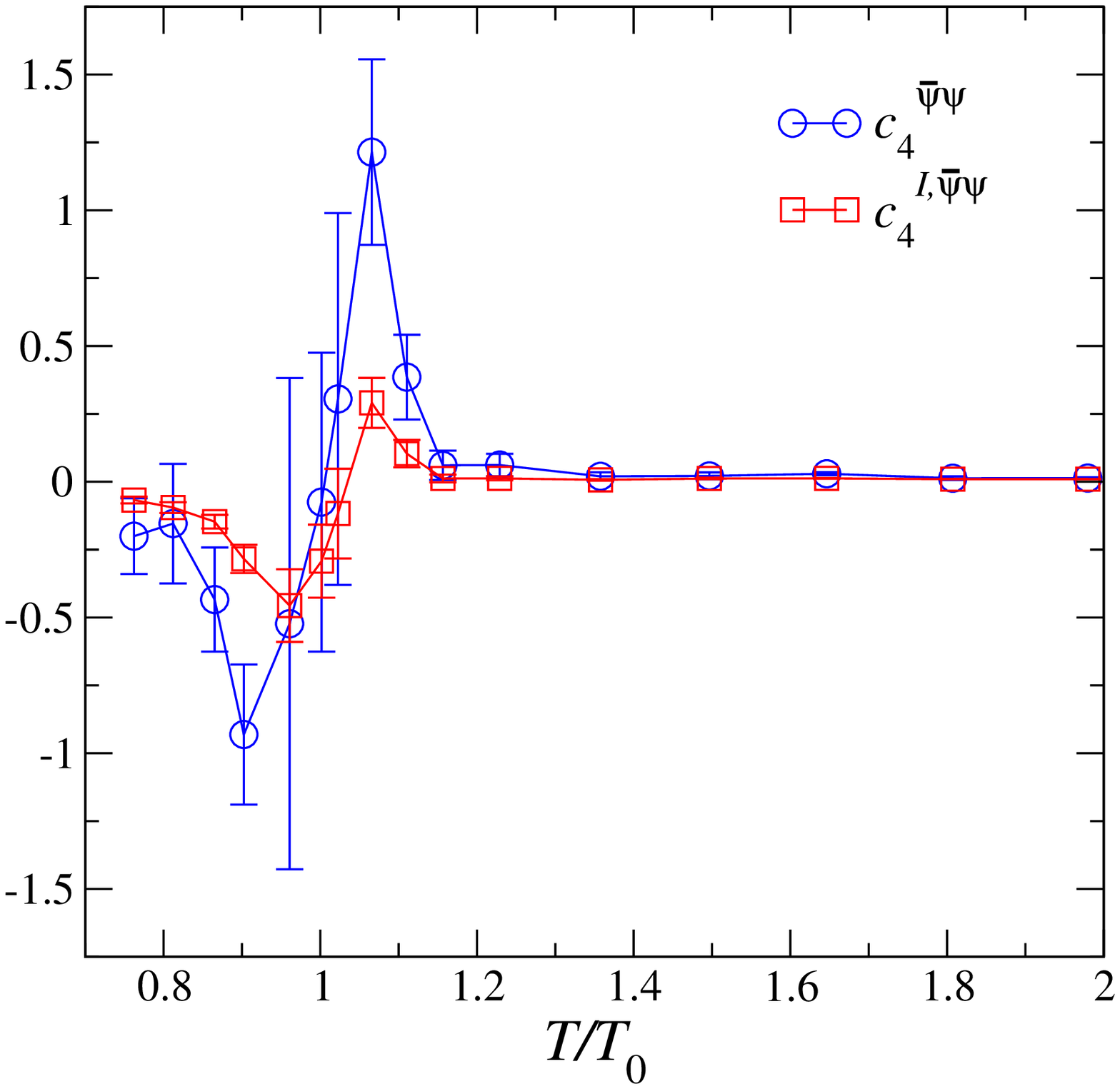, width=5.0cm}\\[-1mm]
\end{center}
\end{minipage}
\caption{The Taylor expansion coefficients $c_n^{\bar{\psi} \psi}$ 
of the chiral condensate for $n=0, 2$ and 4 as functions of $T/T_{0}$.
Also shown are the coefficients $c_n^{I,\bar{\psi} \psi}$ for $n=0, 2$
which define the quark mass derivatives of the isovector susceptibility
in analogy to Eq.~(\ref{eq:chiqm}).} 
\label{fig:dcc}
\end{center}
\end{figure}
We have calculated the derivatives of $c_n$ with respect to the quark mass 
for $n=0$, 2 and 4.  These derivatives are shown in Fig.~(\ref{fig:dcc})
together with the corresponding derivatives for the expansion coefficients 
of the isovector susceptibility, which have a similar temperature dependence,
\begin{equation}
\frac{\partial \chi_I/T^2}{\partial m/T} 
= 2c_2^{I,\bar{\psi} \psi}+
12c_4^{I,\bar{\psi} \psi}\left({\mu_q\over T}\right)^2+
{\cal O}(\mu_q^4) \quad .
\label{eq:chiIm}
\end{equation}
We note that the expansion coefficients $c_n^{\bar{\psi} \psi}$ are negative for 
$n>0$ and $T\le 0.96 T_0$. The chiral condensate thus will drop at fixed 
temperature
with increasing $\mu_q/T$ and the chiral susceptibilities will increase 
in the hadronic phase with decreasing quark mass. This, together with
the change of sign in $c_4^{\bar{\psi} \psi}$ at $T\simeq T_0$, will shift 
the transition point at non-zero $\mu_q/T$ to lower temperatures. In 
Fig.~(\ref{fig:ccn}) we show the chiral condensate and the related chiral 
susceptibility\footnote{The chiral susceptibility introduced here is not 
the complete derivative of the chiral condensate with respect to the quark 
mass. 
As frequently done also at $\mu_q = 0$ we define the chiral susceptibility 
by ignoring a contribution from the connected part which would arise 
in the derivative $\partial \langle \bar{\psi} \psi \rangle / \partial m$. 
Nonetheless $\chi_{\bar{\psi} \psi}$ seems to capture the leading singular 
behavior that should show up at a $2^{nd}$ order critical point 
\cite{FKEL}.} 
obtained from a Taylor expansion up to and including ${\cal O}(\mu_q^4)$,
\begin{eqnarray}
\frac{\chi_{\bar{\psi} \psi}}{T^2} 
&=& \frac{N_\tau}{N_\sigma^3} \left( \frac{n_{\rm f}}{4}\right)^2 
\left[ \left\langle \left( {\rm tr} M^{-1} \right)^2 \right\rangle 
- \left\langle {\rm tr} M^{-1} \right\rangle^2 
\right] \nonumber \\
&=& c_0^{\chi}+c_2^{\chi} \left( \frac{\mu_q}{T}\right)^2
+c_4^{\chi} \left( \frac{\mu_q}{T}\right)^4 +  {\cal O}\left(
\left( \mu_q/T\right)^6 \right) 
\quad .
\label{eq:chisus} 
\end{eqnarray}
Obviously $\chi_{\bar{\psi} \psi}$ develops a much
more pronounced peak for $\mu_q/T >0$ than at vanishing chemical potential
which, moreover, is shifted to smaller temperatures. 
However, as will become clear from the discussion in the next section
the peaks found in $\chi_{\bar{\psi}\psi}$  and also in other susceptibilities 
have to be analyzed and interpreted carefully. They reflect the abrupt 
transition from the
hadronic regime to the high temperature phase in which fluctuations of the 
chiral condensate are suppressed, but do not signal the presence of 
a $2^{nd}$ order phase transition unambiguously. The rapid rise of 
susceptibilities in the hadronic phase 
is strongly correlated to the increase in the pressure and is also
present in a hadron gas which does not show any singular behavior at the
transition temperature.
\begin{figure}
\begin{center}
\epsfig{file=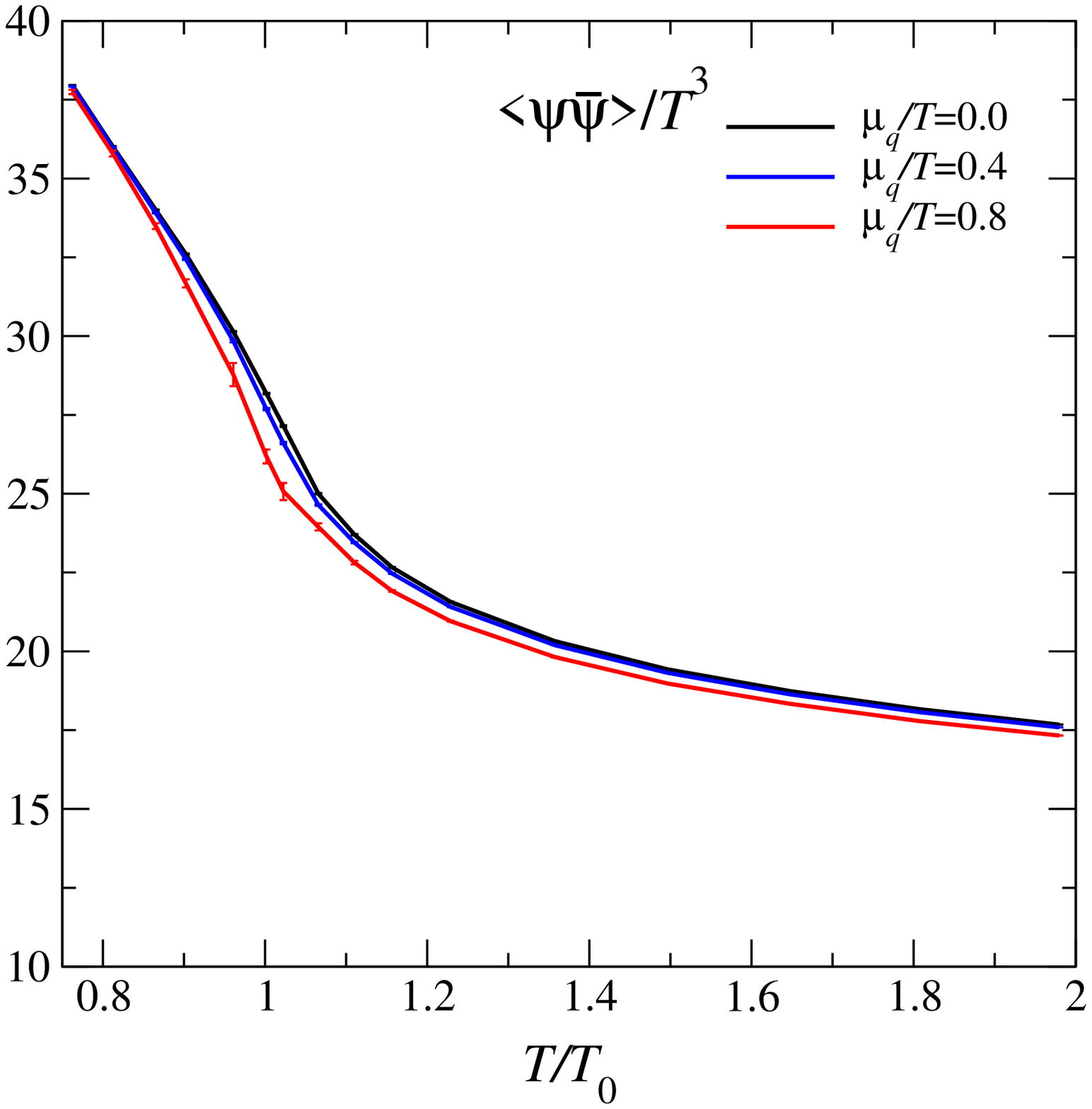, width=7.4cm}
\epsfig{file=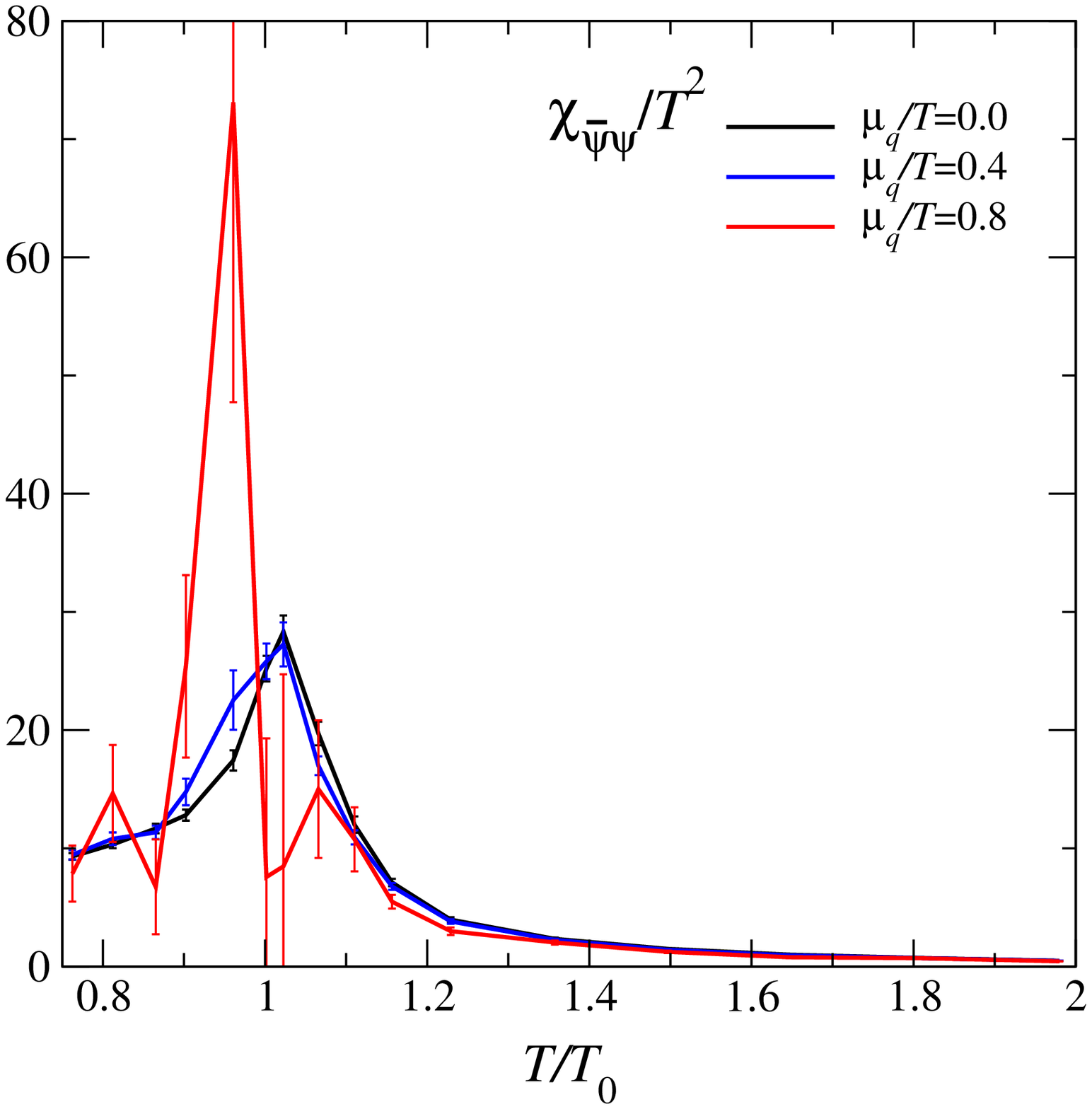, width=7.4cm}
\smallskip
\caption{The chiral condensate $\langle \bar{\psi} \psi \rangle$ (left)
and chiral susceptibility $\chi_{\bar{\psi} \psi}$ (right)
as a function of $T/T_0$ for $\mu_q/T=0,~0.4$ and 0.8. The chiral
condensate drops with increasing $\mu_q/T$ and the peak in
$\chi_{\bar{\psi} \psi}$ becomes more pronounced.}
\label{fig:ccn}
\end{center}
\end{figure}

Also the expansion of the chiral condensate and related observables are 
compatible with the HRG model. In fact, a comparison of the
temperature dependence of $c_n^{\bar{\psi} \psi}$ shown in Fig.~(\ref{fig:dcc})
with that of the expansion coefficients $c_n$ of the grand potential shown 
in Fig.~(\ref{fig:c2c4}) suggests a strong similarity between
$-c_n^{\bar{\psi} \psi}$ and $c_{n+2}$. On the other
hand, for $T<T_0$ the ratio $c_4^{\bar{\psi} \psi} /c_2^{\bar{\psi} \psi}$ 
agrees within errors with the ratio $c_4/c_2$ shown in 
Fig.~(\ref{fig:rc2c4c6}). All this is consistent with the HRG 
model where the quark mass (or spectrum) dependence only enters through 
the functions $F(T)$ and $G(T)$ and does not modify the dependence on 
$\mu_q/T$.

\section{Radius of convergence and the hadron resonance gas}
\label{sec:radius}

So far we have not discussed the range of validity of the Taylor expansion.
In general the Taylor series will only converge for 
$\mu_q/T < \rho$ (or $\mu_q/T \le \rho$) where the radius of convergence,
$\rho$,
is determined by the zero of ${\cal Z} (T,\mu_q,\mu_q)$ closest to the origin
of the complex $\mu_q$ plane. If this zero happens to lie on the 
real axis the radius of convergence coincides with a critical point of the
QCD partition function. A sufficient condition for this is that
all expansion coefficients are positive \cite{Gaunt}. 
Apparently this is the case for
all coefficients $c_n(T)$ with $T/T_0 < 0.96$ that have been calculated so
far by us\footnote{In \cite{Gavai04} it is reported that $c_8(T)$ is 
negative for $T<0.95T_0$, however, the statistical significance of this
result unfortunately is not given.}. 
Above $T_0$, however, we find from the calculation
of $c_6(T)$ that the expansion coefficients do not stay strictly positive.
This is in accordance with our expectation to find a chiral critical point
at some temperature $T < T_0$.

\begin{figure}[tb]
\begin{center}
\begin{minipage}[c][5.2cm][c]{5.0cm}
\begin{center}
\epsfig{file=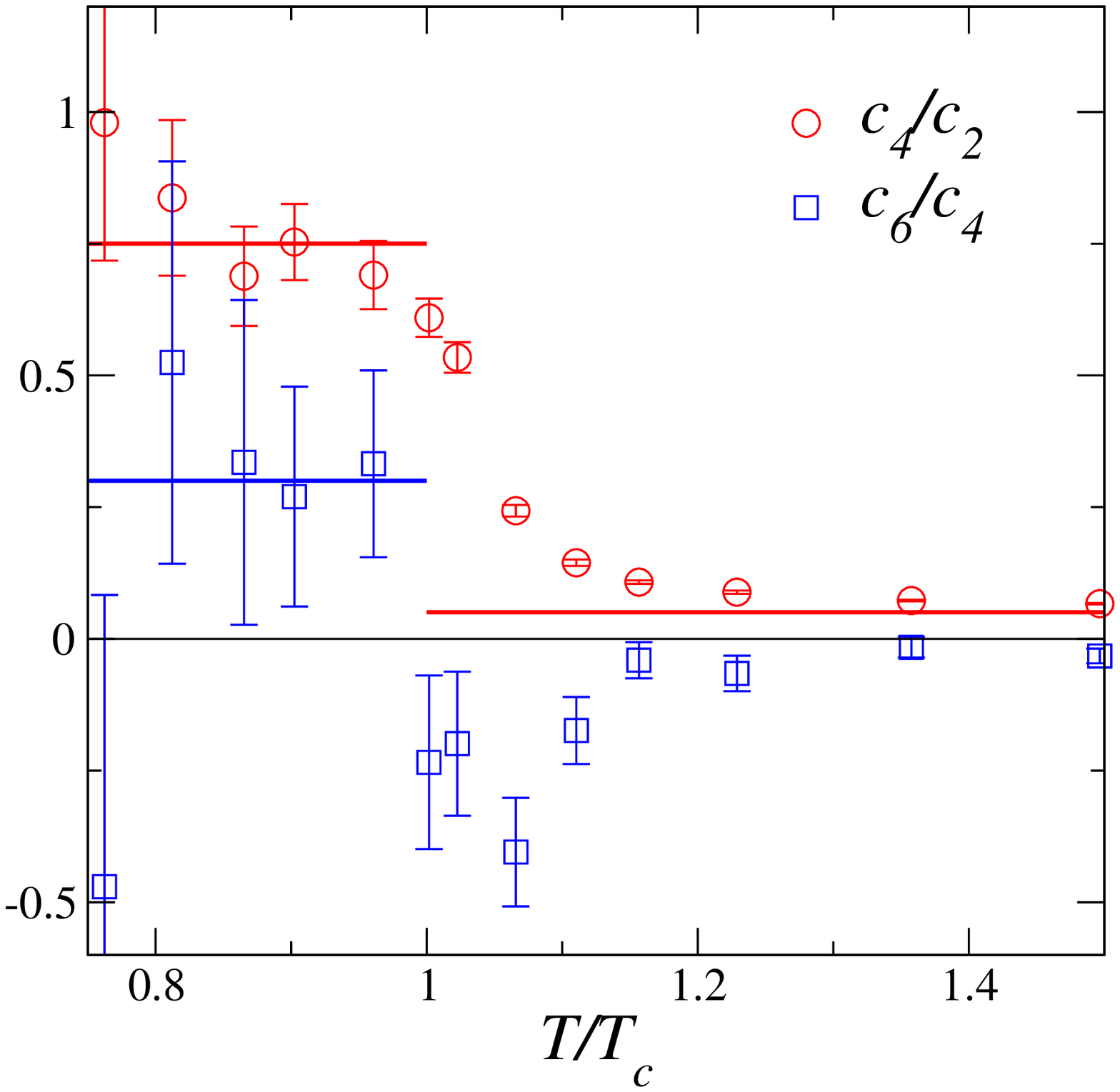, width=5.0cm}\\[-1mm]
(a)
\end{center}
\end{minipage}
\begin{minipage}[c][5.2cm][c]{5.0cm}
\begin{center}
\epsfig{file=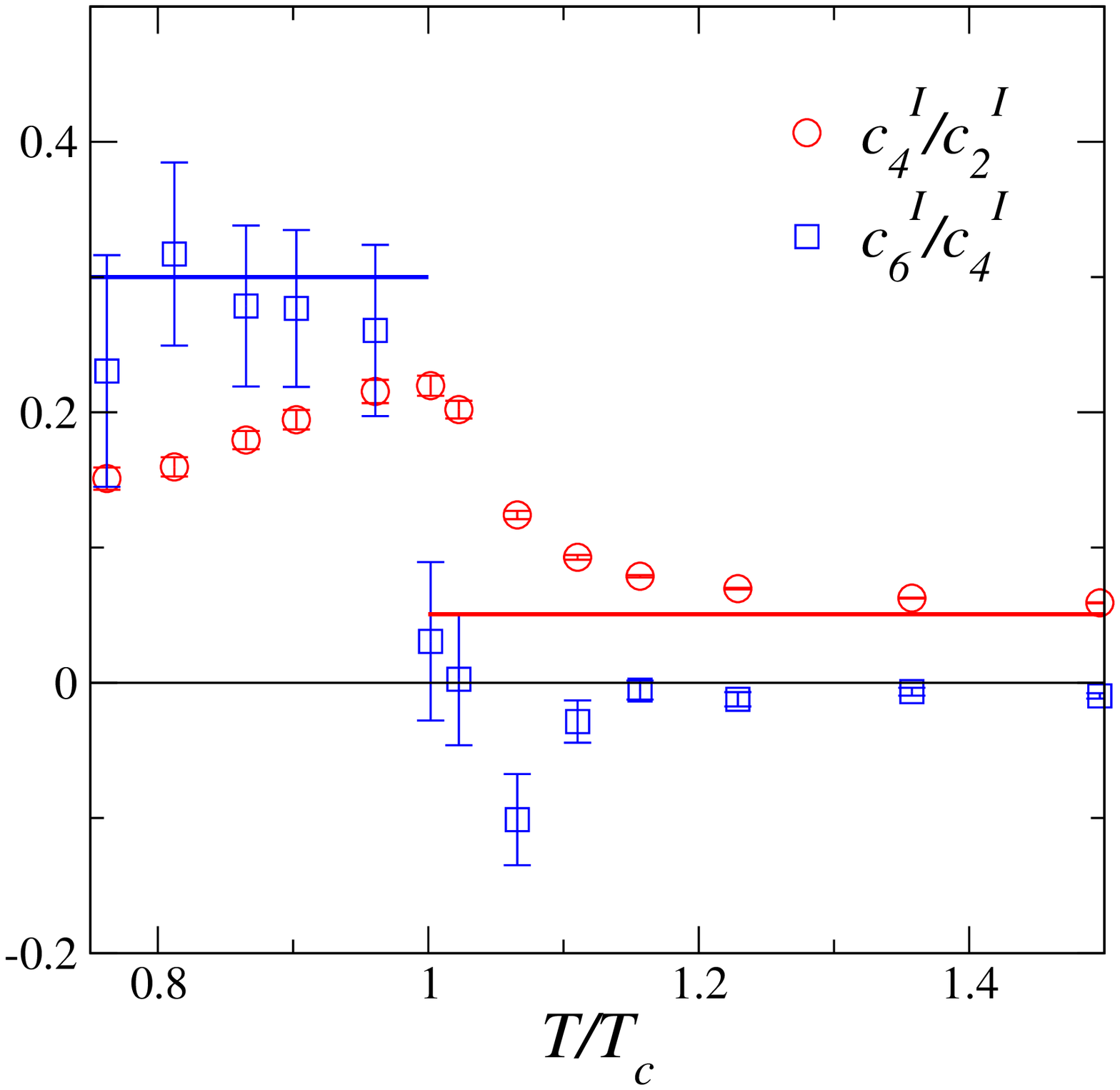, width=5.0cm}\\[-1mm]
(b)
\end{center}
\end{minipage}
\begin{minipage}[c][5.2cm][c]{5.0cm}
\begin{center}
\epsfig{file=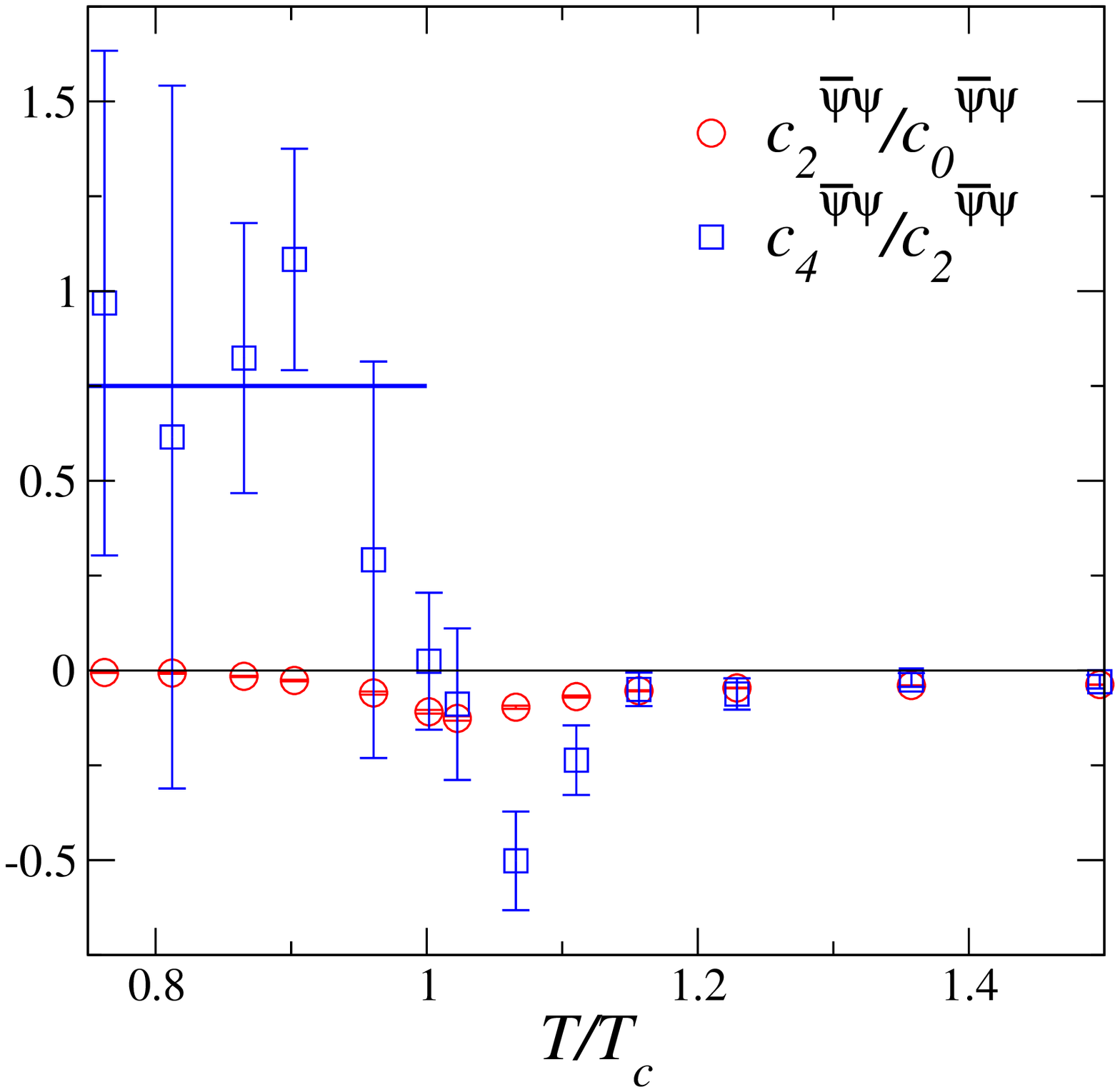, width=5.0cm}\\[-1mm]
(c)
\end{center}
\end{minipage}
\caption{(a) The ratios $c_4/c_2$ and $c_6/c_4$, and
(b) $c^I_4/c^I_2$ and $c^I_6/c^I_4$ 
as well as (c) $c_2^{\bar{\psi}\psi}/c_0^{\bar{\psi}\psi}$ and 
$c_4^{\bar{\psi}\psi}/c_2^{\bar{\psi}\psi}$ as functions of $T$.
Horizontal lines indicate the HRG prediction for $T<T_{0}$, and
the SB prediction for $T>T_{0}$. Note the difference in vertical scale 
between the plots.}
\label{fig:rc2c4c6}
\end{center}
\end{figure}
  
The radius of convergence of the Taylor series
for $\Omega (T,\mu_q,\mu_q)$ can be estimated by inspecting ratios of subsequent
expansion coefficients,
\begin{equation}
\rho=\lim_{n\to\infty}\rho_{2n}\equiv\lim_{n\to\infty}\sqrt{\biggl\vert
{c_{2n}\over c_{2n+2}}\biggr\vert} \quad ,
\end{equation}
where the square root arises because the Taylor expansion of the grand
potential $\Omega$ is an even series in $\mu_q/T$. The ratios $c_{2n+2}/c_{2n}$
are shown in Fig.~(\ref{fig:rc2c4c6}) together with ratios of the expansion 
coefficients $c_n^I$ of the isovector susceptibility
and $c_n^{\bar{\psi}\psi}$ of the chiral condensate.
It is obvious that these ratios rapidly change across $T_0$ and approach the value
of corresponding ratios obtained in the high temperature ideal gas limit.
Another remarkable feature, however, is that below $T_0$ the ratios
involving expansion coefficients of the $\mu_q$-dependent parts of
$\Omega$, $\chi_I$ and $\chi_{\bar{\psi}\psi}$ are almost temperature independent.
In fact, these ratios are consistent with the corresponding ratios deduced
from the grand potential of a hadron resonance gas (Eq.~(\ref{eq:HRGratios})),
{\it i.e.} ${c_4/ c_2}=c_4^{\bar{\psi}\psi}/c_2^{\bar{\psi}\psi}={3/4}$ 
and ${c_6/ c_4}={c^I_6/ c^I_4}={3/10}$. In  
ratios that contain the lowest order expansion coefficients, {\it i.e.}
$c_0$, $c_0^{\bar{\psi}\psi}$ and $c_2^I$, the spectrum dependence
does not cancel because the lowest order expansion coefficients also
depend on the meson sector which is not the case for higher order 
coefficients. These ratios thus show a significant temperature 
dependence as can be seen for $c^I_4/ c^I_2$ and
$c_2^{\bar{\psi}\psi}/  c_0^{\bar{\psi}\psi}$ shown in Fig.~(\ref{fig:rc2c4c6}).

\begin{figure}
\begin{center}
\epsfig{file=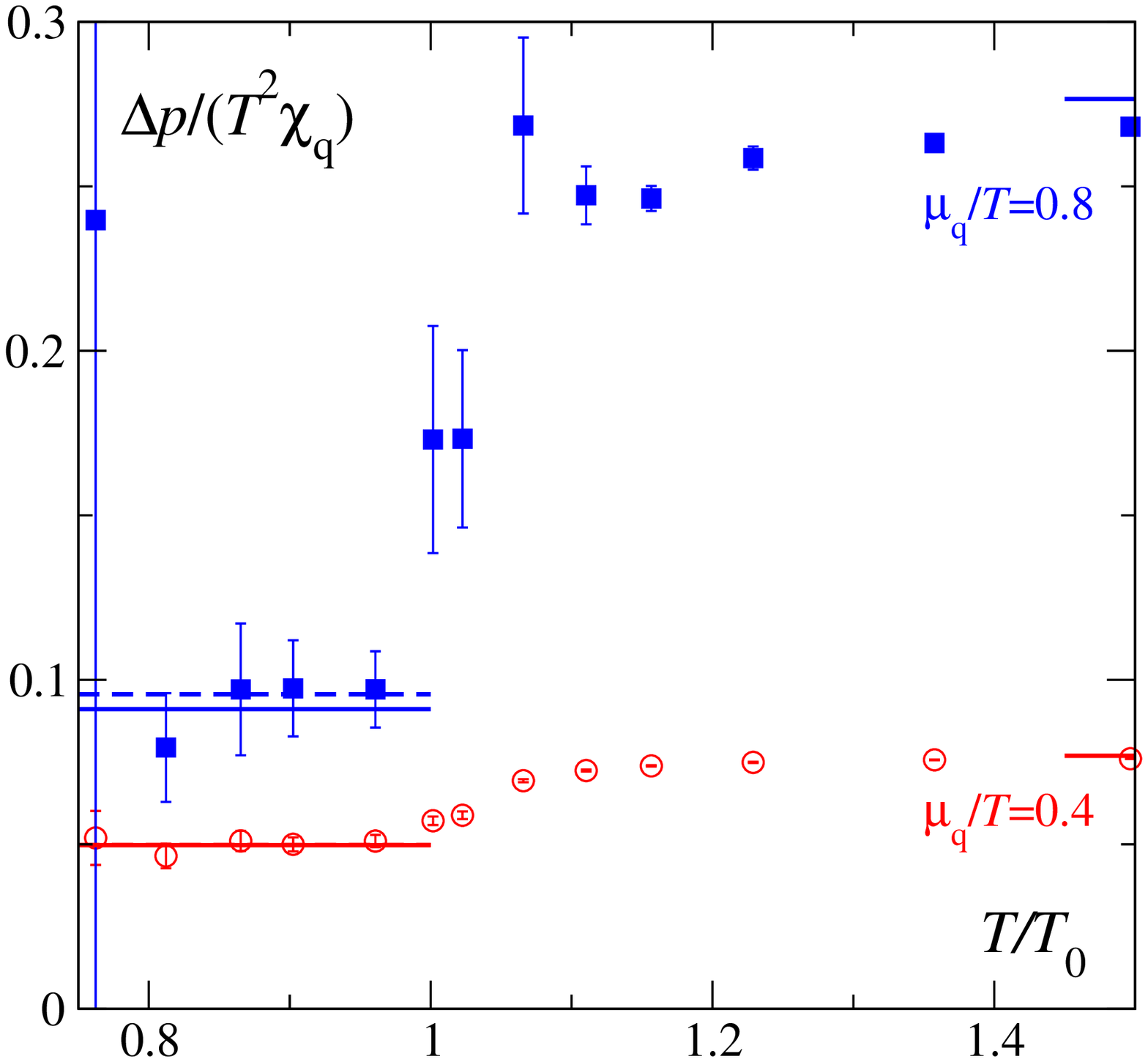, width=7.4cm}
\epsfig{file=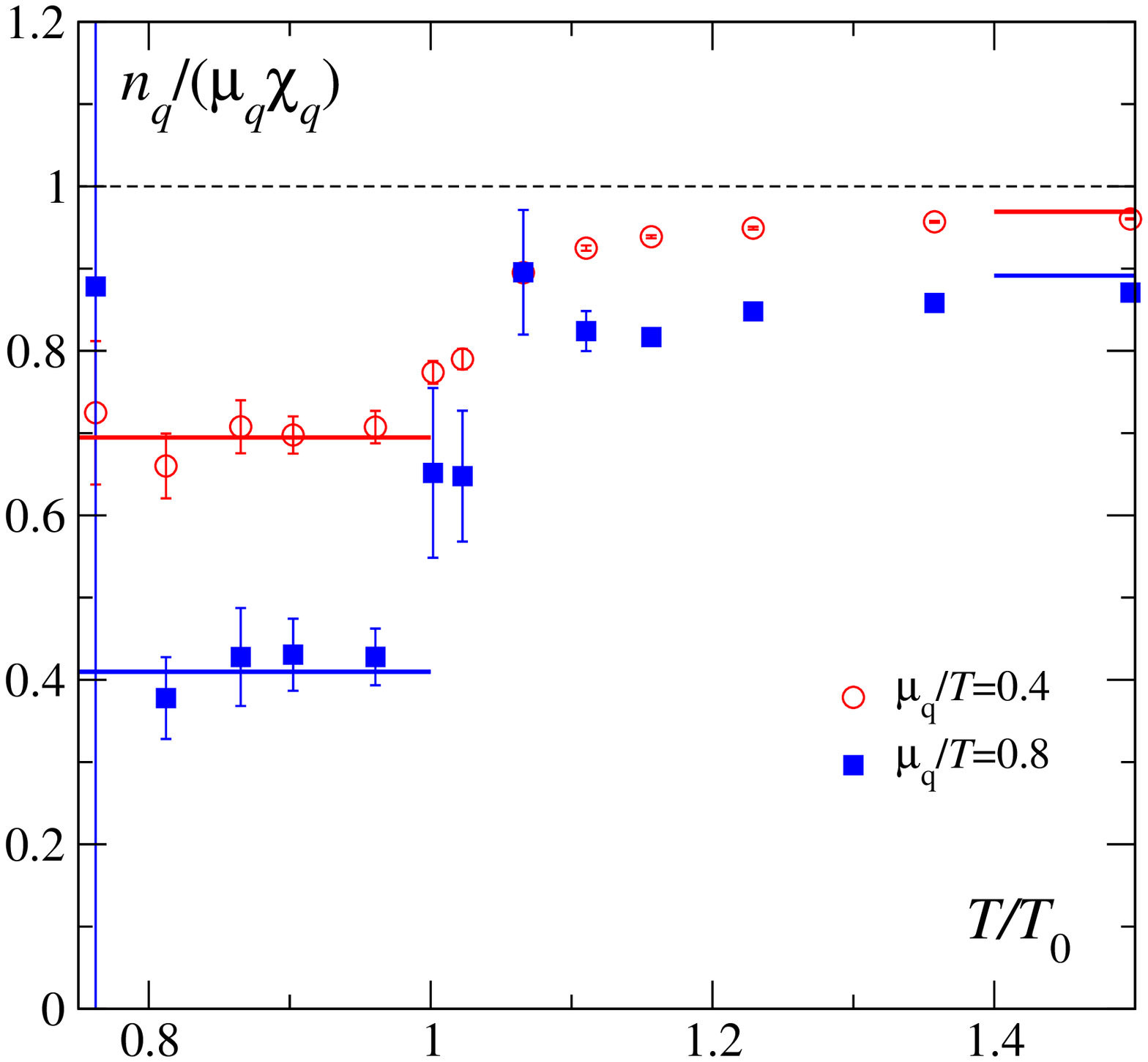, width=7.4cm}
\smallskip
\caption{The baryonic part of the pressure  
divided by the quark number susceptibility, $\Delta p / \chi_q$ (left), 
and the normalized derivative of pressure with respect to quark number 
density, $n_q/\chi_q$ (right), as a function of $\mu_q/T$ for various $T/T_0$.
Horizontal lines show the infinite temperature ideal gas values and
the HRG model prediction ($T\le T_0$) (solid lines) and expanded to 
$6^th$-order in  $\mu_q/T$ (dashed lines). The difference is visible
only for $\Delta p / \chi_q$ at $\mu_q/T=0.8$.}
\label{fig:rnqchiq}
\end{center}
\end{figure}

Similar information is contained in the ratios of  
physical observables, e.g. the quark number density or 
pressure over the quark number susceptibility, introduced in 
Eq.~(\ref{eq:npchiq}).
These ratios are shown in Fig.~(\ref{fig:rnqchiq}). Here 
$n_q/T^3$ and $\chi_q/T^2$ have been calculated 
using Eqs.~(\ref{eq:nq}) and (\ref{eq:chiq}) up to ${\cal O}(\mu_q^6)$. 
The ratio, $n_q/\chi_q=  (\partial p/ \partial \mu_q)/(\partial n_q/ \partial 
\mu_q)= \partial p/ \partial n_q$ is related
to the isothermal compressibility, $\kappa_T = \chi_q/n_q^2$ which diverges
at a $2^{nd}$ order phase transition point, {\it i.e.} at a 
point at which $\partial p/ \partial n_q =0$, the number
of particles is unstable under small changes in the pressure 
(mechanical instability) and large density fluctuations occur. This instability 
leads to a divergence in the quark number susceptibility \cite{Kunihiro}. 
A $2^{nd}$ order phase transition is 
thus expected to be signalled by a zero in both ratios shown in
Fig.~(\ref{fig:rnqchiq}). On the other hand, for $\mu_I=0$  
these ratios are expected to be
constant in an ideal quark-gluon plasma as well as in a hadron resonance gas,
\begin{equation}
\frac{n_q^{SB}}{\mu_q \chi_q^{SB}}
=\frac{1+\frac{1}{\pi^2}\left( \frac{\mu_q}{T}\right)^2}{1+
\frac{3}{\pi^2}\left( \frac{\mu_q}{T}\right)^2} \quad , \quad
\frac{n_q^{HRG}}{\mu_q \chi_q^{HRG}}
=\frac{T}{3\mu_q} \tanh \left( \frac{3\mu_q}{T} \right) \quad .
\end{equation}
The corresponding values are indicated in Fig.~(\ref{fig:rnqchiq})
by horizontal lines. 

As far as the determination of a possible $2^{nd}$ order
critical point at non-vanishing quark chemical potential (chiral
critical point) is concerned
Fig.~(\ref{fig:rc2c4c6}) and Fig.~(\ref{fig:rnqchiq}) contain identical
information.
For $T\le 0.96 T_0$ bulk thermodynamic observables agree with predictions 
based on an HRG model, which in itself does not show any
critical behavior as function of $\mu_q/T$ at fixed $T$. 
In particular, there is no hint for a dip in $\Delta p/\chi_q$ or 
$n_q/\chi_q$ which could 
signal the presence of a second order transition point. 
The same observation, albeit with
larger statistical errors,  holds for ratios involving
the chiral susceptibility $\chi_{\bar{\psi}\psi}$. Nonetheless, all
these quantities change rapidly in the transition from the low temperature
to the high temperature regime and, moreover, at $T=T_0$ the $6^{th}$ order 
expansion
coefficients clearly cannot be described within the HRG 
model. Due to the good agreement with the HRG model and its
Taylor expansion at lower temperature we cannot, however, present
an upper limit for the radius of convergence below $T_0$; 
the ratios shown in
Fig.~(\ref{fig:rc2c4c6}) suggest that a lower limit is given by 
$(\mu_q/T)_c \gsim 1$.  
Also from the analysis of the temperature dependence
of bulk thermodynamic observables we get, at present, no unambiguous
evidence for the existence of a phase transition. At present, therefore,
we cannot
rule out that in the temperature range covered by our analysis 
($T\gsim 0.8T_0$) the transition to the high temperature phase is a rapid 
crossover transition rather than a phase transition. This situation then would be
similar to that at $\mu_q=0$.
In order to exclude this possibility we would need, in the future: 
to consider even 
higher orders in the Taylor series; to scan in more detail the small
temperature interval $[0.95T_0,T_0]$; and to explore systematically 
the quark mass and volume dependence of our results. 
These issues are partially  
addressed already in the next section where we discuss the use of
reweighting techniques to calculate some thermodynamic observables and compare
results obtained within this approach with results from the Taylor expansion.

The good agreement found here for different ratios of Taylor expansion 
coefficients calculated on the lattice and within the HRG model suggests 
that we may use this information for a more detailed analysis of the 
composition of hadronic matter at temperatures below $T_0$. 
In Ref.~\cite{KRT} also the temperature dependence of thermodynamic
observables like the pressure or the quark number susceptibility have 
been compared to the HRG model.\footnote{For these observables
a good functional agreement between lattice data and the leading
$\mu_q$ dependent term in the HRG model has also been noted
in \cite{Lombardoeos} within the imaginary chemical potential approach.}
In order to do so the hadron spectrum
has been adjusted to the conditions realized in the lattice calculations,
{\it i.e.} all masses have been shifted to larger values as the lattice
calculations have been performed with unphysically large quark masses.
This approach can also be turned around.  The HRG model for $T< T_0$ can 
be used as an ansatz to determine the contributions
of the mesonic and baryonic parts of the spectrum without making assumptions
on the distortion of the spectrum due to the unphysical quark mass values.

\begin{figure}[tb]
\begin{center}
\begin{minipage}[c][5.2cm][c]{5.0cm}
\begin{center}
\epsfig{file=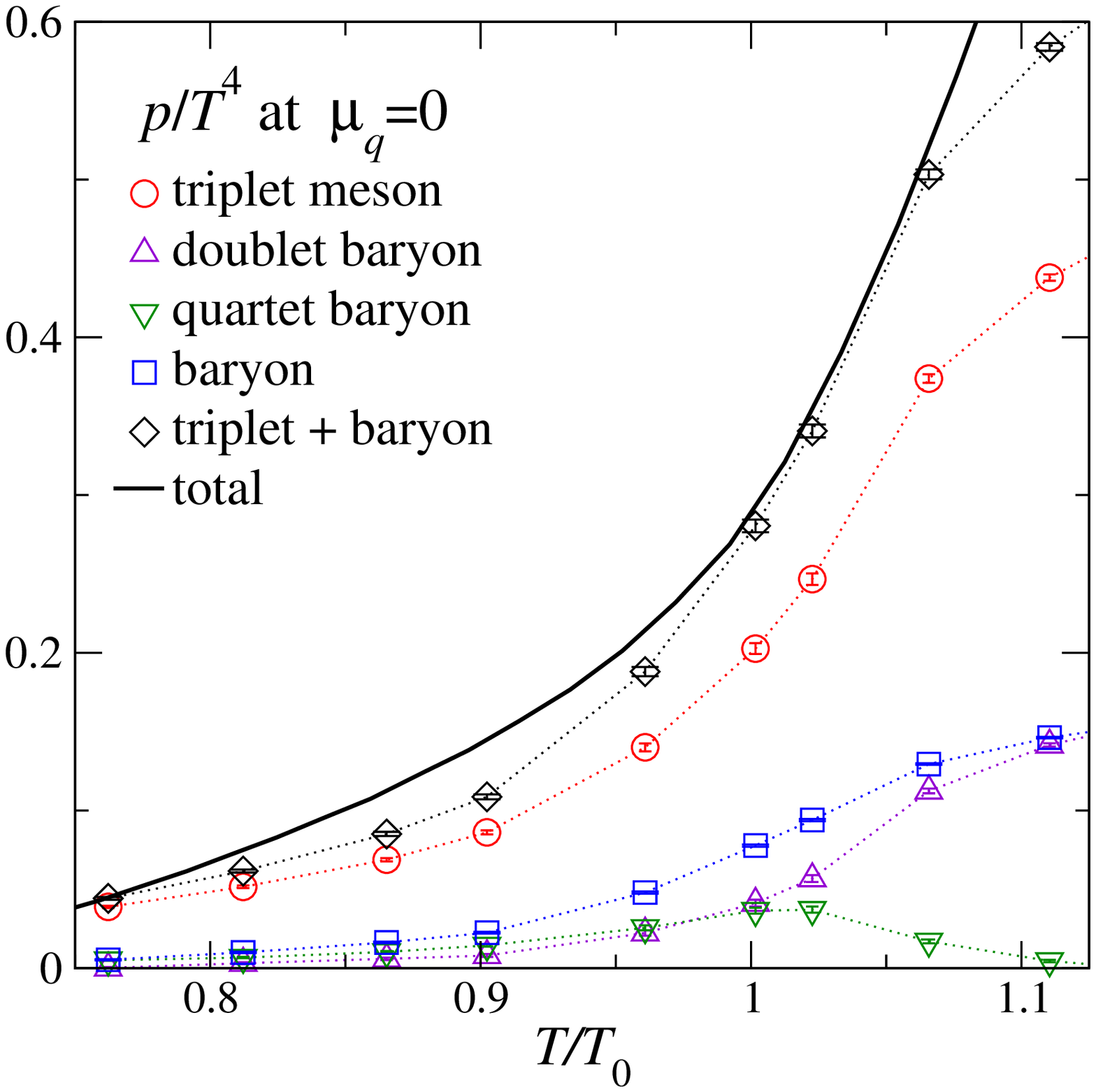, width=5.0cm}\\[-1mm]
(a)
\end{center}
\end{minipage}
\begin{minipage}[c][5.2cm][c]{5.0cm}
\begin{center}
\epsfig{file=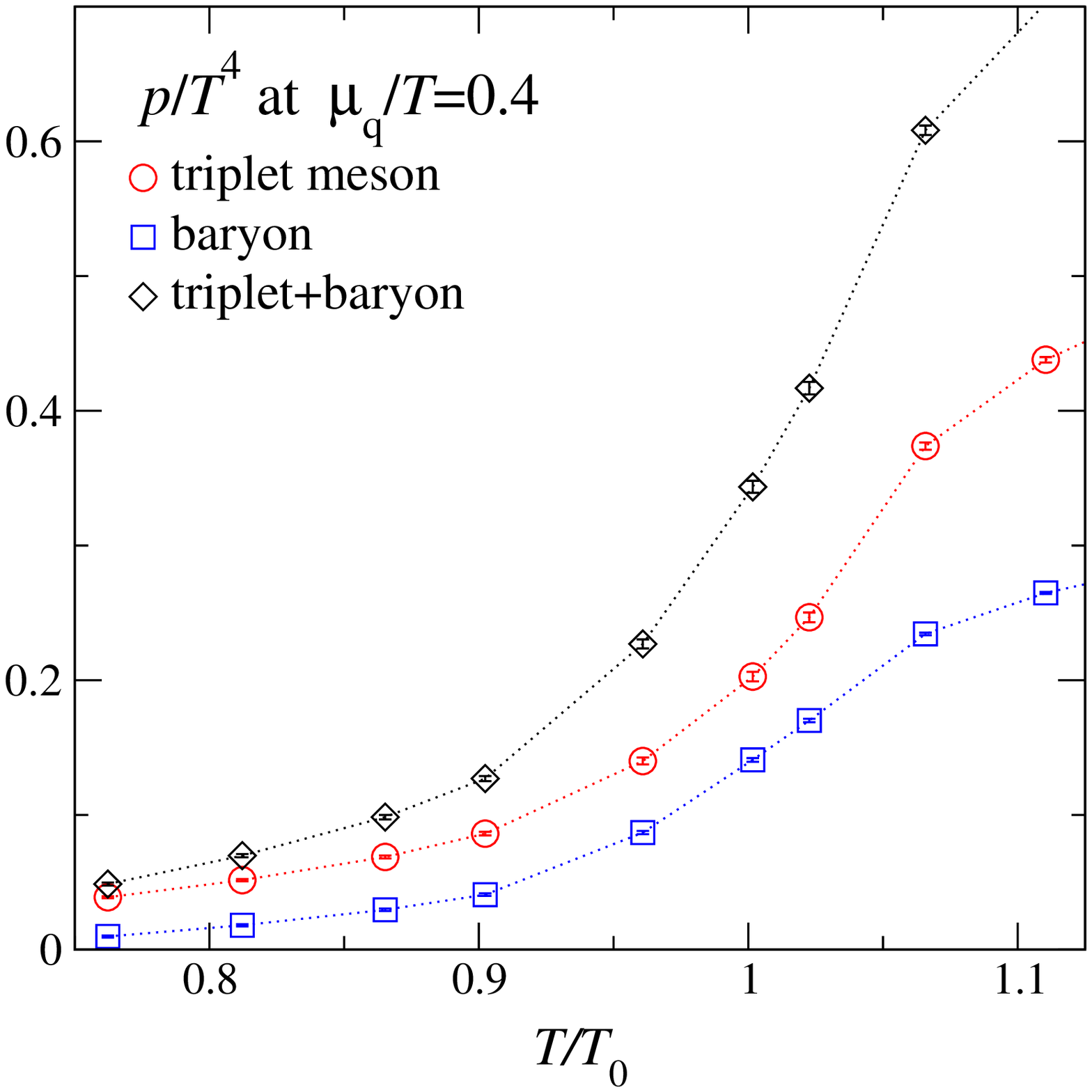, width=5.0cm}\\[-1mm]
(b)
\end{center}
\end{minipage}
\begin{minipage}[c][5.2cm][c]{5.0cm}
\begin{center}
\epsfig{file=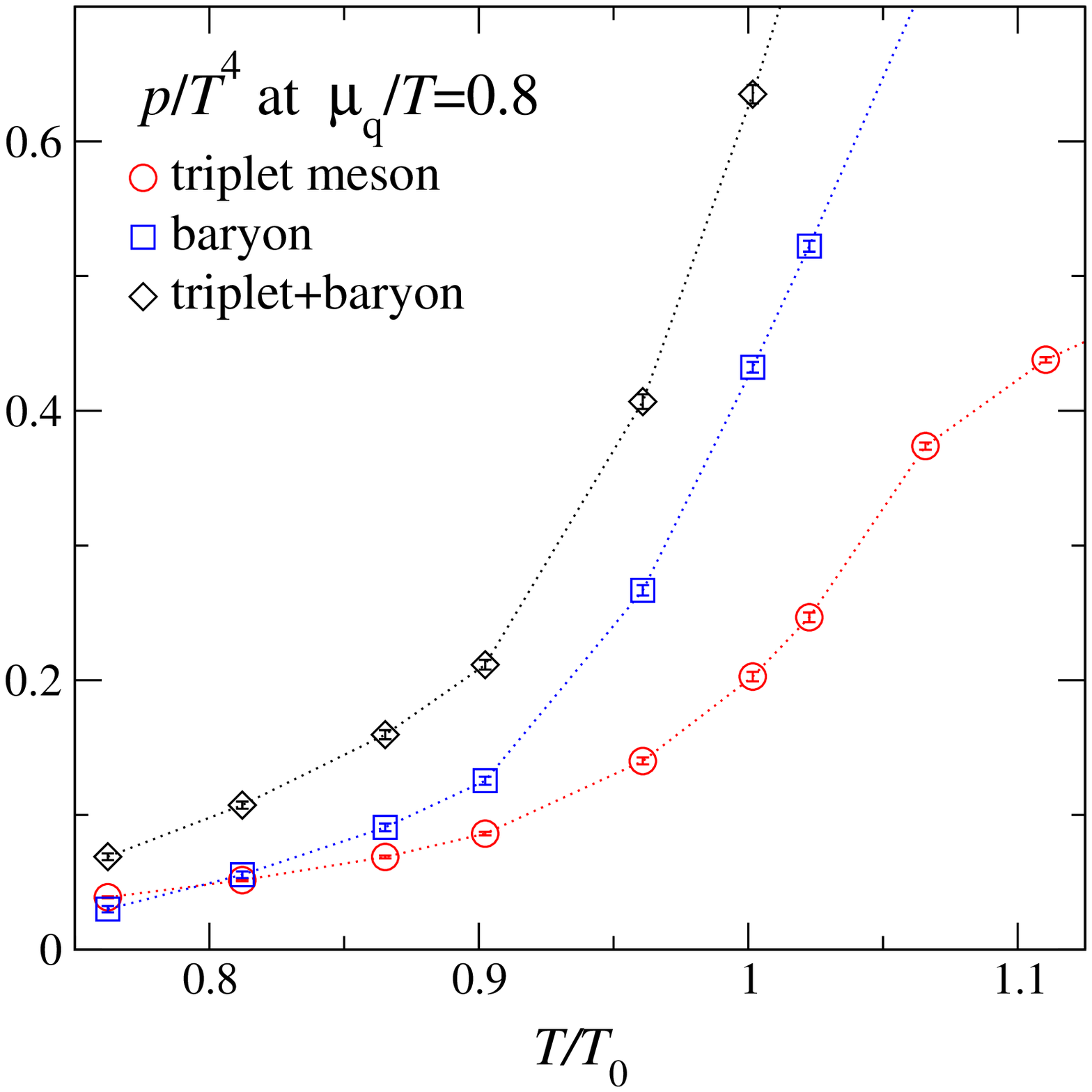, width=5.0cm}\\[-1mm]
(c)
\end{center}
\end{minipage}
\caption{Contribution of different hadronic channels to the total
pressure $p/T^4$ obtained by using the HRG ansatz. Shown are results
for (a) $\mu_q/T=0$,  (b) $\mu_q/T=0.4$ and (c) $\mu_q/T=0.8$.}
\label{fig:pHRGmu}
\end{center}
\end{figure}

As outlined in section 2, 
within the Boltzmann approximation the HRG model yields a simple
dependence of the pressure on the quark chemical potential.
The relation given in Eq.~(\ref{eq:pHRG}) can easily be extended to also
include a non-vanishing isovector chemical potential.
Neglecting the mass difference among isospin partners the pressure can be
written as,
\begin{eqnarray}
\frac{p(T,\mu_q,\mu_I)}{T^4} &\simeq& G^{(1)}(T) 
+G^{(3)}(T) \frac{1}{3} \left(2\cosh\left(\frac{2\mu_I}{T}\right)+1 \right)
\nonumber \\ 
&&  +F^{(2)}(T) \cosh \left( \frac{3\mu_q}{T} \right) 
\cosh \left( \frac{\mu_I}{T} \right) \\ 
&&  
+F^{(4)}(T) \frac{1}{2}\cosh \left( \frac{3\mu_q}{T} \right) \left[ 
\cosh \left( \frac{\mu_I}{T} \right) +\cosh \left( \frac{3\mu_I}{T} \right)
\right]~~, \nonumber
\label{eq:pHRGmu}
\end{eqnarray}
where $G^{(1)}, G^{(3)}, F^{(2)}$ and $F^{(4)}$ are the contributions to 
the pressure at $\mu_q=\mu_I=0$ arising from isosinglet mesons 
$(\eta, \ldots,~ [B_i=0, I_{3i}=0])$, 
isotriplet mesons $(\pi, \ldots,~[B_i=0, I_{3i}=\{0, \pm1\}])$, 
isodoublet baryons $(n, p, \ldots,~[B_i=\pm1, I_{3i}=\{\pm1/2\}])$  and
isoquartet baryons $(\Delta, \ldots,~[B_i=\pm1, I_{3i}=\{\pm1/2, \pm3/2\}])$, 
respectively. These functions contain all the information on the 
hadron spectrum in different quantum number channels. Performing the
Taylor expansion of the pressure as well as quark number and isovector
susceptibilities allows to relate these functions to combinations of the
various Taylor expansion coefficients. This way one finds  
\begin{equation}
G^{(3)}(T)=\frac{3}{4}c_2^I-c_4^I, \hspace{8mm}
F^{(2)}(T)=\frac{5}{18}c_2 -\frac{2}{3}c_4^I, \hspace{8mm}
F^{(4)}(T)= -\frac{1}{18}c_2 +\frac{2}{3}c_4^I~~. 
\label{eq:gf}
\end{equation}
The various contributions to the pressure are shown
in Fig.~(\ref{fig:pHRGmu})a for $\mu_q =0$. With increasing quark chemical
potential the relative weight of hadrons in different quantum number 
channels changes. As expected the baryonic component becomes more
important with increasing $\mu_q$ (Fig.~(\ref{fig:pHRGmu})b  and c). We find
that for $\mu_q/T \gsim 0.6$ the baryonic sector gives the dominant
contribution.

\section{Reweighting Approach}
\label{sec:rew}

\begin{figure}
\begin{center}
\epsfig{file=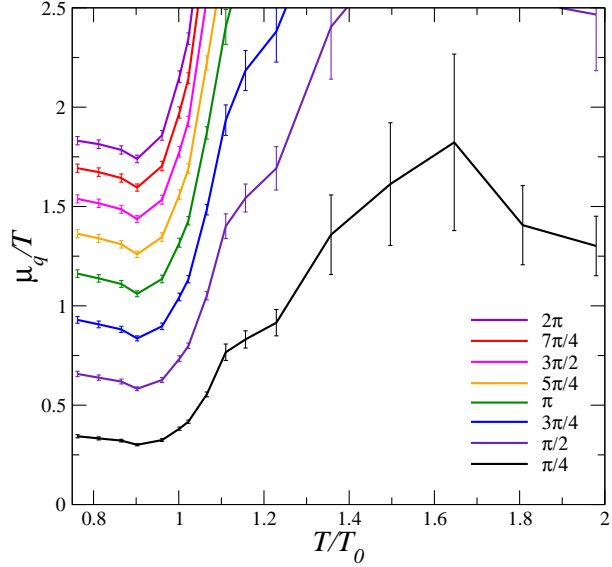, width=8.0cm}
\smallskip
\caption{Contour plot of the variance of the phase of the quark 
determinant, $\sigma (\theta)$ calculated for
$\theta^{(3)}$ in the $(T/T_0, \mu_q/T)$ plane.
Contour lines for  $\sigma ( \theta^{(3)} )$ are 
given in steps of $\pi/4$ ranging from $\pi/4$  (lowest curve) to
$2 \pi$.} 
\label{fig:phasefl}
\end{center}
\end{figure}

\begin{figure}[tb]
\begin{center}
\begin{minipage}[c][7.8cm][c]{7.4cm}
\begin{center}
\epsfig{file=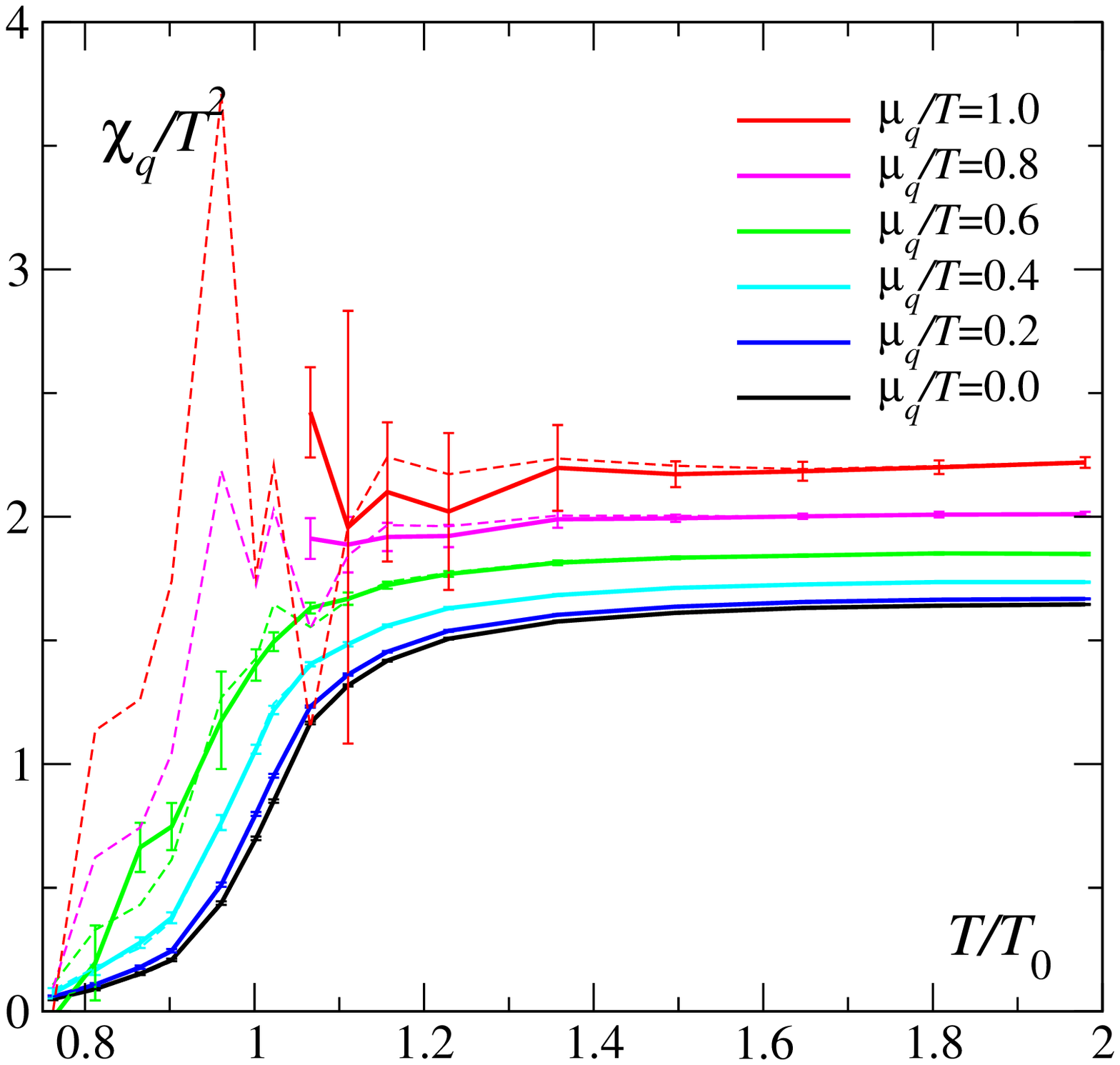, width=7.4cm}\\[-1mm]
(a)
\end{center}
\end{minipage}
\begin{minipage}[c][7.8cm][c]{7.4cm}
\begin{center}
\epsfig{file=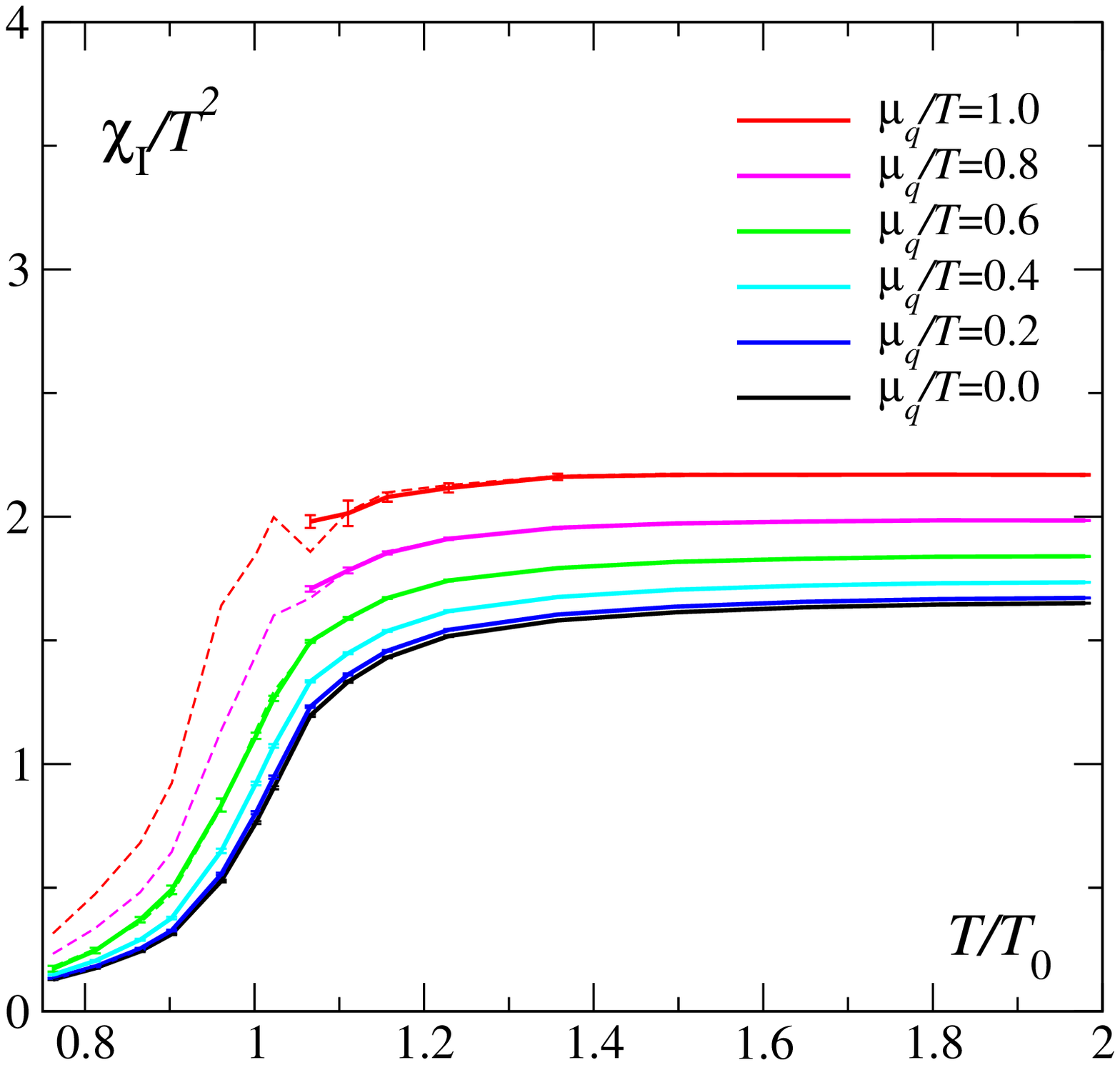, width=7.4cm}\\[-1mm]
(b)
\end{center}
\end{minipage}
\caption{Susceptibilities $\chi_q/T^2$ (left) and $\chi_I/T^2$ (right) for
various $\mu_q/T$ ranging from $\mu_q/T=0$ (lowest curve) rising in
steps of 0.2 to  $\mu_q/T=1$. Results are obtained 
by a combination of reweighting (solid lines and data points) and from 
a $6^{th}$-order Taylor expansion (dashed lines).}
\label{fig:chirew}
\end{center}
\end{figure}

An alternative to a strict Taylor expansion of thermodynamic observables
in terms of $\mu_q/T$ is the reweighting approach. Here the
dependence of the grand potential on the quark chemical potential
is included in the calculation of observables, $X$, by
shifting the $\mu_q$-dependent piece of the QCD action into the 
calculation of expectation values rather than taking it into account
in the statistical
weights used for the generation of gauge field configurations.
This reweighting approach has been used to analyze the 
thermodynamics of QCD at
non-zero chemical potential \cite{FKS,Fodoreos}. Within this approach
thermodynamic observables $X(\beta,\mu)$ are estimated via the expression
\begin{equation}
\langle X\rangle_{(\beta,\mu)}=
{{\langle X\; e^{{n_{\rm f}\over4}\Delta\ln {\rm det}M}
e^{-\Delta S_g}\rangle_{(\beta_0,0)}}\over
{\langle e^{{n_{\rm f}\over4}\Delta\ln {\rm det}M}
e^{-\Delta S_g} \rangle_{(\beta_0,0)}}} \quad ,
\label{eq:reweight}
\end{equation}
where $\Delta\ln\mbox{det}M\equiv \ln{\rm det}M(\mu) -\ln{\rm det}M(0)$
and $\Delta S_g \equiv S_g(\beta)-S_g(\beta_0)$ is the difference of the
gluonic part of the QCD action.
The expectation values on the RHS of Eq.~(\ref{eq:reweight}) are obtained 
in simulations at $(\beta_0,0)$. In \cite{us1} we implemented a version of 
Eq.~(\ref{eq:reweight}) in which the reweighting factor $\Delta\ln {\rm det}M$
as well as the operator $X$ itself have been replaced by a Taylor series 
about $\mu=0$. The advantage over an exact evaluation of $\mbox{det} M$
\cite{Fodor1} clearly
is  that the required expressions are calculable with relatively little 
computational effort even on large lattices. In our initial study we
performed the expansion consistently up to and including ${\cal O}(\mu^2)$. 
Here we extend 
this analysis by expanding $\ln\mbox{det}M$ up to and including
terms of ${\cal O}(\mu^6)$. Unlike the direct
evaluation of thermodynamic observables in terms of a Taylor expansion
up to a certain order the reweighting approach with a Taylor expanded weight
factor also includes effects of higher orders in $\mu_q/T$ which 
are partially resummed in the exponentiated observables 
$\exp \{ {n_{\rm f}\over4}\Delta\ln {\rm det}M \}$.

The effectiveness of any reweighting approach strongly depends on the overlap 
between the ensemble simulated at $(\beta_0, \mu_0=0)$ and that corresponding 
to the true equilibrium state at $(\beta, \mu)$ which one wants to analyze. 
This can be judged by inspecting the average phase factor of the 
complex valued quark determinant, $\langle
e^{i\theta}\rangle_{(\beta_0,0)}$, where the quark determinant is written 
as $\mbox{det}M=\vert\mbox{det}M\vert e^{i\theta}$. 
Reweighting loses its reliability once $\langle e^{i\theta}\rangle\ll1$ 
as both expectation values appearing in the numerator and denominator
of Eq.~(\ref{eq:reweight}) then become difficult to control \cite{Shinji}.
In our approach we estimate the phase factor via the variance
of the phase $\theta$, 
$\sigma(\theta) = \sqrt{\langle \theta^2 \rangle - \langle \theta\rangle^2}$, 
where we approximate the phase by its Taylor expansion up to 
${\cal O}(\mu^{2n-1})$,
\begin{equation}
\theta^{(n)}=
{{n_{\rm f}\over4}}\mbox{Im}\sum_{j=1}^n{\mu^{2j-1}\over{(2j-1)!}}
{{\partial^{2j-1}\ln\mbox{det}M}\over{\partial\mu^{2j-1}}} \quad .
\end{equation}
As discussed and shown in Fig.~6 of \cite{Shinji}, the value of $\mu_q/T$ 
for which
the standard deviation of $\theta^{(n)}$ exceeds $\pi / 2$ is a reasonable 
criterion for judging the applicability of reweighting in our 
simulated systems. In Fig.~(\ref{fig:phasefl}) we show contour lines for
the variance of $\theta^{(3)}$. All contour lines 
drop dramatically in the vicinity of $T_0$; 
the contour corresponding to $\sigma(\theta^{(3)}) = \pi/2$ yields
$\mu_q/T\approx 1.5$ at $T\simeq1.2T_{0}$ and
reaches a minimum value of about $0.6$ at $T\simeq0.9T_{0}$. Reweighting 
is thus much easier to control in
the high temperature phase than in the hadronic phase.

In Fig.~(\ref{fig:chirew}) we plot quark number and isovector 
susceptibilities for various $\mu_q/T$ 
calculated by using reweighting where possible. 
The operators needed for the susceptibilities are calculated with an error
of $O(\mu^6)$,
the reweighting factor $\Delta\ln\mbox{det}M$ with error $O(\mu^8)$.
Dashed lines show a comparison
with the results obtained from direct Taylor expansion of the susceptibilities
up to and including ${\cal O}(\mu_q^4$) (see Fig.~(\ref{fig:chiIq})).

As anticipated above, for $T\lsim T_{0}$ reweighting becomes difficult
to control for $\mu_q/T>0.6$. However, where reweighting appears to be 
statistically under control,
it agrees well with the direct Taylor expansion and even shows similar
features in the statistical errors; {\it i.e.} the signal 
is much noisier in the vicinity of $T_{0}$ for $\chi_q$ than for $\chi_I$.

\section{Conclusions}

We have extended our analysis of the thermodynamics of two flavor QCD at
non-zero quark chemical potential to the $6^{th}$ order in a Taylor
expansion around $\mu_q/T = 0$. We find clear evidence for a rapid 
transition from a low temperature hadronic phase to a high temperature
quark-gluon plasma phase which is signalled by large fluctuations in the
quark number density and the chiral condensate. The transition temperature
shifts to smaller values with increasing quark chemical 
potential. Above $T_0$ the Taylor expansion coefficients and bulk thermodynamic 
observables agree with qualitative features of the perturbative high 
temperature expansion and approach the ideal gas limit to within
$\sim 20\%$ for $T\gsim 1.5 T_0$. Thermodynamics in the low temperature
phase agrees well with predictions based on a hadron resonance gas for
temperatures $T\lsim 0.96 T_0$ and $\mu_q/T\lsim 1$.

From the analysis of bulk thermodynamic observables alone we cannot
provide strong evidence for the existence of a $2^{nd}$ order phase
transition point in the QCD phase diagram. At present we cannot rule out
the transition being a rapid crossover in the entire parameter space
covered by our analysis. In particular, we have
shown that large fluctuations in the quark number and the chiral
condensate are consistent with expectations based on a hadron resonance
gas. The current estimates on the radius of convergence of the Taylor
expansion favor a critical value of the quark chemical potential close
to $\mu_q \approx T_0$. The good agreement of the expansion coefficients
with those of a hadron gas, however, prohibit any firm conclusion 
on the location and even on the existence of the chiral critical point.

Likewise, we cannot rule out that a $2^{nd}$ order transition occurs at
temperatures closer to $T_0$ than the largest value in the hadronic phase,
$T=0.96 T_0$ which we have analyzed here.
In order to improve on the current analysis it would be important to
perform calculations at smaller quark masses in a narrower temperature
interval around $T_0$. An improvement over the current statistical
errors on the $6^{th}$ order coefficient as well as the analysis of
higher order expansion coefficients with high statistics is needed.

\vfill
\newpage

\section*{Acknowledgments}
Numerical work was performed using APEmille computers in Swansea,
supported by PPARC grant PPA/G/S/1999/00026, and Bielefeld.
SJH acknowledges support from PPARC. 
Work of the Bielefeld group has been supported partly through the DFG
under grant FOR 339/2-1, the DFG funded Graduate School GRK 881 and a grant 
of the BMBF under contract no. 06BI102. KR acknowledges partial support by 
KBN under contract 2PO3 (06925).

\appendix
\section{Appendix: Taylor expansion coefficients}
\label{sec:appA}

In this appendix, we derive some equations which are used in the 
calculation of the various thermodynamic quantities and expansion
coefficients of the Taylor series presented in this study. 
The partition function is given by 
\begin{eqnarray}
{\cal Z} = \int 
{\cal D}U (\det M)^{n_{\rm f}/4} e^{-S_g} \quad , 
\label{eq:partition} 
\end{eqnarray}
with $U\in SU(3)$.  The expectation value of a physical quantity, 
$\left\langle {\cal O} \right\rangle$ is then obtained as 
\begin{eqnarray}
\left\langle {\cal O} \right\rangle 
\hspace{-4mm} &=& \hspace{-4mm} 
\frac{1}{\cal Z} \int {\cal D}U {\cal O} 
(\det M)^{n_{\rm f}/4} e^{-S_g} \quad ,
\end{eqnarray}
and its derivatives with respect to quark chemical potential and quark
mass are given by,
\begin{eqnarray}
\frac{\partial \langle {\cal O} \rangle}{\partial \mu} 
\hspace{-4mm} &=& \hspace{-4mm} 
\left\langle 
\frac{\partial {\cal O}}{\partial \mu} \right\rangle 
+ \frac{n_{\rm f}}{4} \left(
\left\langle {\cal O} \; 
\frac{\partial (\ln \det M)}{\partial \mu} \right\rangle 
-\left\langle {\cal O} \right\rangle 
\left\langle  
\frac{\partial (\ln \det M)}{\partial \mu} \right\rangle \right) \quad , 
\\ \nonumber \\
\frac{\partial \langle {\cal O} \rangle}{\partial m} 
\hspace{-4mm} &=& \hspace{-4mm} 
\left\langle 
\frac{\partial {\cal O}}{\partial m} \right\rangle 
+ \frac{n_{\rm f}}{4} \left(
\left\langle {\cal O}  \;
\frac{\partial (\ln \det M)}{\partial m} \right\rangle 
-\left\langle {\cal O} \right\rangle 
\left\langle  
\frac{\partial (\ln \det M)}{\partial m} \right\rangle \right) \nonumber \\
\hspace{-4mm} &=& \hspace{-4mm} 
\left\langle 
\frac{\partial {\cal O}}{\partial m} \right\rangle 
+ \frac{n_{\rm f}}{4} \left(
\left\langle {\cal O}  \;
{\rm tr} M^{-1} \right\rangle 
-\left\langle {\cal O} \right\rangle 
\left\langle {\rm tr} M^{-1} \right\rangle \right) \quad .
\end{eqnarray}
Here we use $m$ as the dimensionless quark mass value instead of $ma$, 
and also $\mu = \mu_q a$ for the dimensionless quark chemical potential. 
The temperature is $T=(N_{\tau} a)^{-1}$ and the volume is 
$V=(N_{\sigma} a)^3$.
Moreover, we introduce for simplification,  
\begin{eqnarray}
{\cal C}_n = \frac{n_{\rm f}}{4} \frac{\partial^n {\rm tr} M^{-1}} 
{\partial \mu^n}
= \frac{n_{\rm f}}{4} \frac{\partial^{n+1} \ln \det M} 
{\partial \mu^n \partial m} \quad , \quad 
{\cal D}_n = \frac{n_{\rm f}}{4} \frac{\partial^n \ln \det M} 
{\partial \mu^n} \quad . 
\label{eq:basic}
\end{eqnarray}
All Taylor expansion coefficients used in this paper can be expressed
in terms of expectation values of certain combinations of ${\cal C}_n$  
and ${\cal D}_n$. The required derivatives of $\ln \det M$ and 
${\rm tr}M^{-1}$ are explicitly given in the following.

\paragraph{Derivatives of \boldmath $\ln \det M$: }
\begin{eqnarray}
\frac{\partial \ln \det M}{\partial \mu} 
\hspace{-4mm} &=& \hspace{-4mm} 
{\rm tr} \left( M^{-1} \frac{\partial M}{\partial \mu} \right) ,
\label{eq:dermu1} \\
\frac{\partial^2 \ln \det M}{\partial \mu^2} 
\hspace{-4mm} &=& \hspace{-4mm} 
{\rm tr} \left( M^{-1} \frac{\partial^2 M}{\partial \mu^2} \right)
 - {\rm tr} \left( M^{-1} \frac{\partial M}{\partial \mu}
                   M^{-1} \frac{\partial M}{\partial \mu} \right) ,
\label{eq:dermu2} \\
\frac{\partial^3 \ln \det M}{\partial \mu^3} 
\hspace{-4mm} &=& \hspace{-4mm} 
{\rm tr} \left( M^{-1} \frac{\partial^3 M}{\partial \mu^3} \right)
 -3 {\rm tr} \left( M^{-1} \frac{\partial M}{\partial \mu}
              M^{-1} \frac{\partial^2 M}{\partial \mu^2} \right) 
\nonumber \\ && \hspace{-4mm}
+2 {\rm tr} \left( M^{-1} \frac{\partial M}{\partial \mu}
        M^{-1} \frac{\partial M}{\partial \mu}
        M^{-1} \frac{\partial M}{\partial \mu} \right) ,
\label{eq:dermu3} \\
\frac{\partial^4 \ln \det M}{\partial \mu^4} 
\hspace{-4mm} &=& \hspace{-4mm} 
{\rm tr} \left( M^{-1} \frac{\partial^4 M}{\partial \mu^4} \right)
 -4 {\rm tr} \left( M^{-1} \frac{\partial M}{\partial \mu}
              M^{-1} \frac{\partial^3 M}{\partial \mu^3} \right) \nonumber \\
&& \hspace{-2cm}
-3 {\rm tr} \left( M^{-1} \frac{\partial^2 M}{\partial \mu^2}
        M^{-1} \frac{\partial^2 M}{\partial \mu^2} \right) 
 +12 {\rm tr} \left( M^{-1} \frac{\partial M}{\partial \mu}
        M^{-1} \frac{\partial M}{\partial \mu}
        M^{-1} \frac{\partial^2 M}{\partial \mu^2} \right) \nonumber \\
&& \hspace{-2cm}
-6 {\rm tr} \left( M^{-1} \frac{\partial M}{\partial \mu}
        M^{-1} \frac{\partial M}{\partial \mu}
        M^{-1} \frac{\partial M}{\partial \mu}
        M^{-1} \frac{\partial M}{\partial \mu} \right) ,
\label{eq:dermu4} \\
\frac{\partial^5 \ln \det M}{\partial \mu^5} 
\hspace{-4mm} &=& \hspace{-4mm} 
{\rm tr} \left( M^{-1} \frac{\partial^5 M}{\partial \mu^5} \right)
 -5 {\rm tr} \left( M^{-1} \frac{\partial M}{\partial \mu}
              M^{-1} \frac{\partial^4 M}{\partial \mu^4} \right) \nonumber \\
&& \hspace{-25mm}
 -10 {\rm tr} \left( M^{-1} \frac{\partial^2 M}{\partial \mu^2}
        M^{-1} \frac{\partial^3 M}{\partial \mu^3} \right) 
 +20 {\rm tr} \left( M^{-1} \frac{\partial M}{\partial \mu}
        M^{-1} \frac{\partial M}{\partial \mu} 
        M^{-1} \frac{\partial^3 M}{\partial \mu^3} \right) \nonumber \\
&& \hspace{-25mm}
 +30 {\rm tr} \left( M^{-1} \frac{\partial M}{\partial \mu}
        M^{-1} \frac{\partial^2 M}{\partial \mu^2}
        M^{-1} \frac{\partial^2 M}{\partial \mu^2} \right) \nonumber \\
&& \hspace{-25mm}
 -60 {\rm tr} \left( M^{-1} \frac{\partial M}{\partial \mu}
        M^{-1} \frac{\partial M}{\partial \mu}
        M^{-1} \frac{\partial M}{\partial \mu}
        M^{-1} \frac{\partial^2 M}{\partial \mu^2} \right) \nonumber \\
&& \hspace{-25mm}
 +24 {\rm tr} \left( M^{-1} \frac{\partial M}{\partial \mu}
        M^{-1} \frac{\partial M}{\partial \mu}
        M^{-1} \frac{\partial M}{\partial \mu}
        M^{-1} \frac{\partial M}{\partial \mu}
        M^{-1} \frac{\partial M}{\partial \mu} \right) ,
\label{eq:dermu5} \\ 
\frac{\partial^6 \ln \det M}{\partial \mu^6} 
\hspace{-4mm} &=& \hspace{-4mm} 
{\rm tr} \left( M^{-1} \frac{\partial^6 M}{\partial \mu^6} \right)
 -6 {\rm tr} \left( M^{-1} \frac{\partial M}{\partial \mu}
              M^{-1} \frac{\partial^5 M}{\partial \mu^5} \right) \nonumber \\
&& \hspace{-25mm}
 -15 {\rm tr} \left( M^{-1} \frac{\partial^2 M}{\partial \mu^2}
        M^{-1} \frac{\partial^4 M}{\partial \mu^4} \right) 
 -10 {\rm tr} \left( M^{-1} \frac{\partial^3 M}{\partial \mu^3} 
        M^{-1} \frac{\partial^3 M}{\partial \mu^3} \right) \nonumber \\
&& \hspace{-25mm}
 +30 {\rm tr} \left( M^{-1} \frac{\partial M}{\partial \mu}
        M^{-1} \frac{\partial M}{\partial \mu} 
        M^{-1} \frac{\partial^4 M}{\partial \mu^4} \right)
 +60 {\rm tr} \left( M^{-1} \frac{\partial M}{\partial \mu}
        M^{-1} \frac{\partial^2 M}{\partial \mu^2}
        M^{-1} \frac{\partial^3 M}{\partial \mu^3} \right) \nonumber \\
&& \hspace{-25mm}
 +60 {\rm tr} \left( M^{-1} \frac{\partial^2 M}{\partial \mu^2}
        M^{-1} \frac{\partial M}{\partial \mu}
        M^{-1} \frac{\partial^3 M}{\partial \mu^3} \right)
 +30 {\rm tr} \left( M^{-1} \frac{\partial^2 M}{\partial \mu^2}
        M^{-1} \frac{\partial^2 M}{\partial \mu^2}
        M^{-1} \frac{\partial^2 M}{\partial \mu^2} \right) \nonumber \\
&& \hspace{-25mm}
 -120 {\rm tr} \left( M^{-1} \frac{\partial M}{\partial \mu}
        M^{-1} \frac{\partial M}{\partial \mu}
        M^{-1} \frac{\partial M}{\partial \mu}
        M^{-1} \frac{\partial^3 M}{\partial \mu^3} \right) \nonumber \\
&& \hspace{-25mm}
 -180 {\rm tr} \left( M^{-1} \frac{\partial M}{\partial \mu}
        M^{-1} \frac{\partial M}{\partial \mu}
        M^{-1} \frac{\partial^2 M}{\partial \mu^2}
        M^{-1} \frac{\partial^2 M}{\partial \mu^2} \right) \nonumber \\
&& \hspace{-25mm}
 -90 {\rm tr} \left( M^{-1} \frac{\partial M}{\partial \mu}
        M^{-1} \frac{\partial^2 M}{\partial \mu^2}
        M^{-1} \frac{\partial M}{\partial \mu}
        M^{-1} \frac{\partial^2 M}{\partial \mu^2} \right) \nonumber \\
&& \hspace{-25mm}
 +360 {\rm tr} \left( M^{-1} \frac{\partial M}{\partial \mu}
        M^{-1} \frac{\partial M}{\partial \mu}
        M^{-1} \frac{\partial M}{\partial \mu}
        M^{-1} \frac{\partial M}{\partial \mu}
        M^{-1} \frac{\partial^2 M}{\partial \mu^2} \right) \nonumber \\
&& \hspace{-25mm}
 -120 {\rm tr} \left( M^{-1} \frac{\partial M}{\partial \mu}
        M^{-1} \frac{\partial M}{\partial \mu}
        M^{-1} \frac{\partial M}{\partial \mu}
        M^{-1} \frac{\partial M}{\partial \mu}
        M^{-1} \frac{\partial M}{\partial \mu}
        M^{-1} \frac{\partial M}{\partial \mu} \right) .
\label{eq:dermu6}
\end{eqnarray}

\paragraph{Derivatives of \boldmath ${\rm tr} M^{-1}$: }
\begin{eqnarray}
\frac{\partial {\rm tr} M^{-1}}{\partial \mu} 
\hspace{-4mm} &=& \hspace{-4mm} 
- {\rm tr} \left( M^{-1} \frac{\partial M}{\partial \mu}
 M^{-1} \right) \\
\frac{\partial^2 {\rm tr} M^{-1}}{\partial \mu^2} 
\hspace{-4mm} &=& \hspace{-4mm} 
- {\rm tr} \left( M^{-1} \frac{\partial^2 M}{\partial \mu^2}
 M^{-1} \right)
 + 2 {\rm tr} \left( M^{-1} \frac{\partial M}{\partial \mu}
    M^{-1} \frac{\partial M}{\partial \mu} M^{-1} \right) \\
\frac{\partial^3 {\rm tr} M^{-1}}{\partial \mu^3} 
\hspace{-4mm} &=& \hspace{-4mm} 
- {\rm tr} \left( M^{-1} \frac{\partial^3 M}{\partial \mu^3}
 M^{-1} \right)
 +3 {\rm tr} \left( M^{-1} \frac{\partial^2 M}{\partial \mu^2}
    M^{-1} \frac{\partial M}{\partial \mu} M^{-1} \right)  \\
&& \hspace{-2cm}
+3 {\rm tr} \left( M^{-1} \frac{\partial M}{\partial \mu}
    M^{-1} \frac{\partial^2 M}{\partial \mu^2} M^{-1} \right)
-6 {\rm tr} \left( M^{-1} \frac{\partial M}{\partial \mu}
    M^{-1} \frac{\partial M}{\partial \mu} M^{-1} 
    \frac{\partial M}{\partial \mu} M^{-1} \right) \nonumber \\
\frac{\partial^4 {\rm tr} M^{-1}}{\partial \mu^4} 
\hspace{-4mm} &=& \hspace{-4mm} 
- {\rm tr} \left( M^{-1} \frac{\partial^4 M}{\partial \mu^4}
 M^{-1} \right)
 +4 {\rm tr} \left( M^{-1} \frac{\partial^3 M}{\partial \mu^3}
    M^{-1} \frac{\partial M}{\partial \mu} M^{-1} \right)  \\
&& \hspace{-2cm}
+6 {\rm tr} \left( M^{-1} \frac{\partial^2 M}{\partial \mu^2}
    M^{-1} \frac{\partial^2 M}{\partial \mu^2} M^{-1} \right) 
+4 {\rm tr} \left( M^{-1} \frac{\partial M}{\partial \mu}
    M^{-1} \frac{\partial^3 M}{\partial \mu^3} M^{-1} \right) \nonumber \\
&& \hspace{-2cm}
-12 {\rm tr} \left( M^{-1} \frac{\partial^2 M}{\partial \mu^2}
    M^{-1} \frac{\partial M}{\partial \mu} M^{-1} 
    \frac{\partial M}{\partial \mu} M^{-1} \right) \nonumber \\
&& \hspace{-2cm}
-12 {\rm tr} \left( M^{-1} \frac{\partial M}{\partial \mu}
    M^{-1} \frac{\partial^2 M}{\partial \mu^2} M^{-1} 
    \frac{\partial M}{\partial \mu} M^{-1} \right) \nonumber \\
&& \hspace{-2cm}
-12 {\rm tr} \left( M^{-1} \frac{\partial M}{\partial \mu}
    M^{-1} \frac{\partial M}{\partial \mu} M^{-1} 
    \frac{\partial^2 M}{\partial \mu^2} M^{-1} \right) \nonumber \\
&& \hspace{-2cm}
+24 {\rm tr} \left( M^{-1} \frac{\partial M}{\partial \mu}
    M^{-1} \frac{\partial M}{\partial \mu} M^{-1} 
    \frac{\partial M}{\partial \mu} M^{-1} 
    \frac{\partial M}{\partial \mu} M^{-1} \right) \nonumber 
\end{eqnarray}

\noindent
Having defined the explicit representation of ${\cal C}_n$ and ${\cal D}_n$
we now can proceed to define the expansion coefficients for various
thermodynamic quantities discussed in this paper.

\paragraph{Pressure \boldmath $(p)$ \unboldmath :}
The pressure is obtained from the logarithm of the QCD partition function.
Its expansion is defined in Eqs.~(\ref{eq:p}) and (\ref{eq:cn}). The leading
expansion coefficient $c_0$ is given by the pressure calculated at $\mu_q=0$.
All higher order expansion coefficients are given in terms of 
derivatives of $\ln {\cal Z}$.
\begin{eqnarray}
\frac{p}{T^4} \equiv \Omega = 
\frac{1}{VT^3} \ln {\cal Z} = \sum_{n=0}^{\infty} c_n 
\left( \frac{\mu_q}{T} \right)^n \quad ,
\end{eqnarray}
with 
\begin{equation}
c_n = \frac{1}{n! VT^3} \frac{\partial^n \ln {\cal Z}}{\partial (\mu_q/T)^n} 
\biggr|_{\mu=0}
\end{equation}
To generate the expansion we first consider derivatives of $\ln {\cal Z}$
for $\mu \ne 0$. For the first derivative we find
\begin{equation}
\frac{\partial \ln {\cal Z}}{\partial \mu} \equiv {\cal A}_1  = 
\left\langle {\cal D}_1 \right\rangle \quad .
\end{equation}
Higher order derivatives are generated using the relation
\begin{eqnarray}
\frac{\partial {\cal A}_n}{\partial \mu} 
= {\cal A}_{n+1} - {\cal A}_n {\cal A}_1 \quad ,
\label{eq:An1}
\end{eqnarray}
where ${\cal A}_n$ is defined as
\begin{equation}
{\cal A}_n \equiv \left\langle 
\exp\{ -{\cal D}_0 \} \frac{\partial^n \exp\{ {\cal D}_0 \} }{\partial \mu^n}
 \right\rangle \quad .
\label{eq:an}
\end{equation}
With this we can generate higher order derivatives of $\ln {\cal Z}$ 
iteratively using 
\begin{equation}
\partial^{n+1} \ln {\cal Z}/\partial \mu^{n+1} =
\partial^{n} {\cal A}_1 /\partial \mu^{n} \quad .
\label{eq:Zn1}
\end{equation}
Explicitly we find from Eq.~(\ref{eq:an})
\begin{eqnarray}
{\cal A}_2 \hspace{-4mm} &=& \hspace{-4mm} 
\left\langle {\cal D}_2 \right\rangle 
+\left\langle {\cal D}_1^2 \right\rangle, 
\\
{\cal A}_3 \hspace{-4mm} &=& \hspace{-4mm} 
\left\langle {\cal D}_3 \right\rangle 
+3\left\langle {\cal D}_2 {\cal D}_1 \right\rangle 
+\left\langle {\cal D}_1^3 \right\rangle, 
\\
{\cal A}_4 \hspace{-4mm} &=& \hspace{-4mm} 
\left\langle {\cal D}_4 \right\rangle 
+4\left\langle {\cal D}_3 {\cal D}_1 \right\rangle 
+3\left\langle {\cal D}_2^2 \right\rangle 
+6\left\langle {\cal D}_2 {\cal D}_1^2 \right\rangle 
+\left\langle {\cal D}_1^4 \right\rangle, 
\\
{\cal A}_5 \hspace{-4mm} &=& \hspace{-4mm} 
\left\langle {\cal D}_5 \right\rangle 
+5\left\langle {\cal D}_4 {\cal D}_1 \right\rangle 
+10\left\langle {\cal D}_3 {\cal D}_2 \right\rangle 
+10\left\langle {\cal D}_3 {\cal D}_1^2 \right\rangle 
+15\left\langle {\cal D}_2^2 {\cal D}_1 \right\rangle 
+10\left\langle {\cal D}_2 {\cal D}_1^3 \right\rangle 
\nonumber \\ && \hspace{-10mm}
+\left\langle {\cal D}_1^5 \right\rangle, 
\\
{\cal A}_6 \hspace{-4mm} &=& \hspace{-4mm} 
\left\langle {\cal D}_6 \right\rangle 
+6\left\langle {\cal D}_5 {\cal D}_1 \right\rangle 
+15\left\langle {\cal D}_4 {\cal D}_2 \right\rangle 
+10\left\langle {\cal D}_3^2 \right\rangle 
+15\left\langle {\cal D}_4 {\cal D}_1^2 \right\rangle 
+60\left\langle {\cal D}_3 {\cal D}_2 {\cal D}_1 \right\rangle 
\nonumber \\ && \hspace{-5mm}
+15\left\langle {\cal D}_2^3 \right\rangle 
+20\left\langle {\cal D}_3 {\cal D}_1^3 \right\rangle 
+45\left\langle {\cal D}_2^2 {\cal D}_1^2 \right\rangle 
+15\left\langle {\cal D}_2 {\cal D}_1^4 \right\rangle 
+\left\langle {\cal D}_1^6 \right\rangle  \quad .
\end{eqnarray}
From Eq.~(\ref{eq:Zn1}) we then obtain through repeated 
application of Eq.~(\ref{eq:An1}),
\begin{eqnarray}
\frac{\partial \ln {\cal Z}}{\partial \mu} 
\hspace{-4mm} &=& \hspace{-4mm} 
{\cal A}_1 ,
\\
\frac{\partial^2 \ln {\cal Z}}{\partial \mu^2} 
\hspace{-4mm} &=& \hspace{-4mm} 
{\cal A}_2 - {\cal A}_1^2 ,
\\
\frac{\partial^3 \ln {\cal Z}}{\partial \mu^3} 
\hspace{-4mm} &=& \hspace{-4mm} 
{\cal A}_3 -3 {\cal A}_2 {\cal A}_1 +2 {\cal A}_1^3 ,
\\
\frac{\partial^4 \ln {\cal Z}}{\partial \mu^4} 
\hspace{-4mm} &=& \hspace{-4mm} 
{\cal A}_4 -4 {\cal A}_3 {\cal A}_1 -3 {\cal A}_2^2 
+12 {\cal A}_2 {\cal A}_1^2 -6 {\cal A}_1^4 ,
\\
\frac{\partial^5 \ln {\cal Z}}{\partial \mu^5} 
\hspace{-4mm} &=& \hspace{-4mm} 
{\cal A}_5 -5 {\cal A}_4 {\cal A}_1 -10 {\cal A}_3 {\cal A}_2 
+20 {\cal A}_3 {\cal A}_1^2 +30 {\cal A}_2^2 {\cal A}_1 
-60 {\cal A}_2 {\cal A}_1^3 +24 {\cal A}_1^5 ,
\\
\frac{\partial^6 \ln {\cal Z}}{\partial \mu^6} 
\hspace{-4mm} &=& \hspace{-4mm} 
{\cal A}_6 -6 {\cal A}_5 {\cal A}_1 -15 {\cal A}_4 {\cal A}_2 
-10 {\cal A}_3^2 +30 {\cal A}_4 {\cal A}_1^2 
\nonumber \\ && \hspace{-9mm}
+120 {\cal A}_3 {\cal A}_2 {\cal A}_1 
+30 {\cal A}_2^3 -120 {\cal A}_3 {\cal A}_1^3 
-270 {\cal A}_2^2 {\cal A}_1^2 
+360 {\cal A}_2 {\cal A}_1^4 -120 {\cal A}_1^6 \; .
\end{eqnarray}
These relations simplify considerably for $\mu=0$ as all odd
expectation values vanish, {\it i.e.} ${\cal A}_n =0$ for $n$ odd. In fact, 
$\partial^n (\ln \det M)/\partial \mu^n$ 
is strictly real for $n$ even and pure imaginary for $n$ odd. 
Using this property, the odd derivatives of the pressure vanish 
and also the even derivatives become rather simple. This defines
the expansion coefficients $c_n$ introduced in Eqs.~(\ref{eq:p}) and 
(\ref{eq:cn}),
\begin{eqnarray}
c_2 &\equiv& \frac{1}{2}\frac{\partial^2 (p/T^4)}{\partial (\mu_{q}/T)^2}
\biggr|_{\mu_q=0} = 
\frac{1}{2} \frac{N_{\tau}}{N_{\sigma}^3}\; {\cal A}_2 \quad , \nonumber \\
c_4 &\equiv& \frac{1}{4!} \frac{\partial^4 (p/T^4)}{\partial (\mu_{q}/T)^4}
\biggr|_{\mu_q=0} = 
\frac{1}{4!} \frac{1}{N_{\sigma}^3 N_{\tau}} ({\cal A}_4 -3 {\cal A}_2^2) \quad , 
\nonumber \\  
c_6 &\equiv& \frac{1}{6!} \frac{\partial^6 (p/T^4)}{\partial (\mu_{q}/T)^6} 
\biggr|_{\mu_q=0} = 
\frac{1}{6!} \frac{1}{N_{\sigma}^3 N_{\tau}^3} 
({\cal A}_6 -15 {\cal A}_4 {\cal A}_2 +30 {\cal A}_2^3) \quad .
\label{eq:dpmu0}
\end{eqnarray}
Here all expectation values ${\cal A}_n$ are now meant to be evaluated at
$\mu =0$.
\paragraph{
Isovector susceptibility $(\chi_I)$\unboldmath :}
While the Taylor expansion of the quark number susceptibility is easily
obtained from that of the pressure we need to introduce the expansion
of the isovector susceptibility. This has been done in Eq.~(\ref{eq:chiI}). 
More explicitly the isovector  susceptibility is given by
\begin{eqnarray}
\frac{\chi_{I}}{T^2} 
\hspace{-4mm} &=& \hspace{-4mm} 
\frac{N_{\tau}}{N_{\sigma}^3} \left( 
\frac{\partial^2 \ln {\cal Z}}{\partial \bar{\mu}_u^2} 
-\frac{\partial^2 \ln {\cal Z}}{\partial \bar{\mu}_u 
\partial \bar{\mu}_d} 
- \frac{\partial^2 \ln {\cal Z}}{\partial \bar{\mu}_u 
\partial \bar{\mu}_d} 
+ \frac{\partial^2 \ln {\cal Z}}{\partial \bar{\mu}_d^2} 
\right) \nonumber \\
&& \hspace{-2cm}
= \frac{N_{\tau}}{N_{\sigma}^3} \left[ \sum_{f=u,d} \left\langle 
\frac{1}{4} 
\frac{\partial^2 (\ln \det M_{f})}{\partial \bar{\mu}_{f}^2} 
\right\rangle 
+ \left\langle \left( \frac{1}{4} 
\frac{\partial (\ln \det M_u)}{\partial \bar{\mu}_u} 
- \frac{1}{4} \frac{\partial (\ln \det M_d)}
{\partial \bar{\mu}_d} \right)^2 \right\rangle \right.
\nonumber \\
&& \hspace{-5mm}
- \left. \left\langle \frac{1}{4}  
\frac{\partial (\ln \det M_u)}{\partial \bar{\mu}_u} 
- \frac{1}{4} 
\frac{\partial (\ln \det M_d)}{\partial \bar{\mu}_d} 
\right\rangle^2 \right] \quad . 
\end{eqnarray}
Throughout this paper we have considered the case of 
$\bar{\mu}_u = \bar{\mu}_d \equiv \mu_q a \equiv \mu$. The isovector
chemical potential $\mu_I$ has been set equal to zero after appropriate
derivatives have been taken.
In the Taylor expansions of $\chi_I$, which is performed 
at $\mu=0$ in terms of $\mu_q/T$, the derivatives with respect to $\mu_u$
and $\mu_d$ then become identical, {\it i.e.} 
$[\partial^n (\ln \det M_{u(d)})/\partial \bar{\mu}_{u(d)}^n] 
(\bar{\mu}_{u(d)}) \equiv [\partial^n (\ln \det M)/\partial \mu^n] (\mu)$.

The calculation of the isovector susceptibility thus reduces to the
calculation of
\begin{equation}
\frac{\chi_{I}}{T^2} 
= \frac{N_{\tau}}{2N_{\sigma}^3}  \left\langle
\frac{\partial^2 (\ln \det M)}{\partial \mu^2}
\right\rangle \quad .
\end{equation}
To define the expansion of the isovector susceptibility,
$\chi_I$ at fixed $\mu_I=0$ around $\mu_q=0$ we set up an iterative 
scheme similar to that introduced for the pressure.
We introduce the additional kernel ${\cal D}_2$ in Eq.~(\ref{eq:an})
to generate ${\cal B}_n$ for arbitrary $\mu_q \ne 0$,
\begin{equation}
{\cal B}_{n+2} \equiv \left\langle 
\exp\{ -{\cal D}_0 \} 
\frac{\partial^n {\cal D}_2\exp\{ {\cal D}_0 \} }{\partial \mu^n}
 \right\rangle \quad .
\label{eq:bn}
\end{equation}
This yields, 
\begin{eqnarray}
{\cal B}_2 
\hspace{-4mm} &=& \hspace{-4mm} 
\left\langle {\cal D}_2 \right\rangle ,
\\
{\cal B}_3 
\hspace{-4mm} &=& \hspace{-4mm} 
\left\langle {\cal D}_3 \right\rangle 
+ \left\langle {\cal D}_2 {\cal D}_1 \right\rangle ,
\\
{\cal B}_4 
\hspace{-4mm} &=& \hspace{-4mm} 
\left\langle {\cal D}_4 \right\rangle 
+2 \left\langle {\cal D}_3 {\cal D}_1 \right\rangle 
+ \left\langle {\cal D}_2^2 \right\rangle 
+ \left\langle {\cal D}_2 {\cal D}_1^2 \right\rangle ,
\\
{\cal B}_5 
\hspace{-4mm} &=& \hspace{-4mm} 
\left\langle {\cal D}_5 \right\rangle 
+3 \left\langle {\cal D}_4 {\cal D}_1 \right\rangle 
+4 \left\langle {\cal D}_3 {\cal D}_2 \right\rangle 
+3 \left\langle {\cal D}_3 {\cal D}_1^2 \right\rangle 
+3 \left\langle {\cal D}_2^2 {\cal D}_1 \right\rangle 
+ \left\langle {\cal D}_2 {\cal D}_1^3 \right\rangle 
\\
{\cal B}_6 
\hspace{-4mm} &=& \hspace{-4mm} 
\left\langle {\cal D}_6 \right\rangle 
+4 \left\langle {\cal D}_5 {\cal D}_1 \right\rangle 
+7 \left\langle {\cal D}_4 {\cal D}_2 \right\rangle 
+4 \left\langle {\cal D}_3^2 \right\rangle 
+6 \left\langle {\cal D}_4 {\cal D}_1^2 \right\rangle 
+16 \left\langle {\cal D}_3 {\cal D}_2 {\cal D}_1 \right\rangle 
\nonumber \\ && 
+3 \left\langle {\cal D}_2^3 \right\rangle 
+4 \left\langle {\cal D}_3 {\cal D}_1^3 \right\rangle 
+6 \left\langle {\cal D}_2^2 {\cal D}_1^2 \right\rangle 
+ \left\langle {\cal D}_2 {\cal D}_1^4 \right\rangle .
\end{eqnarray}
The derivative of ${\cal B}_n$ with respect to the chemical potential
satisfies,
\begin{equation}
\frac{\partial {\cal B}_n}{\partial \mu} 
= {\cal B}_{n+1} - {\cal B}_n {\cal A}_1 \quad .
\end{equation}
Starting with $(N_{\tau}/N_{\sigma}^3) {\cal B}_2 = \chi_I/T^2$, 
we obtain the equations for 
$\partial^n (\chi_I/T^2) / \partial (\mu_q/T)^n$ iteratively, 
and then use again the CP symmetry  for $\mu=0$, 
i.e. ${\cal A}_n$ and ${\cal B}_n$ are zero for $n$ odd. 
Therefore the odd derivatives in the expansion of $\chi_I/T^2$ 
at $\mu=0$ are zero and
the even derivatives define the expansion coefficients $c_n^I$ used in
Eq.~(\ref{eq:chiI}), 
\begin{eqnarray}
c_2^I &\equiv& \frac{1}{2} \frac{\chi_I}{T^2} 
\biggr|_{\mu_q=0} =
\frac{1}{2} \frac{N_{\tau}}{N_{\sigma}^3} {\cal B}_2 \quad , \nonumber \\ 
c_4^I &\equiv& \frac{1}{4!} \frac{\partial^2 (\chi_I/T^2)}{\partial (\mu_{q}/T)^2} 
\biggr|_{\mu_q=0} =
\frac{1}{4!}\frac{1}{N_{\sigma}^3 N_{\tau}} ({\cal B}_4 - {\cal B}_2 {\cal A}_2) ,
\nonumber \\ 
c_6^I &\equiv& \frac{1}{6!} \frac{\partial^4 (\chi_I/T^2)}{\partial (\mu_{q}/T)^4} 
\biggr|_{\mu_q=0} =
\frac{1}{6!} \frac{1}{N_{\sigma}^3 N_{\tau}^3} 
({\cal B}_6 -6 {\cal B}_4 {\cal A}_2 - {\cal B}_2 {\cal A}_4 
+6 {\cal B}_2 {\cal A}_2^2 ) \quad .
\end{eqnarray}

\paragraph{ Chiral condensate \boldmath $(\langle \bar{\psi} \psi \rangle)$ 
and disconnected chiral susceptibility $(\chi_{\bar{\psi} \psi})$\unboldmath :}
We also discuss the Taylor expansion of the chiral condensate and the related
chiral susceptibility, 
\begin{eqnarray}
\frac{\langle \bar{\psi} \psi \rangle}{T^3} 
\hspace{-4mm} &=& \hspace{-4mm} 
\frac{N_\tau^2}{N_{\sigma}^3 } \frac{\partial \ln {\cal Z}}{\partial m} 
= \frac{N_\tau^2}{N_{\sigma}^3 } \frac{n_{\rm f}}{4} \left\langle
\frac{\partial \ln \det M}{\partial m} \right\rangle 
= \frac{N_\tau^2}{N_{\sigma}^3 } \frac{n_{\rm f}}{4} \left\langle
{\rm tr} M^{-1} \right\rangle \quad ,
\label{eq:chicon} \\
\frac{\chi_{\bar{\psi} \psi}}{T^2} 
\hspace{-4mm} &=& \hspace{-4mm}
\frac{N_\tau}{N_{\sigma}^3 } \left( \frac{n_{\rm f}}{4} \right)^2
\left[ \left\langle \left( 
{\rm tr} M^{-1} \right)^2 \right\rangle 
- \left\langle  {\rm tr} M^{-1} \right\rangle^2 
\right] \quad .
\label{eq:chisusapp} 
\end{eqnarray}
The iterative scheme is similar to that introduced for the isovector
susceptibility but with the generating kernel ${\cal D}_2$ in Eq.~(\ref{eq:bn})
replaced by ${\cal C}_0$. This yields
\begin{eqnarray}
{\cal F}_0 
\hspace{-4mm} &=& \hspace{-4mm} 
\left\langle {\cal C}_0 \right\rangle, \\ 
{\cal F}_1 
\hspace{-4mm} &=& \hspace{-4mm} 
\left\langle {\cal C}_1 \right\rangle 
+ \left\langle {\cal C}_0 {\cal D}_1 \right\rangle, \\ 
{\cal F}_2 
\hspace{-4mm} &=& \hspace{-4mm} 
\left\langle {\cal C}_2 \right\rangle 
+ 2 \left\langle {\cal C}_1 {\cal D}_1 \right\rangle 
+ \left\langle {\cal C}_0 {\cal D}_2 \right\rangle 
+ \left\langle {\cal C}_0 {\cal D}_1^2 \right\rangle, \\ 
{\cal F}_3 
\hspace{-4mm} &=& \hspace{-4mm} 
\left\langle {\cal C}_3 \right\rangle 
+ 3 \left\langle {\cal C}_2 {\cal D}_1 \right\rangle 
+ 3 \left\langle {\cal C}_1 {\cal D}_2 \right\rangle 
+ \left\langle {\cal C}_0 {\cal D}_3 \right\rangle 
+ 3 \left\langle {\cal C}_1 {\cal D}_1^2 \right\rangle 
+ 3 \left\langle {\cal C}_0 {\cal D}_2 {\cal D}_1 \right\rangle 
\nonumber \\ && \hspace{-5mm}
+ \left\langle {\cal C}_0 {\cal D}_1^3 \right\rangle \quad , \\ 
{\cal F}_4 
\hspace{-4mm} &=& \hspace{-4mm} 
\left\langle {\cal C}_4 \right\rangle 
+ 4 \left\langle {\cal C}_3 {\cal D}_1 \right\rangle 
+ 6 \left\langle {\cal C}_2 {\cal D}_2 \right\rangle 
+ 4 \left\langle {\cal C}_1 {\cal D}_3 \right\rangle 
+ \left\langle {\cal C}_0 {\cal D}_4 \right\rangle 
+ 6 \left\langle {\cal C}_2 {\cal D}_1^2 \right\rangle 
+ 12 \left\langle {\cal C}_1 {\cal D}_2 {\cal D}_1 \right\rangle 
\nonumber \\ && \hspace{-5mm}
+ 4 \left\langle {\cal C}_0 {\cal D}_3 {\cal D}_1 \right\rangle 
+ 3 \left\langle {\cal C}_0 {\cal D}_2^2 \right\rangle 
+ 4 \left\langle {\cal C}_1 {\cal D}_1^3 \right\rangle 
+ 6 \left\langle {\cal C}_0 {\cal D}_2 {\cal D}_1^2 \right\rangle 
+ \left\langle {\cal C}_0 {\cal D}_1^4 \right\rangle \quad . 
\end{eqnarray}
These expectation values again satisfy
\begin{equation}
\frac{\partial {\cal F}_n}{\partial \mu} 
= {\cal F}_{n+1} - {\cal F}_n {\cal A}_1 \quad .
\end{equation}
This gives the expansion coefficients $c_n^{\bar{\psi}\psi}$ for the 
chiral condensate defined in Eqs.~(\ref{eq:pbp}) and (\ref{eq:pbpcoef}),
\begin{eqnarray}
c_0^{\bar{\psi}\psi} &\equiv& 
\frac{\left\langle \bar{\psi} \psi \right\rangle}{T^3} 
\biggr|_{\mu_q =0}
= \frac{N_\tau^2}{N_{\sigma}^3 } {\cal F}_0 \quad , \\ 
c_2^{\bar{\psi}\psi} &\equiv& \frac{1}{2}
\frac{\partial^2 \left\langle \bar{\psi} \psi \right\rangle /T^3
}{\partial (\mu_{q}/T)^2}  \biggr|_{\mu_q=0} =
\frac{1}{2} \frac{1}{N_{\tau}^2} \frac{\partial^2 
\left\langle \bar{\psi} \psi \right\rangle/T^3}{\partial \mu^2} \biggr|_{\mu=0}
\nonumber \\
&=& \frac{1}{2}\frac{1}{N_{\sigma}^3 } 
\left( {\cal F}_2 - {\cal F}_0 {\cal A}_2 \right) ,
\\
c_4^{\bar{\psi}\psi} &\equiv& \frac{1}{4!}
\frac{\partial^4 \left\langle \bar{\psi} \psi \right\rangle /T^3
}{\partial (\mu_{q}/T)^4} \biggr|_{\mu_q=0} =
\frac{1}{4!} \frac{1}{N_{\tau}^4} \frac{\partial^4 
\left\langle \bar{\psi} \psi \right\rangle/T^3}{\partial \mu^4} \biggr|_{\mu=0}
\nonumber \\ 
&=& \frac{1}{4!} \frac{1}{N_{\sigma}^3 N_{\tau}^2} 
\left( {\cal F}_4 -6 {\cal F}_2 {\cal A}_2 - {\cal F}_0 {\cal A}_4 
+6 {\cal F}_0 {\cal A}_2^2 \right) ,
\end{eqnarray}
where we used again that
${\cal A}_n$ and ${\cal F}_n$ are zero for $n$ odd ant $\mu =0$.
Hence the odd derivatives in the expansion vanish. Note that these
expansion coefficients also control the quark mass dependence of the
quark number susceptibility given in Eq.~(\ref{eq:chiqm}). The first
derivative of the isovector susceptibility with respect to the quark mass,
\begin{eqnarray}
\frac{\partial (\chi_I/T^2)}{\partial m/T}
= \frac{1}{N_{\sigma}^3} \left[ \left\langle
\frac{n_{\rm f}}{4}
\frac{\partial^3 (\ln \det M)}{\partial \mu^2 \partial m}
\right\rangle
+ \left\langle \left( \frac{n_{\rm f}}{4} \right)^2
\frac{\partial^2 (\ln \det M)}{\partial \mu^2}
{\rm tr} M^{-1} \right\rangle \right.
\nonumber \\ \left.
- \left\langle \frac{n_{\rm f}}{4}
\frac{\partial^2 (\ln \det M)}{\partial \mu^2} \right\rangle
\left\langle  \frac{n_{\rm f}}{4} {\rm tr} M^{-1} \right\rangle \right]
\quad ,
\end{eqnarray}
for which we have introduced the Taylor expansion in Eq.~(\ref{eq:chiIm})
requires the calculation of further expectation values ${\cal I}_n$,
\begin{eqnarray}
{\cal I}_2
\hspace{-4mm} &=& \hspace{-4mm}
\left\langle {\cal C}_2 \right\rangle
+ \left\langle {\cal C}_0 {\cal D}_2 \right\rangle ,
\\
{\cal I}_3
\hspace{-4mm} &=& \hspace{-4mm}
\left\langle {\cal C}_3 \right\rangle
+ \left\langle {\cal C}_2 {\cal D}_1 \right\rangle
+ \left\langle {\cal C}_1 {\cal D}_2 \right\rangle
+ \left\langle {\cal C}_0 {\cal D}_3 \right\rangle
+ \left\langle {\cal C}_0 {\cal D}_2 {\cal D}_1 \right\rangle ,
\\
{\cal I}_4
\hspace{-4mm} &=& \hspace{-4mm}
\left\langle {\cal C}_4 \right\rangle
+2 \left\langle {\cal C}_3 {\cal D}_1 \right\rangle
+2 \left\langle {\cal C}_1 {\cal D}_3 \right\rangle
+2 \left\langle {\cal C}_2 {\cal D}_2 \right\rangle
+ \left\langle {\cal C}_2 {\cal D}_1^2 \right\rangle
+2 \left\langle {\cal C}_1 {\cal D}_2 {\cal D}_1 \right\rangle
\nonumber \\ && \hspace{-4mm}
+\left\langle {\cal C}_0 {\cal D}_4 \right\rangle
+2 \left\langle {\cal C}_0 {\cal D}_3 {\cal D}_1 \right\rangle
+ \left\langle {\cal C}_0 {\cal D}_2^2 \right\rangle
+ \left\langle {\cal C}_0 {\cal D}_2 {\cal D}_1^2 \right\rangle \quad .
\end{eqnarray} 
With these and the coefficients ${\cal B}_n$ and ${\cal F}_n$ 
one finds for the expansion coefficients,
\begin{eqnarray}
c_2^{I,\bar{\psi}\psi} &=& \frac{1}{2}
\frac{\partial (\chi_I/T^2)}{\partial m/T} \biggr|_{\mu=0}
= \frac{1}{2} \frac{1}{N_{\sigma}^3} ({\cal I}_2 - {\cal B}_2 {\cal F}_0 )
\quad ,
\nonumber \\
c_4^{I,\bar{\psi}\psi} &=& \frac{1}{4!}
\frac{\partial^3 (\chi_I/T^2)}{\partial (\mu_{q}/T)^2 \partial m/T}
\biggr|_{\mu=0
}
\nonumber \\
&=&  \frac{1}{4!}
\frac{1}{N_{\sigma}^3 N_{\tau}^2}
({\cal I}_4 - {\cal I}_2 {\cal A}_2
- {\cal B}_4 {\cal F}_0 - {\cal B}_2 {\cal F}_2
+2 {\cal B}_2 {\cal F}_0 {\cal A}_2 ) \quad .
\end{eqnarray}
\noindent
Finally we present the expansion of $\chi_{\bar{\psi} \psi}$ which is 
generated in analogy to the isovector susceptibility by 
using ${\cal C}_0^2$ as a kernel in Eq.~(\ref{eq:an}). We find
\begin{eqnarray}
{\cal G}_0 
\hspace{-4mm} &=& \hspace{-4mm} 
\left\langle {\cal C}_0^2 \right\rangle, \\ 
{\cal G}_1 
\hspace{-5mm} &=& \hspace{-5mm} 
2 \left\langle {\cal C}_1 {\cal C}_0 \right\rangle 
+ \left\langle {\cal C}_0^2 {\cal D}_1 \right\rangle, \\ 
{\cal G}_2 
\hspace{-4mm} &=& \hspace{-4mm} 
2 \left\langle {\cal C}_2 {\cal C}_0 \right\rangle 
+ 2 \left\langle {\cal C}_1^2 \right\rangle 
+ 4 \left\langle {\cal C}_1 {\cal C}_0 {\cal D}_1 \right\rangle 
+ \left\langle {\cal C}_0^2 {\cal D}_2 \right\rangle 
+ \left\langle {\cal C}_0^2 {\cal D}_1^2 \right\rangle, \\ 
{\cal G}_3 
\hspace{-4mm} &=& \hspace{-4mm} 
2 \left\langle {\cal C}_3 {\cal C}_0 \right\rangle 
+ 6 \left\langle {\cal C}_2 {\cal C}_1 \right\rangle 
+ 6 \left\langle {\cal C}_2 {\cal C}_0 {\cal D}_1 \right\rangle 
+ 6 \left\langle {\cal C}_1^2 {\cal D}_1 \right\rangle 
+ 6 \left\langle {\cal C}_1 {\cal C}_0 {\cal D}_2 \right\rangle 
+ \left\langle {\cal C}_0^2 {\cal D}_3 \right\rangle 
\nonumber \\ && \hspace{-5mm}
+ 6 \left\langle {\cal C}_1 {\cal C}_0 {\cal D}_1^2 \right\rangle 
+ 3 \left\langle {\cal C}_0^2 {\cal D}_2 {\cal D}_1 \right\rangle 
+ \left\langle {\cal C}_0^2 {\cal D}_1^3 \right\rangle, \\ 
{\cal G}_4 
\hspace{-4mm} &=& \hspace{-4mm} 
2 \left\langle {\cal C}_4 {\cal C}_0 \right\rangle 
+ 8 \left\langle {\cal C}_3 {\cal C}_1 \right\rangle 
+ 6 \left\langle {\cal C}_2^2 \right\rangle 
+ 8 \left\langle {\cal C}_3 {\cal C}_0 {\cal D}_1 \right\rangle 
+ 24 \left\langle {\cal C}_2 {\cal C}_1 {\cal D}_1 \right\rangle 
\nonumber \\ && \hspace{-5mm}
+ 12 \left\langle {\cal C}_2 {\cal C}_0 {\cal D}_2 \right\rangle 
+ 12 \left\langle {\cal C}_1^2 {\cal D}_2 \right\rangle 
+ 8 \left\langle {\cal C}_1 {\cal C}_0 {\cal D}_3 \right\rangle 
+ \left\langle {\cal C}_0^2 {\cal D}_4 \right\rangle 
+ 12 \left\langle {\cal C}_2 {\cal C}_0 {\cal D}_1^2 \right\rangle 
\nonumber \\ && \hspace{-5mm}
+ 12 \left\langle {\cal C}_1^2 {\cal D}_1^2 \right\rangle 
+ 24 \left\langle {\cal C}_1 {\cal C}_0 {\cal D}_2 {\cal D}_1 \right\rangle 
+ 4 \left\langle {\cal C}_0^2 {\cal D}_3 {\cal D}_1 \right\rangle 
+ 3 \left\langle {\cal C}_0^2 {\cal D}_2^2 \right\rangle 
+ 8 \left\langle {\cal C}_1 {\cal C}_0 {\cal D}_1^3 \right\rangle 
\nonumber \\ && \hspace{-5mm}
+ 6 \left\langle {\cal C}_0^2 {\cal D}_2 {\cal D}_1^2 \right\rangle 
+ \left\langle {\cal C}_0^2 {\cal D}_1^4 \right\rangle. 
\end{eqnarray}
These expectation values satisfy,
\begin{eqnarray}
\frac{\partial {\cal G}_n}{\partial \mu} 
= {\cal G}_{n+1} - {\cal G}_n {\cal A}_1 \quad .
\end{eqnarray}
The expansion of the disconnected chiral susceptibility defined in 
Eq.~(\ref{eq:chisus}) is then given by, 
\begin{eqnarray}
c_0^{\chi} &\equiv& \frac{\chi_{\bar{\psi} \psi}}{T^2} \biggr|_{\mu=0}
= \frac{N_\tau}{N_{\sigma}^3} \left({\cal G}_0 - {\cal F}_0^2 \right), 
\\
c_2^{\chi} &\equiv& \frac{1}{2}
\frac{\partial^2 \chi_{\bar{\psi} \psi}/T^2}{\partial (\mu_{q}/T)^2} 
\biggr|_{\mu=0}
= \frac{1}{2N_{\sigma}^3 N_{\tau}} \left[ 
{\cal G}_2  - {\cal G}_0 {\cal A}_2 
-2 \left( {\cal F}_2 - {\cal F}_0 {\cal A}_2 \right) {\cal F}_0 
\right] ,
\\
c_4^{\chi} &\equiv& \frac{1}{4!}
\frac{\partial^4 \chi_{\bar{\psi} \psi}/T^2}{\partial (\mu_{q}/T)^4} 
\biggr|_{\mu=0}
\nonumber \\
&=& \frac{1}{4!\; N_{\sigma}^3 N_{\tau}^3} \left[ 
{\cal G}_4 -6 {\cal G}_2 {\cal A}_2 - {\cal G}_0 {\cal A}_4 
+6 {\cal G}_0 {\cal A}_2^2 
-6 \left( {\cal F}_2 - {\cal F}_0 {\cal A}_2 \right)^2 
\right. 
\nonumber \\ 
&& \hspace{15mm} \left.
-2 \left({\cal F}_4 -6 {\cal F}_2 {\cal A}_2 - {\cal F}_0 {\cal A}_4 
+6 {\cal F}_0 {\cal A}_2^2 \right) {\cal F}_0 
\right] .
\end{eqnarray}

\end{document}